\pdfoutput=1
\documentclass[a4paper]{article}
\usepackage{jheppub}
\usepackage{dsfont}
\usepackage{amsmath,amssymb,amscd,amsfonts,mathtools,graphicx}

\usepackage[toc,page]{appendix}
\usepackage{epsfig}
\usepackage{epstopdf}
\usepackage{latexsym}
\usepackage{enumerate} 
\usepackage{placeins}
\usepackage{floatrow}
\usepackage{caption}
\usepackage{booktabs}
\usepackage{bbm}
\usepackage{color}
\usepackage{physics}
\usepackage{tensor}
\usepackage{tikz}
\usepackage{cancel}
\usetikzlibrary{matrix}
\usetikzlibrary{decorations.markings,calc,shapes,decorations.pathmorphing,patterns,decorations.pathreplacing,arrows.meta}
\usetikzlibrary{positioning}
\usepackage{hyperref}
\usepackage{subcaption}
\usepackage{tabularx}
\newcolumntype{Y}{>{\centering\arraybackslash}X}

\newcommand{\be}{\begin{equation}}
	\newcommand{\ee}{\end{equation}}
\newcommand{\bea}{\begin{eqnarray}}
	\newcommand{\eea}{\end{eqnarray}}

\usepackage{soul}

\definecolor{mycolor}{RGB}{255,238,140}

\newcommand{\Rconf}{r_{\text{c}}}
\newcommand{\mut}{\tilde \mu}

\newcommand{\zs}{\zeta_*}

\newcommand{\parent}[1]{\left(#1\right)}
\newcommand{\rt}{\varrho}
\newcommand{\LL}{\mathcal{L}}
\newcommand{\NN}{\mathcal{N}}
\newcommand{\htt}{\mathbf{h}}
\newcommand{\eUV}{\varepsilon}
\newcommand{\zUV}{\zeta_{\text{\tiny UV}}}
\newcommand{\eKS}{\epsilon}
\newcommand{\Lrec}{L}

\newcommand{\Lads}{\ell}

\newcommand{\mt}[1]{{\text{\scriptsize{#1}}}}

\def\cUV{c_{\mt{UV}}}
\def\CLM{\mathcal{C}_\mt{LM}}

\def\pIR{\phi_\mt{IR}}

\def\rhoIR{\rho_*}
\def\Sct{S_\mt{ct}}
\def\Lct{{\cal L}_\mt{ct}}

\DeclareMathOperator\vol{Vol}

\begin{document}

	\begin{flushright}HIP-2025-16/TH\end{flushright}
	
	\title{On entanglement c-functions \\ in confining gauge field theories}
	
	\author[a,b]{Niko Jokela,}
	\affiliation[a]{Department of Physics, University of Helsinki, P.O. Box 64, FI-00014, University of Helsinki, Finland}
	\affiliation[b]{Helsinki Institute of Physics, P.O. Box 64, FIN-00014 University of Helsinki, Finland}
	\emailAdd{niko.jokela@helsinki.fi}
	\author[c]{Jani Kastikainen,}
	\affiliation[c]{Institute for Theoretical Physics and Astrophysics and W\"urzburg-Dresden Cluster of Excellence
		ct.qmat, Julius-Maximilians-Universit\"at W\"urzburg, Am Hubland, 97074 W\"urzburg, Germany}
	\emailAdd{jani.kastikainen@uni-wuerzburg.de}
	\author[d]{Carlos Nunez,}
	\affiliation[d]{Department of Physics, Swansea University, Swansea SA2 8PP, United Kingdom}
	\emailAdd{c.nunez@swansea.ac.uk}
	\author[e]{Jos\'e Manuel Pen\'in,}
	\affiliation[e]{INFN, Sezione di Firenze; Via G. Sansone 1; I-50019 Sesto Fiorentino (Firenze), Italy}
	\emailAdd{jmanpen@gmail.com} 
	\author[a,b]{\\ Helime Ruotsalainen,}
	\emailAdd{helime.ruotsalainen@helsinki.fi}
	\author[f]{and Javier G. Subils}
	\affiliation[f]{Institute for Theoretical Physics, Utrecht University, 3584 CC Utrecht, The Netherlands}
	\emailAdd{j.gomezsubils@uu.nl}
	
	\abstract{Entanglement entropy has proven to be a powerful tool for probing renormalization group (RG) flows in quantum field theories, with c-functions derived from it serving as candidate measures of the effective number of degrees of freedom. While the monotonicity of such c-functions is well established in many settings, notable exceptions occur in theories with a mass scale. In this work, we investigate entanglement c-functions in the context of holographic RG flows, with a particular focus on flows across dimensions induced by circle compactifications. We argue that in spacetime dimensions $d \geq 4$, standard constructions of c-functions, which rely on higher derivatives of the entanglement entropy of either a ball or a cylinder, generically lead to non-monotonic behavior. Working with known dual geometries, we argue that the non-monotonicity stems not from any pathology or curvature singularity, but from a transition in the holographic Ryu--Takayanagi surface. In compactifications from four to three dimensions, we propose a modified construction that restores monotonicity in the infrared, although a fully monotonic ultraviolet extension remains elusive. Furthermore, motivated by entanglement entropy inequalities, we conjecture a bound on the cylinder entanglement c-function, which holds in all our examples.}
	
	\dedicated{Dedicated to the memory of Umut G\"ursoy}
	
	\maketitle
	\flushbottom
	\setcounter{page}{2}
	\newpage
	
	\section{Introduction}\label{sec:introduction}
	
	In the study of confinement in quantum chromodynamics (QCD)-like gauge theories at large-$N$, the holographic entanglement entropy (EE) has emerged as a powerful tool to probe the underlying dynamics \cite{Nishioka:2006gr,Klebanov:2007ws}. Previous investigations using slab-shaped entangling regions have demonstrated that at large scales, the (finite part of the) EE saturates due to a finite correlation length \cite{Jokela:2020wgs}. Partial evidence for this saturation has been obtained through direct lattice computations of R\'enyi entropies in SU($N$) Yang--Mills theories in four dimensions \cite{Nakagawa:2009jk,Rabenstein:2018bri,Rindlisbacher:2022bhe}. Nevertheless, some exotic confining holographic field theories do not seem to conform with this saturation notion \cite{Kol:2014nqa} and it is justified to ask if curvature singularities of the dual geometry could play a role \cite{Barbosa:2024pyn}.
	
	This paper instead aims to extend the analysis of slab-shaped entangling regions by considering regions where both the volume and the interface area with the complementary region grow. Unlike slabs, where the area remains constant as the volume increases, these new regions provide a more nuanced understanding of EE in the context of confinement. We explore whether definitive statements about EE within theories having a mass gap, possibly also confining, can still be made under these conditions. 
	
	Particularly promising entanglement measures are \emph{entanglement c-functions} in various dimensions. These c-functions are non-uniquely defined, but they can be constructed to be free of renormalization scheme ambiguities and to match with the central charge (the type A Weyl anomaly coefficient) at the conformal fixed point of the underlying quantum field theory (QFT).
	Any such non-increasing function along renormalization group flows toward the infrared can serve as a measure of the effective number of degrees of freedom at a given energy scale. For scalable entangling regions, balls with radii $R$, in a QFT in flat space with conformal fixed points, the proposal by Liu and Mezei~\cite{Liu:2012eea,Liu:2013una} (LM) for a candidate c-function, $\CLM$ hereafter, has received notable attention. In particular, it is noteworthy that $\CLM$ has been proven to be monotonic using strong subadditivity of EE for intervals \cite{Casini:2004bw} and for disks \cite{Casini:2012ei,Casini:2015woa,Casini:2017vbe} in ambient ($1+1$)- and ($2+1$)-dimensional QFTs, respectively. In spacetime dimensions higher than four, monotonicity is not guaranteed by strong subadditivity \cite{Casini:2017vbe}, and in fact, there are holographic counterexamples \cite{Liu:2012eea}. Analogous c-functions constructed from entanglement entropies of slabs in spacetimes of dimension $d\geq 3$ (also called strips in $d = 3$) have been shown to be monotonic in \cite{Hirata:2006jx,Myers:2012ed}. 
	For a discussion of challenges in identifying suitable candidate c-functions, see~\cite{Jensen:2015swa}.
	
	In this work we restrict our analysis on entanglement c-functions to two type of scenarios that lead to confining infrared (IR) dynamics. In the first type of scenario, the RG flow is initiated by a deformation of an ultraviolet (UV) theory leading to a gapped theory of the same dimension in the IR. Examples in $(3+1)$ dimensions are the Klebanov--Strassler (KS) model \cite{Klebanov:2000hb}, which can be thought of as a `quasi-marginal' deformation of the Klebanov--Witten theory \cite{Klebanov:1998hh}, and the Girardello--Petrini--Porrati--Zaffaroni (GPPZ) \cite{Girardello:1999bd} model, in which case the UV CFT is deformed by a relevant operator.
	The characteristic property of the holographic dual geometries is the shrinking of an internal compact cycle to zero size deep in the bulk: the geometry ends at a finite distance measured by the gap. The KS solution is technically gapless due to the presence of a massless Goldstone mode linked to a broken global symmetry, as detailed in \cite{Gubser:2004qj}. However, for our purposes, the crucial aspect is the existence of an IR scale that truncates the flow of information, as discussed in \cite{Jokela:2020wgs}.
	
	The LM c-function in the $(3+1)$-dimensional GPPZ model was already studied in \cite{Liu:2012eea}, where it was found to be non-monotonic (as a function of the ball radius $R$) and unphysical as a measure of number of degrees of freedom: starting from the UV fixed point, it decreases monotonically until it crosses zero and turns around to approach zero again from below in the IR. The mass gap manifests in the entanglement entropy as a transition of the Ryu--Takayanagi (RT) surface between two topologically distinct configurations. The second-order nature of this transition makes the c-function reach large negative values at the critical point. One explanation for this pathological behavior, proposed in \cite{Liu:2012eea}, is that the GPPZ solution is singular, a feature that also persists in its embedding into ten-dimensional type IIB supergravity \cite{Petrini:2018pjk,Bobev:2018eer}. A regular top-down solution that avoids this issue is provided by the KS solution.
	
	Previous studies \cite{Klebanov:2012yf,Klebanov:2007ws} of the KS model focused on slab-shaped entangling regions and their associated c-functions \cite{Hirata:2006jx,Myers:2012ed}. In this work, we extend the analysis to ball-shaped regions and the LM c-function. We will find that the LM c-function is again non-monotonic, exhibiting qualitatively the same behavior as in the GPPZ case: there is a turning point at large negative values due to a second-order phase transition in the entanglement entropy.\footnote{The main difference from the GPPZ case is that the KS solution does not describe a UV fixed point. As a result, the c-function diverges to positive values in the UV, whereas in the GPPZ case, it asymptotes to the type A Weyl anomaly constant. This divergence can be interpreted as accounting for an infinite number of degrees of freedom arising in the UV theory.} This shifts the focus away from IR singularities in the dual geometry as the primary source of non-monotonic behavior, instead highlighting the phase transition itself as the key factor. Interestingly, analogous KS-like theories in $(2+1)$ dimensions \cite{Cvetic:2000db,Faedo:2017fbv} also feature such a transition, yet still yield well-behaved, monotonic, albeit non-differentiable, LM c-functions \cite{Klebanov:2012yf,Jokela:2020wgs}, in agreement with the field theory monotonicity theorem \cite{Casini:2012ei,Casini:2015woa,Casini:2017vbe}.
	
	The second kind of scenario involves gapped theories arising from circle compactifications and has been studied holographically in various works \cite{Pakman:2008ui,Ishihara:2012jg,Fujita:2023bdk,Fujita:2024tzc}. For analysis of entanglement entropy and entanglement c-functions for flows induced by compactifications on higher-dimensional spaces, see~\cite{GonzalezLezcano:2022mcd,Deddo:2022wxj,de-la-Cruz-Moreno:2023mew,Deddo:2024rde}. The presence of a compact spatial direction complicates the definition of the LM c-function, since a $(d-1)$ dimensional solid ball with an $S^{d-2}$ entangling surface can wrap around the circle and intersect itself when its radius becomes sufficiently large \cite{Deddo:2022wxj}. A more appropriate entangling region, one that preserves the symmetries of the compactified Cauchy slice, is a solid $(d-2)$-dimensional ball of radius $R$ times the circle; topologically, its entangling surface is $S^{d-3} \times S^1$. As shown in \cite{Ishihara:2012jg}, a modified version of the LM c-function exists for this compactified setting: the Ishihara--Lin--Ning (ILN) \mbox{c-function}. It is free from renormalization scheme ambiguities, and its RG flow is now governed by the radius of the lower-dimensional ball.
	
	Along a compactified flow from $3+1$ to $2+1$ dimensions, the ILN c-function approaches zero from negative values in the IR \cite{Ishihara:2012jg,Fujita:2023bdk}. 
	In this paper we will demostrate that that the non-monotonicity of the ILN c-function persists even in IR-regular top-down constructions. In particular, we focus on a family of  4d ${\cal N}=1$ superconformal field theories (SCFT) compactified on a circle (with a twist to preserve four supercharges). The dual QFT is effectively a confining $(2+1)$-dimensional at low energies, growing an extra dimension as we move towards the UV \cite{Chatzis:2024top,Chatzis:2024kdu}.
	
	Because these theories are effectively three-dimensional in the IR, we will construct an ``IR-adapted'' entanglement c-function by applying the three-dimensional LM operator for a disk to the entropy of the solid torus in the compactified four-dimensional theory. This approach is justified by the fact that, at large radii (or equivalently, small compactification circles) the solid torus effectively resembles a disk in the three-dimensional IR theory. This c-function depends on the renormalization scheme and thus involves an arbitrary reference scale. Still, in our top-down models, we find that with a suitable choice of this scale, the function approaches zero from above in the IR and remains monotonic deep into the UV, even across the phase transition. However, for any choice of reference scale, monotonicity is eventually violated close to the UV.
	
	In light of the aforementioned issues with entanglement c-functions, and to complement our analysis, we turn to the so-called holographic (or flow) c-functions, originally introduced in \cite{Macpherson:2014eza, Bea:2015fja, Merrikin:2022yho}, as alternative measures of degrees of freedom. Unlike entanglement c-functions, their definition relies explicitly on the existence of a holographic dual geometry and cannot be defined purely using field theory observables. However, they satisfy desired properties of reducing to universal CFT data (type A Weyl anomaly coefficient) at fixed points, and in many situations, they can be shown to be monotonic under bulk energy conditions. The holographic c-functions can be defined and computed in both of the two above scenarios involving gapped field theories, and we find that they are monotonic in all top-down examples we consider, including flows across dimensions.
	
	The paper is organized as follows. We begin by surveying entanglement inequalities and c-functions in quantum field theories in Sec.~\ref{sec:liu-mezei}. In Sec.~\ref{sec:3d-non-monotonic}, we consider entanglement entropy of ball-shaped regions in holographic CFTs dual to various supersymmetry-preserving gravity backgrounds and compute corresponding type A and B Weyl anomaly coefficients. In Sec.~\ref{sec:flow}, we use a top-down holographic background to study various c-functions in a gapped field theory obtained via circle compactification. In Sec.~\ref{sec:4d-non-monotonic}, we return to the first scenario in four spacetime dimensions and calculate c-functions in the GPPZ and KS models. We conclude with a discussion in Sec.~\ref{sec:discussion}. A review of entanglement c-theorems for strips and slabs is presented for completeness in Appendix \ref{app:strips}. In Appendix~\ref{app:Weyl_conventions} we briefly comment on our conventions in making contact with the Weyl anomaly. In Appendix \ref{app:cylinders}, a bottom-up holographic model illustrating non-monotonicity of cylinder c-functions is presented. Details regarding various supergravity backgrounds dual to gapped field theories are relegated to Appendices \ref{app:confining} and \ref{app:quiver-geometry}. 
	
	\section{Survey of entanglement c-functions}\label{sec:liu-mezei}
	
	In this section, we first review inequalities satisfied by entanglement entropies of ball-shaped subregions presented in \cite{Casini:2017vbe}. These generalize previous inequalities \cite{Casini:2004bw,Casini:2012ei,Casini:2015woa} valid in $d=2$ and $d = 3$ to higher dimensions. Then we present the construction of c-functions from entanglement entropies of balls and cylinders using the Liu--Mezei (LM) \cite{Liu:2012eea,Liu:2013una} and the Ishihara--Lin--Ning (ILN) \cite{Ishihara:2012jg} prescriptions respectively. The entropy inequalities of \cite{Casini:2017vbe} imply monotonicity of the LM c-function in $d = 2$ and $d = 3$ along the RG flow. In $d = 4$, monotonicity is not implied, however, a constraint on the type A Weyl anomaly coefficients at the fixed points arises (the a-theorem). Motivated by the inequalities for balls, we conjecture an analogous inequality for the entanglement entropy of a cylinder. This imposes a novel inequality on the ILN c-function along the flow together with a constraint on the type B Weyl anomaly coefficients at fixed points. For completeness, a review of similar inequalities for strips and slabs is given in Appendix~\ref{app:strips}.
	
	\subsection{Entanglement entropy inequalities in QFT}\label{subsec:ent_inequalities}
	
	Let us consider the action
	\begin{equation}
		I_{\lambda} = I_0 + \lambda\int \dd^{d}x\,\mathcal{O}(x) \ ,
	\end{equation}
	where $I_0$ is the action of a CFT and $\mathcal{O}(x)$ is a relevant operator with scaling dimension $\Delta < d$ and $\lambda$ a coupling constant. Let $\ket{\lambda}$ be the ground state of the theory $I_\lambda$ prepared by the Euclidean path integral over the half-space $\mathbb{R}^{d-1}\times \mathbb{R}_-$ where Euclidean time runs over $\mathbb{R}_-$. Consider a spacelike subregion $\mathcal{A}$ of dimension $d-2$ and let $\rho_{\lambda}$ be its reduced density matrix in the state $\ket{\lambda}$. Usually in practice, one takes the subregion $\mathcal{A}\subset \mathbb{R}^{d-2}$ to lie on a constant-time slice $\mathbb{R}^{d-2}$, but this is not necessary. The entanglement entropy of the subregion is then defined by
	\begin{equation}
		S(\mathcal{A}) = -\tr{(\rho_{\lambda}\log{\rho_{\lambda}})}
	\end{equation}
	such that $S(\mathcal{A})\vert_{\lambda = 0}\equiv S_0(\mathcal{A})$. In general, entanglement entropy is a UV divergent quantity in quantum field theory (see for example the CFT formulae below).
	
	Given another arbitrary spacelike subregion $\mathcal{B}$, the entanglement entropy satisfies the strong subadditivity inequality \cite{Lieb:1973cp}
	\begin{equation}
		S(\mathcal{A})+S(\mathcal{B})\geq S(\mathcal{A}\cap \mathcal{B})+S(\mathcal{A}\cup \mathcal{B})\,.
		\label{eq:SSA}
	\end{equation}
	Consider now the UV fixed point $\lambda = 0$ and assume that $\mathcal{B}$ is such that its boundary (entangling surface) lies on the light-cone of the boundary of $\mathcal{A}$. In this special case, the inequality \eqref{eq:SSA} is saturated \cite{Casini:2017roe}
	\begin{equation}
		S_0(\mathcal{A})+S_0(\mathcal{B})- S_0(\mathcal{A}\cap \mathcal{B})-S_0(\mathcal{A}\cup \mathcal{B}) = 0\,,
		\label{eq:SSA_saturated}
	\end{equation}
	which is known as the Markov property of the CFT vacuum. Therefore it is useful to define the UV subtracted entropy \cite{Casini:2017vbe}
	\begin{equation}
		\Delta S(\mathcal{A}) \equiv S(\mathcal{A})-S_0(\mathcal{A})\,,
		\label{eq:DeltaS_A}
	\end{equation}
	which measures the difference in entanglement entropy of the subregion $\mathcal{A}$ evaluated in the ground state of the deformed CFT and the vacuum of the UV CFT. When $\partial \mathcal{B}$ lies on the light-cone of $\partial\mathcal{A}$, \eqref{eq:DeltaS_A} also satisfies the strong subadditivity \eqref{eq:SSA} due to the Markov property \eqref{eq:SSA_saturated}. The subtracted entropy is finite as long as the scaling dimension satisfies $\Delta < (d+2)\slash 2< d$ in $d\geq 2$ \cite{Casini:2017vbe}.
	
	\paragraph{Solid balls.} In this work, we are mainly interested in subregions which are solid $d-1$-balls $\mathcal{A} = B_{d-1}(R)$ of radius $R$ (with $S^{d-2}_R$ boundary) whose entanglement entropy is denoted by
	\begin{equation}
		S(R) \equiv S(B_{d-1}(R))
	\end{equation}
	and the vacuum subtracted entropy \eqref{eq:DeltaS_A} by $\Delta S(R)$. Using the Markov property \eqref{eq:SSA_saturated} together with strong subadditivity \eqref{eq:SSA}, one can prove the inequality \cite{Casini:2017vbe}
	\begin{equation}
		R\,\Delta S''(R)-(d-3)\,\Delta S'(R)\leq 0\, ,
		\label{eq:Markov_inequality}
	\end{equation}
	where the prime denotes differentiation with respect to $R$.
	We define the quantity
	\begin{equation}
		C(R) \equiv R\,S'(R)-(d-2)\, S(R) = \begin{cases}
			R\,S'(R)\,,\quad &d=2\\
			R\,S'(R) -S(R)\,,\quad &d=3\\
			R\,S'(R)-2 S(R)\,,\quad &d=4
		\end{cases}
		\label{eq:capital_C_function}
	\end{equation}
	and the difference
	\begin{equation}
		\Delta C(R) \equiv C(R)-C_0(R) =R\,\Delta S'(R)-(d-2)\,\Delta S(R)\ ,
	\end{equation}
	where $C_0(R)\equiv C(R)\vert_{\lambda=0}$. Then the inequality \eqref{eq:Markov_inequality} may be written as
	\begin{equation}
		\Delta C'(R)\leq 0
		\label{eq:DeltaC_inequality}
	\end{equation}
	so that $\Delta C(R)$ is a monotonically decreasing function of $R$.
	
	Let us introduce a short-distance (UV) cutoff $\varepsilon$ with a dimension of length that can be understood to be the lattice spacing of the underlying discrete theory when it exists. In $d=2$, a ball of radius $R$ on the spatial slice is simply an interval of length $2R$ whose entanglement entropy in the vacuum state of the UV CFT is logarithmically divergent
	\begin{equation}
		S_0(R) = \frac{\cUV}{3}\log{\frac{2R}{\eUV}} + \mathcal{O}(1)\ ,\quad \varepsilon\rightarrow 0\ , \quad d = 2 \ ,
		\label{eq:ball_EE_2D}
	\end{equation}
	which gives $C_0(R) =\frac{c_{\mt{UV}}}{3}$. 
	Therefore, $\Delta C'(R) = C'(R)  \leq 0$ and $C(R) = R\,S'(R)$ decreases monotonically in $d=2$, which is Zamolodchikov's c-theorem \cite{Zamolodchikov:1986gt}. The finite term $\mathcal{O}(1)$ is dependent on the regularization scheme, but in the scheme where one cuts out disks of radius $\varepsilon$ around the entangling points, it coincides with the boundary entropy \cite{Affleck:1991tk} of the CFT in the presence of two boundaries \cite{Cardy:2016fqc}. In this regularization scheme, $\varepsilon$ has an interpretation as the lattice spacing of the underlying lattice theory.
	
	In $d=3$, the entanglement entropy of a disk in the vacuum state of the UV CFT is given by
	\begin{equation}\label{eq:diskEE}
		S_0(R) = \frac{R}{\eUV} - F_{\text{UV}} + \mathcal{O}(\varepsilon)\,,\quad \varepsilon\rightarrow 0\,,\quad d = 3\,,
	\end{equation}
	where $F_{\text{UV}} $ is the Euclidean partition function of the UV CFT on a round $S^3$ and it is positive if the theory is unitary. It follows that $C_0(R) = F_{\text{UV}}$ and $\Delta C'(R) = C'(R)  \leq 0$ and $C(R) = R\,S(R)-S(R)$ decreases monotonically in $d=3$.
	
	The situation becomes more complicated in $d =4$ where \cite{Solodukhin:2008dh}
	\begin{equation}
		S_{0}(R) = \frac{R^{2}}{\eUV^{2}} - a_{\text{UV}}\log{\frac{R}{\eUV}} + \mathcal{O}(1)\,,\quad \varepsilon\rightarrow 0\,,\quad d = 4 \,,
		\label{eq:d_4_EE}
	\end{equation}
	where $a_{\text{UV}}$ is the coefficient of the type A Weyl anomaly of the effective action of the CFT when it is coupled to a curved background metric (see Appendix \ref{app:Weyl_conventions} for our conventions). In this case, $ C_0'(R) = \frac{2a_{\text{UV}}}{R}$ is not a constant as a function of $R$, but it is UV finite. Therefore the inequality \eqref{eq:DeltaC_inequality} becomes
	\begin{equation}
		C'(R)\leq \frac{2a_{\text{UV}}}{R}\,,
		\label{eq:d_4_C_prime_inequality}
	\end{equation}
	which does not imply the monotonicity of $C(R)$ as in lower dimensions. However, it does imply the weaker a-theorem $a_{\text{IR}}\leq a_{\text{UV}}$ when the deformed theory flows to an IR fixed point CFT for which the coefficient of the logarithmic divergence in the entanglement entropy \eqref{eq:d_4_EE} equals $a_{\text{IR}}$ \cite{Casini:2017vbe}.
	
	In $d = 2, 3$, we see that the inequality \eqref{eq:Markov_inequality} is also obeyed by the entropy $S(R)$ itself. However, this is no longer true in $d = 4$, where $R\, S''(R)-S'(R) = \frac{2a_{\text{UV}}}{R} \geq 0$, because the type A anomaly coefficient is always positive \cite{Casini:2012ei,Hofman:2008ar}. The reason behind this is that in proving the inequality \eqref{eq:Markov_inequality}, strong subadditivity is applied to boosted balls that contain cusps. These cusps produce additional UV divergences in dimensions $d\geq 4$ that violate the inequality for $S(R)$, but cancel in $\Delta S(R)$ leading to \eqref{eq:Markov_inequality} (see \cite{Casini:2012ei,Casini:2017vbe} for details).
	
	\paragraph{Solid cylinders.} In dimensions $d\geq 3$, a finite solid cylinder of radius $R$ and length $2L$ is defined as the region $B_{d-1}(R)\times (-L,L)\subset \mathbb{R}^{d-1}$ of the spatial slice. The implications of strong subadditivity for the entanglement entropy of such a cylinder have not been studied so far in the literature in dimensions $d\geq 4$. In $d = 3$, a cylinder is simply a strip $(-L,L)\times (-R,R)$ which satisfies an inequality similar to the interval in $d = 2$ \cite{Hirata:2006jx,Myers:2012ed} (see Appendix \ref{app:strips} for a review). Instead of attempting to prove an inequality in full using strong subadditivity in general dimensions, we will focus on $d = 4$, and merely study implications of inequalities of the form
	\begin{equation}
		(R^2\partial_R^2+s_1R\,\partial_R+s_0)\,S^{\text{cyl}}(R)\leq 0\,,
		\label{eq:cylinder_guess}
	\end{equation}
	where $S^{\text{cyl}}(R)$ is the entanglement entropy of the cylinder and $s_{0,1}$ are coefficients to be determined. Derivatives higher than second-order are not expected to appear from strong subadditivity.
	
	To fix the coefficients, we should first consider the UV fixed point $\lambda = 0$. In $d = 4$, the entropy of the cylinder has the form \cite{Solodukhin:2008dh}
	\begin{equation}
		S_0^{\text{cyl}}(R) = d_1\,\frac{RL}{\varepsilon^2} - b_{\text{UV}}\,\frac{L}{R}\log{\frac{R}{\varepsilon}} + \mathcal{O}(\varepsilon)\,,\quad \varepsilon\rightarrow 0\,,\quad d = 4\,,
		\label{eq:UV_cylinder_entropy}
	\end{equation}
	where $L \rightarrow \infty$ is the length of the cylinder (understood here as an IR regulator), $d_1$ is a constant and $b_{\text{UV}}$ is the type B Weyl anomaly coefficient as defined in Appendix \ref{app:Weyl_conventions}. We also assume that there is no finite $\mathcal{O}(1)$ term at the fixed point.
	
	Demanding cancelations of quadratic and logarithmic UV divergences fixes the coefficients uniquely to $s_0 = -1$ and $s_1 = 1$ in $d = 4$, as has been found in \cite{Ishihara:2012jg,Fujita:2023bdk}. Thus, we are led to propose an inequality involving the unique combination of derivatives of the entropy that yields a UV-finite result at a fixed point
	\begin{equation}
		\text{naive proposal}\colon \quad (R^2\partial_R^2+R\,\partial_R-1)\,S^{\text{cyl}} \leq 0\,.
		\label{eq:cylinder_guess_fixed}
	\end{equation}
	At the fixed point, we obtain explicitly
	\begin{equation}
		(R^2\partial_R^2+R\,\partial_R-1)\,S^{\text{cyl}}_0 = 2b_{\text{UV}}\frac{L}{R}\,,
		\label{eq:cylinder_fixed_point_c_function}
	\end{equation}
	which always violates the proposal \eqref{eq:cylinder_guess_fixed} in a unitary theory in which $b_{\text{UV}}> 0$ by the positivity of the two-point function of the stress tensor \cite{Hofman:2008ar,Nakayama:2017oye}. This is analogous to the entanglement entropy of a ball in $d = 4$ which violates the inequality $R\, S''(R)-S'(R)\leq 0$ due to $a_{\text{UV}}> 0$, as explained above. In that case, a valid inequality was found by considering the UV-subtracted entropy due to the Markov property. Therefore the first proposal \eqref{eq:cylinder_guess_fixed} is false, but we may propose that the UV subtracted cylinder entropy satisfies the inequality
	\begin{equation}
		\text{proposal}\colon\quad (R^2\partial_R^2+R\,\partial_R-1)\,\Delta S^{\text{cyl}} \leq 0\,.
		\label{eq:second_proposal}
	\end{equation}
	Together with \eqref{eq:cylinder_fixed_point_c_function} this is equivalent to
	\begin{equation}
		\text{proposal}\colon\quad (R^2\partial_R^2+R\,\partial_R-1)\,S^{\text{cyl}}\leq 2b_{\text{UV}}\frac{L}{R}\,,
		\label{eq:cylinder_inequality_conjecture}
	\end{equation}
	which is analogous to the inequality \eqref{eq:d_4_C_prime_inequality} for ball subregions above. Notice though that unlike for balls, the left-hand side of \eqref{eq:cylinder_inequality_conjecture} cannot be written as a total $R$-derivative. If the theory flows to an IR fixed point with type B anomaly coefficient $b_{\text{IR}}$ then \eqref{eq:cylinder_inequality_conjecture} would imply $b_{\text{IR}} \leq b_{\text{UV}}$ to which there are known field theory \cite{Anselmi:1997am,Anselmi:1997ys} counterexamples; it is also straightforward to construct bottom-up holographic counterexamples as in \cite{Myers:2010xs}. This is just the usual statement that the type B coefficient is not a good measure of number of degrees of freedom of the theory, because it can be larger in the IR. Therefore, also the second proposal \eqref{eq:second_proposal} is generically false. 
	
	However, there are special theories for which $b_{\text{IR}} \leq b_{\text{UV}}$ is always true approximately: these are large-$N$ theories that have a holographic dual in terms of classical Einstein gravity. For such theories, the type A and B coefficients coincide at a fixed point at leading order in large-$N$ \cite{Henningson:1998gx,Henningson:1998ey} which implies that $b_{\text{IR}} \leq b_{\text{UV}}$ follows from the weak a-theorem $a_{\text{IR}} \leq a_{\text{UV}}$. Therefore in holographic theories, the inequality \eqref{eq:cylinder_inequality_conjecture} might hold in the strict large-$N$ limit.
	
	\subsection{C-functions from entanglement entropy}\label{subsec:cfunction_review}
	
	As shown in the previous section, entanglement entropies of certain subregions in quantum field theory obey monotonicity inequalities as functions of the size of the subregion. This makes the use of entanglement entropy a prime candidate for the construction of measures of degrees of freedom known as c-functions. In this context, a c-function should satisfy the following two properties:
	\begin{enumerate}[(i)]
		\item  It should be a renormalization scheme independent constant at the UV fixed point.
		\item It should monotonically decrease from the UV to the IR as a function of the size of the subregion.
	\end{enumerate}
	We will now review the LM \cite{Liu:2012eea,Liu:2013una} and ILN \cite{Ishihara:2012jg} c-function proposals obtained from entropies of balls and cylinders respectively. They satisfy the property (i) and we review how the entropy inequalities of the previous section imply also the monotonicity condition (ii) in certain cases. In $d=4$, we also conjecture an inequality for the ILN c-function.
	
	\paragraph{C-functions from balls.} The LM c-function is constructed from the entanglement entropy of a solid ball by using a differential operator which cancels out the UV divergent parts of the entropy. For $d\geq 2$, the entropy of the ball in the deformed theory $\lambda\neq 0$ generically diverges as \cite{Liu:2012eea}
	\begin{equation}
		S(R) = \begin{dcases}   
			p_{d}\frac{R^{d-2}}{\varepsilon^{d-2}}+\ldots +p_3\frac{R}{\varepsilon} +F + \mathcal{O}(\varepsilon)\,,\quad &d = \text{odd}\\
			p_{d}\frac{R^{d-2}}{\varepsilon^{d-2}}+\ldots + p_4\frac{R^{2}}{\varepsilon^{2}} +A\log{\frac{R}{\varepsilon}}+f+\mathcal{O}(\varepsilon^2)\,,\quad &d = \text{even}
		\end{dcases}\,,
		\label{eq:ball_divergence_structure}
	\end{equation}
	where the coefficients $p_i,A$ are dimensionless constants while $F,f$ are dimensionless functions of the dimensionless radius $\lambda^{1\slash (d-\Delta)} R$. At the fixed point $\lambda = 0$, $ F\vert_{\lambda = 0}$ coincides with the Euclidean partition function on a round $S^d$ (up to a $d$-dependent factor) and $A$ coincides with the higher-dimensional type A Weyl anomaly coefficient of the UV CFT (also up to a factor). In addition, in a gapped theory in $d = 3$, the large radius limit $\lim_{R\rightarrow \infty}F$ is called topological entropy \cite{Kitaev:2005dm,Levin:2006zz}. For $d=2,3,4$, the form of these expansions coincide with the ones given in Section~\ref{subsec:ent_inequalities}.
	
	The LM c-function is defined as \cite{Liu:2012eea,Liu:2013una}
	\begin{equation}
		\CLM(R) = \lim_{\varepsilon\rightarrow 0}\mathcal{D}^{(d)}_{\text{ball}}(R\,\partial_R)\,S(R)\,,
		\label{eq:LM-c-function}
	\end{equation}
	where the differential operator is given by
	\begin{equation}
		\mathcal{D}^{(d)}_{\text{ball}}(R\,\partial_R) \equiv \frac{1}{(d-2)!!}\begin{dcases} \left(R\,\partial_R-1\right)\left(R\,\partial_R - 3\right) \cdots \left(R\,\partial_R-(d-2)\right) \ , &\quad d = \text{odd} \\
			R\,\partial_R\left(R\,\partial_R-2\right)\, \cdots\, \left(R\,\partial_R-(d-2)\right)\ , &\quad d = \text{even}
		\end{dcases}\ . \label{eq:LM-c-function_operator}
	\end{equation}
	In $d = 2,3,4$, the LM c-function equals the function \eqref{eq:capital_C_function} introduced above
	\begin{equation}\label{eq:LM-c-function_bis}
		\CLM(R) = \begin{cases}
			R\,S'(R)\,,\quad &d=2\\
			R\,S'(R) -S(R) \,,\quad &d=3\\
			\frac{1}{2}\,R\,(R\,S'(R)-2S(R))'\,,\quad &d=4
		\end{cases} \ .
	\end{equation}
	We see that for $d = 2$ and $d = 3 $ we have $\CLM(R) = C(R)$. Therefore the entropy inequalities of the previous section imply that $\CLM(R)$ decrease monotonically. In four dimensions, we have
	\begin{equation}
		\CLM(R) = \frac{1}{2}\,R\,C'(R)\,,\quad d = 4\,.
		\label{eq:C_LM_bound_d_4}
	\end{equation}
	At the UV fixed point in $d = 4$ using \eqref{eq:d_4_EE}, it picks up the type A anomaly coefficient
	\begin{equation}
		\CLM(R) = a_{\text{UV}}\,.
		\label{eq:CLM_is_aUV}
	\end{equation}
	Thus \eqref{eq:d_4_C_prime_inequality} implies the inequality
	\begin{equation}\label{eq:CLM_ball_inequality}
		\CLM(R)\leq a_{\text{UV}}\,,\quad d = 4\ ,
	\end{equation}
	which does not constrain the monotonicity properties of the LM c-function which would amount to the inequality $\CLM'(R)\leq 0$. By the fact that
	\begin{equation}
		\CLM'(R) = \frac{1}{2}\,R^2S'''(R)+\frac{\CLM(R)}{R}
	\end{equation}
	monotonicity is equivalent to
	\begin{equation}
		S'''(R)\leq -\frac{2\,\CLM(R)}{R^3}\,,
	\end{equation}
	which is a bound on the third derivative of the entanglement entropy. Notice that $S'''(R)$ is UV finite at least at the UV fixed point $\lambda = 0$ where the entropy is given by \eqref{eq:d_4_EE}. We emphasize that this bound does not follow from the use of strong subadditivity.
	
	\paragraph{C-functions from cylinders.} In general dimensions $d\geq 3$, the entropy $S^{\text{cyl}}(R)$ of a cylinder has been argued to have the same divergence structure \eqref{eq:ball_divergence_structure} as the entropy of a ball, but multiplied by a factor of $L\slash R$ and with a different set of coefficients \cite{Ishihara:2012jg}
	\begin{equation}
		S^{\text{cyl}}(R) = \frac{L}{R}\times \begin{dcases}   
			\widetilde{p}_{d}\frac{R^{d-2}}{\varepsilon^{d-2}}+\ldots +\widetilde{p}_3\frac{R}{\varepsilon} +\widetilde{F} + \mathcal{O}(\varepsilon)\,,\quad &d = \text{odd}\\
			\widetilde{p}_{d}\frac{R^{d-2}}{\varepsilon^{d-2}}+\ldots + \widetilde{p}_4\frac{R^{2}}{\varepsilon^{2}} +B\log{\frac{R}{\varepsilon}}+\widetilde{f}+\mathcal{O}(\varepsilon^2)\,,\quad &d = \text{even}
		\end{dcases}\,.
		\label{eq:cylinder_divergence_structure}
	\end{equation}
	Here the notation $\mathcal{O}(\varepsilon)$ denotes terms linear and higher-order in $\varepsilon$ and the $\varepsilon\rightarrow 0$ limit is understood in the sense of $\varepsilon$ being smaller than all other length scales of the problem. This divergence structure is valid for an infinitely long cylinder $L\rightarrow \infty$ or in the case when $L$ is finite, but its ends are periodically identified by compactifying the full spatial slice $\mathbb{R}^{d-1}$ on an $S^1$ as in \cite{Ishihara:2012jg} (see Sec.~\ref{eq:c_compactification} for a detailed discussion). The coefficients $\widetilde{p}_i,B$ are dimensionless constants, and in the $L\rightarrow \infty$ limit, the coefficients $\widetilde{F},\widetilde{f}$ are functions of the dimensionless radius $\lambda^{1\slash (d-\Delta)}R$ only, but in the compactified case with finite $L$, they can also depend on $R\slash L$. In even dimensions, the coefficient $B$ of the logarithmic divergence coincides up to a constant factor with the type B Weyl anomaly coefficient.
	
	The ILN c-function is defined as \cite{Ishihara:2012jg}
	\begin{equation}
		\mathcal{C}_{\text{ILN}}(R) = \lim_{\varepsilon\rightarrow 0}\mathcal{D}^{(d)}_{\text{cyl}}(R\,\partial_R)\,S^{\text{cyl}}(R)\,,
		\label{eq:cylinder-c-function}
	\end{equation}
	where the differential operator is defined in terms of the LM one \eqref{eq:LM-c-function_operator} as
	\begin{equation}
		\mathcal{D}^{(d)}_{\text{cyl}}(R\,\partial_R) \equiv \frac{1}{L}\,\mathcal{D}^{(d)}_{\text{ball}}(R\,\partial_R)\,R\,,
		\label{eq:cylinder_operator}
	\end{equation}
	which cancels all the UV divergences in \eqref{eq:cylinder_divergence_structure}. Note that compared to \cite{Ishihara:2012jg}, we have multiplied the operator by an additional factor of $R\slash L$: this ensures that \eqref{eq:cylinder-c-function} picks up exactly the type B Weyl anomaly coefficient at the UV fixed point. Note also that the differential operator $ \frac{1}{R}\,\mathcal{D}^{(d)}_{\text{ball}}(R\,\partial_R)\,R$ used in \cite{Ishihara:2012jg} is invariant under a rescaling $R\rightarrow \lambda R$ of the radius of the cylinder alone, while the operator \eqref{eq:cylinder_operator} we use is only invariant under a simultaneous rescaling $R\rightarrow \lambda R$ and $L\rightarrow \lambda L$.
	
	In $d = 4$, \eqref{eq:cylinder-c-function} takes the form (see also \cite{Fujita:2023bdk})
	\begin{equation}
		\mathcal{C}_{\text{ILN}}(R) =\frac{R}{2L}\,(R\,\partial_R-1)(R\,\partial_R+1)\, S^{\text{cyl}}(R) = \frac{R}{2L}\,(R^2\partial_R^2+R\,\partial_R-1)\,S^{\text{cyl}}(R)\,.
		\label{eq:Ccyl}
	\end{equation}
	At the fixed point it equals $b_{\text{UV}}$ as defined in \eqref{eq:UV_cylinder_entropy} and if our second proposal \eqref{eq:cylinder_inequality_conjecture} is true, it satisfies
	\begin{equation}
		\mathcal{C}_{\text{ILN}}(R) \leq b_{\text{UV}}\,,\quad d = 4\,,\quad (\text{conjecture})\,,
		\label{eq:Ccyl_inequality_conjecture}
	\end{equation}
	which is analogous to the inequality \eqref{eq:CLM_ball_inequality} for the LM c-function. As we have explained above, we expect this inequality to be violated in general except possibly in holographic theories in the strict large-$N$ limit where the type A and B Weyl anomaly coefficients coincide. Indeed, we have evidence for this in two distinct holographic examples. In Sec.~\ref{eq:c_compactification}, we compute $\mathcal{C}_{\text{ILN}}(R)$ in a top-down holographic model dual to a gapped field theory, while in Appendix \ref{app:cylinders}, we consider a bottom-up holographic model with a flow between two fixed points. In both cases, we find that the bound \eqref{eq:Ccyl_inequality_conjecture} is satisfied. Note that monotonicity as a function of $R$ is not guaranteed by the conjecture.
	
	\section{Entanglement for balls in conformal field theories and c-functions}\label{sec:3d-non-monotonic}
	
	In this section we investigate EE in holographic backgrounds that are dual to $d$-dimensional CFTs. These backgrounds always include an external AdS${}_{d+1}$ spacetime, ensuring compatibility with the global conformal symmetry of the boundary theory. Additionally, they feature an internal space, which remains unspecified and encodes other global symmetries of the CFT. Various examples of supersymmetry-preserving gravity backgrounds of this type exist for integer dimensions \mbox{$d+1 = 2,3,\ldots,7$}; see \cite{Lozano:2020txg, Lozano:2020bxo, Assel:2011xz, Lozano:2019zvg, Akhond:2021ffz, DHoker:2016ysh, Legramandi:2021uds, Nunez:2019gbg, Gaiotto:2009tk, Apruzzi:2015wna} for a representative sample of backgrounds preserving eight Poincar\'e SUSYs. This corresponds to the existence of ${\cal N}=2$ superconformal field theories in dimensions \mbox{$d =1,2,\ldots,6$}. The non-supersymmetric case of AdS${}_8$ is discussed in \cite{Cordova:2018eba}. 
	
	We focus on entanglement entropy across a $(d-1)$-dimensional radius-$R$ ball subregion in an underlying CFT${}_{d}$. The computation follows the standard RT prescription, where the eight-dimensional dual embedding surface $\Sigma_8$ is anchored to the boundary of the $(d-1)$-dimensional ball: $\partial B=S^{d-2}$. Armed with expressions for EE in generic dimension, we will apply the LM operator and display results for $\CLM(R)$. We note that same computations in the case where the internal space is omitted exist in the literature~\cite{Ryu:2006ef,Liu:2012eea}, but we find it useful to review them as they serve as a springboard for computations that follow in the coming sections.
	A comparison with other measures of the number of degrees of freedom in CFTs is also presented.
	
	\subsection{Ten-dimensional background geometries}
	
	The string frame metric associated with the string theory backgrounds of interest is,
	\begin{eqnarray}
		\dd s^2 
		& = & f_1(\vec{y})\left[ r^2(-\dd t^2+ \dd{\vec x}_{d-1}^2) + \frac{\dd r^2}{r^2} \right] +\dd s^2_{\text{int}, 9-d}(\vec{y}) \label{eq:general_metric} \\
		& = & f_1(\vec{y})\left[ r^2(-\dd t^2+ \dd \rho^2 +\rho^2 \dd\Omega_{d-2}^2) + \frac{\dd r^2}{r^2} \right] +\dd s^2_{\text{int}, 9-d}(\vec{y}) \ .\label{eq:backAdSQ}
	\end{eqnarray}
	Generically, there is a dilaton $\Phi(\vec{y})$ together with Ramond and Neveu--Schwarz fluxes that complete the background. The subscript `int' denotes the internal part of the metric.  See references \cite{Lozano:2020txg, Lozano:2020bxo,Assel:2011xz, Lozano:2019zvg,Akhond:2021ffz, DHoker:2016ysh, Legramandi:2021uds, Nunez:2019gbg, Gaiotto:2009tk,Apruzzi:2015wna} for details that do not play a significant role in the discussion that follows: omitted details contribute to an overall normalization. Note that we have written the AdS$_{d+1}$ spacetime part inside the square brackets in Poincar\'e coordinates and in the second row we have written the spatial field theory directions in terms of spherical coordinates in order to isolate the $(d-1)$-dimensional ball.
	
	It is important to note that the warp factor $f_1(\vec{y})$, the dilaton, and any fluxes present, can only depend on the coordinates of the internal space, $\vec{y}$. Otherwise, we would be breaking the isometries of AdS$_{d+1}$. In what follows, we focus on backgrounds in Type II string theory; the extension to eleven-dimensional supergravity is straightforward, requiring the minimization of a nine-dimensional surface homologous to the entangling region instead of an eight-dimensional one. In Sec.~\ref{sec:flow}  and in Appendix \ref{app:confining}, we generalize the metric \eqref{eq:general_metric} to encompass situations in which there is a nontrivial RG flow away from a UV fixed point to a gapped IR theory. 
	
	\subsection{Holographic entanglement for ball-shaped regions}\label{subsec:ball_hol_ent}
	
	To proceed with the calculation of the holographic entanglement entropy, we need to minimize the area of a codimension-two surface that hangs from the asymptotic AdS boundary, anchored on the ball-shaped entangling region. We parameterize the eight manifold $\Sigma_8$ by coordinates $\{\Omega_{d-2}, r, \vec{y}\}$, with the embedding function $\rho=\rho(r)$. The boundary condition is $\rho(r\to\infty)=R$, with $R$ the radius of the ball. The induced metric on $\Sigma_8$ is
	\be
	ds_{8}^2= f_1(\vec{y})\left[ r^2\rho^2\dd \Omega_{d-2}^2 + \frac{\dd r^2}{r^2}(1+ r^4 \rho'^2) \right] +\dd s^2_{\text{int},9-d}(\vec{y})\ ,\nonumber
	\ee
	where the prime indicates the derivative with respect to the holographic coordinate $r$.
	The area functional that computes the holographic entanglement entropy reads
	\be
	S  =  \frac{1}{4G_{10}}\int_{\Sigma_8} \dd^8 x\, e^{-2 \Phi}\sqrt{ \det[g_8]} =  \frac{\vol{(S^{d-2})}\vol_{\text{int}}}{4G_{10}}\int \dd r\, r^{d-3}\rho^{d-2}\sqrt{1+r^4\rho'^2} \ ,\label{eq:cuti}
	\ee
	where $\vol{(S^{d-2})}=\int d\Omega_{d-2} = 2\pi^{\frac{d-1}{2}}/\Gamma(\frac{d-1}{2})$ and we further defined the integral
	\be\label{eq:volint}
	\vol_{\text{int}} \equiv \int_{M_{\text{int}}} \prod_{i=1}^{9-d}\dd y_i\, e^{-2\Phi}\sqrt{ f_1^{d-1} \det[g_{\text{int}}] } \ . 
	\ee
	over the $(9-d)$-dimensional internal manifold $M_{\text{int}}$ which is equipped with the metric $g_\text{int}$,
	\be\label{eq:relation_newton_constants}
	\frac{1}{4G_{d+1}} \equiv \frac{\vol_\text{int}}{4G_{10}} \ ,
	\ee
	where $G_{10} = 8\pi^6l_s^8g_s^2$. We also collectively denote
	\be\label{eq:def_curly_N}
	{\cal N} \equiv \frac{\vol{(S^{d-2})}}{4G_{d+1}}=\frac{\vol{(S^{d-2})}\vol_\text{int}}{4G_{10}} \ .
	\ee
	The equation of motion following from \eqref{eq:cuti} and its solution that fulfills the boundary condition are
	\begin{equation}
		\rho'' +(d-1)r^3 \rho'^3 -(d-2)\frac{\rho'^2}{\rho} +(d+1)\frac{ \rho'}{r} -\frac{d-2}{r^4 \rho}  =  0\,, \qquad 
		\rho(r)  =  \frac{\sqrt{r^2 R^2-1}}{r} \ .\label{eq:AdsQ}
	\end{equation}
	Here one of the integration constants have been fixed by anchoring the RT surface to the boundary of the entangling region and the other one by demanding that the embedding is regular at the tipping point. 
	We replace the solution in the expression for $S$ of \eqref{eq:cuti}, to obtain
	\be
	S={\cal N}\, R \int^{1/\eUV}_{1/R} \dd r \left( R^2 r^2-1\right)^{\frac{d-3}{2}}
	\ ,
	\label{eq:montiel}
	\ee
	where we regulate the integral from above by $1/\eUV$.
	We have identified the nonuniversal $\eUV$ with the UV cutoff appearing in preceding Sec.~\ref{sec:liu-mezei}.
	We note that (\ref{eq:montiel}) can formally be written in terms of a hypergeometric function. Interestingly, however, by manipulation a simpler form can be found
	\bea 
	S & = & \NN \sum_{j=0}^{\lfloor\frac{d-3}{2}\rfloor}\frac{\left(\frac{3-d}{2}\right)_j}{j!(d-2(j+1))}\bigg(\frac{R}{\eUV}\bigg)^{d-2(j+1)} \nonumber \\
	& & -\NN\frac{\Gamma\left(\frac{d-1}{2}\right)}{\Gamma\left({\frac{d}{2}}\right)} \times \begin{dcases}
		\frac{(-1)^{(d+1)/2}\sqrt{\pi}}{2}   \,,\quad & d=\rm{odd}\\
		\frac{(-1)^{d/2}}{\sqrt{\pi}}\left(\log\frac{2R}{\eUV}+\frac{1}{2}\mathcal{H}_{\frac{d-2}{2}}\right) \,,\quad &d=\rm{even} 
	\end{dcases}  \ ,
	\eea
	where $(a)_j=\frac{\Gamma(a+j)}{\Gamma(a)}$ is the Pochhammer symbol, $\lfloor\cdots\rfloor$ rounds down to an integer, and $\mathcal{H}_{\frac{d-2}{2}}$ denotes the harmonic number. Note that the exponent in the first term is never negative and that the full expression is real-valued. 
	For odd dimensions, the finite constant term in the brackets is universal and for even dimensions, the coefficient of the logarithmic term is universal. 
	
	It is now straightforward to obtain the LM entanglement c-function $\CLM$ (\ref{eq:LM-c-function}),
	\be\label{eq:CLM_AdS} 
	\CLM=\NN \times \frac{\left[\pi(1-(-1)^d)+2(1+(-1)^d)\right]\Gamma(\frac{d-1}{2})}{4\sqrt{\pi}\,\Gamma(\frac{d}{2})} \ .
	\ee
	The formula above gives the correct values as obtained from evaluation of (\ref{eq:LM-c-function}) when $d$ is integer. It would be interesting to understand if there is any meaning beyond integer dimensions. It is worth noting that (\ref{eq:LM-c-function}) also removes the $\varepsilon$ divergence in the case where $d$ takes non-integer values, for which (\ref{eq:CLM_AdS}) needs to be adapted. As expected, there is no dependence on any scale and hence the resulting value is to be interpreted as the central charge of the underlying conformal field theory; recall that for even $d$, it is proportional to the type A Weyl anomaly coefficient. For convenience, we collect all the results in Table \ref{table:differentQs} for known AdS backgrounds. 
	\begin{table*}
		\begin{tabularx}{0.85\textwidth}{YYY} \toprule
			Theory & Entanglement entropy $ S_\text{EE}/{\cal N}$ & Central charge \qquad $ \CLM/{\cal N}$ \\
			\midrule
			CFT${}_3$ & $\frac{R}{\eUV} - {\bf{1}}$ & ${\bf{1}}$ \\
			CFT${}_4$ & $\frac{R^2}{2 \eUV^2} - {\bf{\frac{1}{2}}}\log \left( \frac{2R}{\eUV} \right) - \frac{1}{4}$ & ${\bf{1/2}}$ \\
			CFT${}_5$ & $ \frac{R^3}{3 \eUV^3} - \frac{R}{\eUV}+{\bf\frac{2}{3}}$ & $ {\bf{2/3}}$ \\
			CFT${}_6$ & $\frac{R^4}{4 \eUV^4}- \frac{3R^2}{4 \eUV^2} + {\bf\frac{3}{8}}\log \left( \frac{2R}{\eUV} \right) + \frac{9}{32} $ & $\bf{3/8}$ \\
			CFT${}_7$ & $ \frac{R^5}{5 \eUV^5} - \frac{2 R^3}{3 \eUV^3} + \frac{R}{\eUV} - \bf{\frac{8}{15}}$ & $\bf{8/15}$ \\
			\bottomrule
		\end{tabularx}
		\caption{We collect the results for the entanglement entropies (second column) for balls with radii $R$ in various dimensions (first column). We also display the central charge as extracted using the LM function $\CLM$ (third column). For visualization we found it useful to use bold font to highlight the term entering the central charge.
		}
		\label{table:differentQs}
	\end{table*}
	
	We stress that the computations leading to (\ref{eq:CLM_AdS}) in spirit exist in the literature~\cite{Ryu:2006ef,Liu:2012eea}. Here we found it useful to recapitulate on this, in particular to emphasize that also the prefactor $1/(4G_{d+1})$ appearing in~\cite{Ryu:2006ef,Liu:2012eea}, or ${\cal N}$ can be written in terms of field theory quantities, as we discuss in Sec.~\ref{qft-cft-calculo}. 
	
	\subsection{Comparison to holographic c-function}\label{sec:cholosubsec}
	
	Having obtained values for the central charge using the entanglement entropy, it is interesting to ask how it compares with other means of defining it. To this end, we recall that the free energy  (or the central charge) of linear quiver field theories has been obtained in \cite{Nunez:2023loo}.
	
	Therefore, it is interesting to compare the expressions in Table \ref{table:differentQs} with the holographic c-function $c_\text{hol}(r)$ defined in \cite{Macpherson:2014eza,Bea:2015fja}. This quantity is defined for a ten-dimensional background geometry dual to a $d$-dimensional QFT with a metric and a dilaton profile of a generic form (in string frame),
	\begin{equation}
		\dd s^2= a(r,y^i)\left[-\dd t^2+\dd \vec x_{d-1}^2 +b(r)\, \dd r^2 \right]+ g_{ij}(r,y^i)\, \dd y^i \dd y^j \ ,\qquad \Phi = \Phi(r,y^i) \ ,\label{eq:genericmetriccc}
	\end{equation}
	where $i,j=1,\ldots,9-d$. The definition of the holographic central charge is given by \cite{Macpherson:2014eza,Bea:2015fja}
	\be
	\frac{c_{\text{hol}}(r)}{\vol{(\mathbb{R}^{d-1})}}= \frac{(d-1)^{d-1}}{G_{10}}\, b(r)^{\frac{d-1}{2}} \frac{H(r)^{\frac{2d-1}{2}}}{H'(r)^{d-1}} \ ,\label{eq:chol}
	\ee
	where we have defined the integral
	\be\label{eq:H_integral}
	\sqrt {H(r)} = \int_{M_{\text{int}}} \prod_{i=1}^{9-d}\dd y^i\, e^{-2\Phi(r,y^i)}\,\sqrt{\det[g_{ij}(r,y^i)] \left(\frac{a(r,y^i)}{r^2}\right)^{d-1} } \  r^{d-1}
	\ee
	over the internal manifold $M_{\text{int}}$ and $ r$ plays the role of the ``energy scale''.
	
	The definition \eqref{eq:chol} indicates that the meaningful quantity is the density of degrees freedom due to the division over the infinite volume $\vol{(\mathbb{R}^{d-1})}$ of the field theory directions. Also, notice that (\ref{eq:chol}) is covariant under a reparametrization of the  holographic radial coordinate $r$, respects the SO$(1,d-1)$ isometry, and that $\prod_i\dd y^i\sqrt{\det[g_{ij}(r,y^i)]}$ is the invariant volume element under general coordinate transformations of the internal manifold $M_{\text{int}}$. It would be interesting to analyze how a consistent truncation of the ten-dimensional geometry down to $(d+1)$ dimensions would project (\ref{eq:chol}) to effective holographic c-functions defined directly on the reduced theory \cite{Freedman:1999gp,Myers:2010tj}. This is interesting because the monotonicity properties of the effective holographic c-functions can be linked with the null energy conditions (NECs), while from the ten-dimensional perspective violations of the NECs appear admissible~\cite{Hoyos:2021vhl}. 
	Indeed, while the function (\ref{eq:chol}) at face value seems to satisfy the desired properties of a c-function, in particular being monotonic in all the geometries studied in this work, it nevertheless ceases to be monotonic in the supersymmetric ten-dimensional geometries in \cite{Hoyos:2021vhl}.
	These backgrounds are not known to admit string theory realizations in a strict sense, however.
	
	Let us calculate the holographic central charge in the background of \eqref{eq:general_metric}. Comparing \eqref{eq:general_metric} and (\ref{eq:volint}) with the formulas  \eqref{eq:genericmetriccc}--\eqref{eq:chol}, we find 
	\be
	a(r,y^i)=f_1(\vec y)\,r^2 \ \ , \ \ b(r)=\frac{1}{r^4} \ \ , \ \ \sqrt H = \vol_{\text{int}}\, r^{d-1} \ ,
	\ee
	where the integral $\vol_{\text{int}}$ is defined in \eqref{eq:volint}. We obtain \be\label{eq:centralcharge}  
	\frac{c_\text{hol}}{\vol(\mathbb{R}^{d-1})} = \frac{\NN}{\vol(S^{d-2})} \times\frac{1}{2^{d-3}} \ .
	\ee
	Although \eqref{eq:centralcharge} does not precisely match with the values in Table~\ref{table:differentQs} in general, it is interesting to note that both the value from the LM entanglement c-function and the holographic central charge are measuring intimately related aspects of the CFT. 
	We note that
	\be\label{eq:4dmatching}
	\frac{c_\text{hol}}{\vol(\mathbb{R}^{3})} = \frac{\CLM}{\vol(S^{2})} \ , \ \  d=4 \ .
	\ee
	
	In the next subsection, we focus on this particular $d=4$ by performing a perturbative calculation of the central charge/free energy and compare it with the non-perturbative results.
	
	\subsection{Field theory and holographic derivation of \texorpdfstring{$\CLM$}{} for \texorpdfstring{$d=4$}{}}\label{qft-cft-calculo}
	
	In this subsection we deliver a microscopic, weakly coupled, field theory derivation of the central charge. This complements the holographic (strongly coupled) description of the preceding discussion, but we will provide further details. Notice that we focus on the special case of four-dimensional ${\cal N}=2$ linear quiver CFTs.
	
	\subsubsection{CFT approach}
	
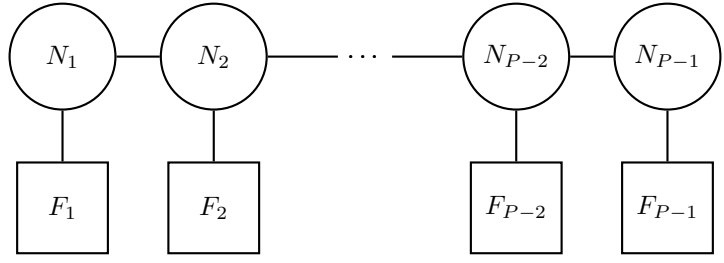
\begin{figure}[ht]
		\begin{center}
			\begin{tikzpicture}
				\node (1) at (-4,0) [circle,draw,thick,minimum size=1.4cm] {$N_1$};
				\node (2) at (-2,0) [circle,draw,thick,minimum size=1.4cm] {$N_2$};
				\node (3) at (0,0)  {$\dots$};
				\node (5) at (4,0) [circle,draw,thick,minimum size=1.4cm] {$N_{P-1}$};
				\node (4) at (2,0) [circle,draw,thick,minimum size=1.4cm] {$N_{P-2}$};
				\draw[thick] (1) -- (2) -- (3) -- (4) -- (5);
				\node (1b) at (-4,-2) [rectangle,draw,thick,minimum size=1.2cm] {$F_1$};
				\node (2b) at (-2,-2) [rectangle,draw,thick,minimum size=1.2cm] {$F_2$};
				\node (3b) at (0,0)  {$\dots$};
				\node (5b) at (4,-2) [rectangle,draw,thick,minimum size=1.2cm] {$F_{P-1}$};
				\node (4b) at (2,-2) [rectangle,draw,thick,minimum size=1.2cm] {$F_{P-2}$};
				\draw[thick] (1) -- (1b);
				\draw[thick] (2) -- (2b);
				\draw[thick] (4) -- (4b);
				\draw[thick] (5) -- (5b);
			\end{tikzpicture}
		\end{center}
		\caption{Long quiver of length $P-1$ with gauge nodes $N_i$ and flavor nodes $F_i$. The quiver is \textit{balanced and conformal} if $F_i = 2 N_i - N_{i-1}-N_{i+1}$.}
		\label{fig:quiverfiga}
	\end{figure}

	We focus our attention on the case of four-dimensional ${\cal N}=2$ linear quiver SCFTs. The quasi-particle, weakly coupled field theory description consists of $n_v$ vector multiples  and $n_h$ hypermultiplets. A generic linear quiver is indicated in Fig.~\ref{fig:quiverfiga}. The condition for conformality in these systems is that the number of flavors on each node equals twice the number of colors of that node. For each node, it must be $2N_i=N_{i+1}+N_{i-1} +F_i$ (with $N_0=N_P=0$).
	
	The information about the number of colors (gauge groups) and flavors (global groups) is encoded in the so-called rank function,
	\begin{equation}
		{\cal R}(\eta) = \begin{cases} 
			N_1 \eta & \ , \quad  0\leq \eta \leq 1 \\
			N_l+ (N_{l+1} - N_l)(\eta-l) & \ , \quad  l \leq \eta\leq l+1\ ,\;\;\; l:=1,\ldots, P-2\\\label{rankfull}
			N_{P-1}(P-\eta) & \ , \quad  (P-1)\leq \eta\leq P 
		\end{cases} \ .
	\end{equation}
	In fact, the rank of the $j^{th}$ gauge group is the value of the rank function at $\eta=j$. The number of flavors $F_i$ is encoded in the second derivative  ${\cal R}''(\eta)$,
	\begin{equation}
		{\cal R}''(\eta)=\sum_{j=1}^{P-1} (2N_j -N_{j-1} -N_{j+1})\delta(\eta-j)= \sum_{j=1}^{P-1} F_j\delta(\eta-j) \ .\nonumber
	\end{equation}
	The four-dimensional ${\cal N}=2$ SCFTs are characterized by two numbers $A_{\text{UV}} $ and $B_{\text{UV}} $, which are related to the coefficients $a_{\text{UV}}$, $b_{\text{UV}}$ introduced in Sec.~\ref{sec:liu-mezei} as (see Appendix \ref{app:Weyl_conventions})
	\begin{equation}
		A_{\text{UV}} = \frac{a_{\text{UV}}}{4}\,,\quad B_{\text{UV}} = 2b_{\text{UV}}\,.
		\label{eq:aUV}
	\end{equation}
	In terms of the microscopic degrees of freedom they are given by \cite{Shapere:2008zf,Nunez:2023loo},
	\begin{equation}
		A_{\text{UV}}= \frac{1}{24}\,(5 n_v +n_h)\,, \qquad B_{\text{UV}}= \frac{1}{24}\,(4n_v+2n_h)\,, \label{eq:centralcharges}  
	\end{equation}
	where $n_v$ and $n_h$ are the number of vector- and hypermultiplets. It was shown in \cite{Nunez:2023loo} that in the limit of long quivers with large ranks ($P\to\infty, N_j\to\infty$) we have
	\be
	A_{\text{UV}} = B_{\text{UV}} = \frac{P}{8}\sum_{k=1}^\infty R_k^2 \qquad \mbox{(large $P$ and $N_j$)} . \label{eq:longlimit}
	\ee
	We have defined $R_k=\frac{2}{P}\int_0^{P} \mathcal{R}(\eta)\sin \bigl(\frac{k\pi \eta}{P} \bigr) \dd\eta$. We find it clarifying to provide a few detailed examples.
	\\
	\\
	{\bf $\blacksquare$ Example I:}
	Let us consider the following quiver:
	\begin{center}
		\begin{tikzpicture}
			\node (1) at (-6,0) [circle,draw,thick,minimum size=1.4cm] {\footnotesize 
				$N$};
			\node (2) at (-4,0) [circle,draw,thick,minimum size=1.4cm] {\footnotesize $2N$};
			\node (3) at (-2,0) [circle,draw,thick,minimum size=1.4cm] {\footnotesize $3N$};	
			\node (4) at (0,0)  {$\dots$};
			\node (6) at (4,0) [rectangle,draw,thick,minimum size=1.2cm] {\footnotesize $PN$};
			\node (5) at (2,0)  {\footnotesize $(P-1)N$};
			\draw[thick] (1) -- (2) -- (3) -- (4) -- (5)-- (6);
			\draw[thick] (2,0) circle (0.67cm) ;
			\draw[thick] (1,0) -- (1.3,0);
			\draw[thick] (2.7,0) -- (3.3,0);
		\end{tikzpicture}
	\end{center}
	The rank function associated with this quiver is
	\[ {\cal R}(\eta) = \begin{cases} 
		N\eta & \ , \quad  0\leq \eta \leq (P-1) \\
		N(P-1) (P-\eta)&  \ , \quad  (P-1)\leq \eta\leq P 
	\end{cases} \ .
	\]
	We calculate the following useful quantities
	\begin{eqnarray}
		n_v & = & \sum_{j=1}^{P-1} (jN)^2-1= \frac{N^2 P^3}{3}\left(1-\frac{3}{2P} +\frac{1}{2P^2}-\frac{3}{N^2P^2}+\frac{3}{N^2P^3}\right) \\ 
		n_h & = & \sum_{j=1}^{P-1}j(j+1)N^2= \frac{N^2P^3}{3}\left(1-\frac{1}{P^2}\right) \nonumber\\
		A_{\text{UV}} & = & \frac{N^2P^3}{12}\left(1+\frac{1}{4P^2}+\frac{5}{2N^2P^3}-\frac{5}{2 N^2P^2} -\frac{5}{4P} \right), ~~B_{\text{UV}}=\frac{N^2P^3}{12}\left(1+\frac{2}{N^2P^3}-\frac{2}{ N^2P^2} -\frac{1}{P} \right) \nonumber\\
		R_k & = & \frac{2 N P^2}{k^2\pi^2}\sin\left( k\pi -\frac{k\pi}{P}\right)\ .\label{danielpassarella}
	\end{eqnarray}
	Following \eqref{eq:longlimit}, we calculate
	\begin{equation}
		\frac{P}{8}\sum_{k=1}^\infty R_k^2= \frac{N^2P^3}{12}\bigg(1-\frac{1}{P}\bigg)^2\approx \frac{N^2P^3}{12}\ . 
	\end{equation}
	Showing that for this long quiver with large gauge ranks, the formula in \eqref{eq:longlimit} coincides at leading order with \eqref{eq:centralcharges}. Let us see this at work in another example.
	\\
	\\
	{\bf $\blacksquare$ Example II:}
	As our second example, the CFT we now characterize by a quiver and a rank function as follows:
	\begin{center}
		\begin{tikzpicture}
			\node (1) at (-4,0) [rectangle,draw,thick,minimum size=1.2cm] {$N$};
			\node (2) at (-2,0) [circle,draw,thick,minimum size=1.4cm] {$N$};
			\node (3) at (0,0)  {$\dots$};
			\node (5) at (4,0) [rectangle,draw,thick,minimum size=1.2cm] {$N$};
			\node (4) at (2,0) [circle,draw,thick,minimum size=1.4cm] {$N$};
			\draw[thick] (1) -- (2) -- (3) -- (4) -- (5);
			\draw [decorate,decoration={brace,amplitude=15pt,mirror},thick,yshift=-1.5em]
			(-2.8,0) -- (2.8,0) node[midway,yshift=-2.5em]{$P-1$};
		\end{tikzpicture}
	\end{center}
	\[ {\cal R}(\eta) = \begin{cases} 
		N\eta & \ , \quad 0\leq \eta \leq 1 \\
		N & \ , \quad 1\leq \eta\leq (P-1)\\
		N (P-\eta) & \ , \quad (P-1)\leq \eta\leq P 
	\end{cases} \ .
	\]
	Similar calculations as in the previous example reveal,
	\begin{eqnarray}
		n_v & = & (N^2-1)(P-1) \ ,\ n_h= P N^2 \label{maritokempes}\\
		A_{\text{UV}} & =& \frac{N^2P}{4}\left(1-\frac{5}{6N^2} -\frac{5}{6P} +\frac{5}{6N^2 P}\right) \ , \ B_{\text{UV}} =\frac{N^2P}{4}\left(1-\frac{2}{3N^2} -\frac{2}{3P} +\frac{2}{3N^2 P}\right)\nonumber\\
		R_k & = & \frac{2 P N}{k^2\pi^2}\left[ \sin\left(\frac{k\pi}{P} \right) +\sin\left( k\pi -\frac{k\pi}{P}\right) \right]\ .\nonumber
	\end{eqnarray}
	It is straightforward to check that \eqref{eq:longlimit} is satisfied at leading order in $P, N$.
	
	Let us return to the holographic calculation presented in Sec.~\ref{sec:cholosubsec}, now equipped with all the field theory details to be inserted.
	
	\subsubsection{Holographic approach}
	
	Recall that the holographic description of the four-dimensional linear quiver ${\cal N}=2$ SCFTs is given in \cite{Gaiotto:2009gz}, see also \cite{Aharony:2012tz, Reid-Edwards:2010vpm, Nunez:2018qcj, Nunez:2019gbg, Macpherson:2024frt} for further discussion. The holographic background and dilaton are of the form in eqs. \eqref{eq:backAdSQ}. In the conventions from \cite{Aharony:2012tz}, the string frame metric and the dilaton read
	\begin{align}\label{eq:Gaiotto_Maldacena}
		ds_{10}^2 &= \kappa^{\frac{2}{3}} \sqrt{ \frac{2\dot{V} - \ddot{V}}{V''} } \left[ 4 r^2\left( -\dd t^2+ \dd\vec{x}^2 \right)+\frac{4\dd r^2}{r^2} + \frac{2V'' \dot{V}}{{\Delta}} \dd\hat{\Omega}_2 + \frac{2V''}{\dot{V}}(\dd\sigma^2 + \dd\eta^2) + \frac{4\sigma^2 V''}{2\dot{V} - \ddot{V}} \dd\chi^2 \right]\,,\nonumber \\
		e^{4\phi} &= \frac{4(2\dot{V} - \ddot{V})^3}{V'' \dot{V}^2 {\Delta}^2}\,,\qquad \Delta(\sigma,\eta) \equiv (2 \dot{V} - \ddot{V}) V'' + V'(\sigma, \eta)^2 \,.
	\end{align}
	Remarkably, all the metric functions can be written in terms of a `potential' $V(\sigma,\eta)$ and its derivatives, since primed and dotted quantities denote $f' \equiv \partial_\eta f$ and $\dot{f} \equiv \sigma \partial_\sigma f$, respectively. See \cite{Gaiotto:2009gz,Nunez:2018qcj, Nunez:2019gbg, Macpherson:2024frt} for details.
	
	Importantly, \eqref{eq:Gaiotto_Maldacena} is obtained from the eleven-dimensional supergravity solutions in \cite{Gaiotto:2009gz}. In these conventions, $\kappa = \pi l_p^3/2$ is imposed by flux quantization, with Planck length $l_p = g_s^ {1/3} l_s$. Because the circumference of the M-theory circle used in \cite{Gaiotto:2009gz} is $\ell_{11} = 2\pi \kappa^{1/3}$, the ten- and eleven-dimensional Newton's constant are related through
	\begin{equation}
		\tilde{G}_{10} = G_{11} \ell_{11}^{-1} = 8\pi^ 6 g_s^3l_s^9/\kappa^{\frac{1}{3}} = G_{10}\, (2\pi g_s l_s)/(2\pi \kappa^{\frac{1}{3}})\,,
	\end{equation}
	since $G_{11} = 16\pi^ 7l_p^ 9$ as usual and $G_{10} = 8\pi^6l_s^8g_s^2$. The subtlety here is that $\ell_{11}=(4\pi^4 g_s)^{1/3} l_s$ instead of $\ell_{11} = 2\pi g_sl_s$ which would be the usual convention in the literature (and which we used in Sec.~\ref{subsec:ball_hol_ent}).
	
	We compute the holographic entanglement entropy for a two-ball as we did in the previous section. We first rewrite the spatial slices 
	$\dd\vec{x}^2 = \dd \rho^ 2 + \rho^ 2\dd\Omega_2$. The eight manifold that is attached to the surface of the ball at the boundary extends in the coordinates $[\Omega_2,r,\hat{\Omega}_2,\chi,\sigma,\eta]$ at fixed $t$, and the only non-trivial dependence is captured by the embedding function $\rho(r)$. The induced metric on the surface becomes
	\begin{equation}
		\dd s_{8}^2 =\kappa^{\frac{2}{3}} \sqrt{ \frac{2\dot{V} - \ddot{V}}{V''} }\, \Big[4  r^2 \rho^2\dd\Omega_2 +\frac{4\dd r^2}{r^2}\left(1+ r^4 (\partial_r\rho)^2 \right) + \frac{2V'' \dot{V}}{{\Delta}} \dd\hat{\Omega}_2 + \frac{2V''}{\dot{V}}(\dd\sigma^2 + \dd\eta^2) + \frac{4\sigma^2 V''}{2\dot{V} - \ddot{V}} \dd\chi^2 \Big].\\
	\end{equation}
	With this information, it is now easy to compute the entanglement entropy, which takes a form analogous to \eqref{eq:cuti} (with prefactor $\tilde G_{10}^{-1}$ instead of $G_{10}^{-1}$) and reads
	\begin{equation}
		S = \mathcal{N}\int\dd r r \rho^2 \sqrt{1+r^4 (\partial_r\rho)^2} \ ,
	\end{equation}
	with
	\begin{equation}
		\mathcal{N} = \frac{8\kappa^{8/3}}{\tilde G_{10}}\vol{S^2}\vol{\hat S^2}\vol{S_ \beta^1}\int \dd\sigma\dd\eta \, \sigma\dot V V'' = 4\int \dd\sigma\dd\eta\, \sigma\dot V V''= P \sum_{k=1}^\infty R_k^2\,,
	\end{equation}
	where in the last equality we have used the expression of the potential function in terms of the rank function of the quiver, worked out in \cite{Nunez:2019gbg,Macpherson:2024frt}. We obtain
	\begin{equation}
		\CLM = \frac{\mathcal{N}}{2} = \frac{P}{2} \sum_{k=1}^\infty R_k^2 = {a_{\text{UV}}} \ ,
	\end{equation}
	which is consistent with \eqref{eq:CLM_is_aUV} and with the field theory result \eqref{eq:longlimit} for the type A Weyl anomaly coefficient $A_{\text{UV}} = a_{\text{UV}}\slash 4$ \eqref{eq:aUV}. In this way, we have shown that the holographic calculation of $A_{\text{UV}}$ using the Ryu--Takayanagi formula matches the microscopic CFT calculation using \eqref{eq:centralcharges} in the large $P$ and large $N_i$ limits.
	
	In what follows, we apply a similar treatment to holographic duals to a large class of confining field theories.
	For this purpose, we consider a family of backgrounds dual to generic CFTs as written in \eqref{eq:general_metric}, \emph{i.e.}, that their large-$r$ asymptotics will be of the form \eqref{eq:general_metric}. There we have set $d=4$ to be more specific, while a general case is relegated to Appendix~\ref{app:confining}. 
	
	\section{Flows to three-dimensional gapped field theories}\label{sec:flow}
	
	In this section, we explore holographic entanglement measures in three-dimensional field theories with an infrared energy scale. Recall that a key result in three dimensions is a theorem concerning the LM c-function, defined in \eqref{eq:LM-c-function}, which guarantees that it remains constant at fixed points and monotonically decreases along an RG flow towards the IR. This behavior has been verified in various
	holographic top-down models, see for instance \cite{Klebanov:2012yf}.
	When the theory does not flow from a UV CFT but from a more complicated QFT, the c-function diverges at high energies, signalizing the growing in the number of degrees of freedom. This has also been observed holographically in \cite{Jokela:2020wgs}, where the UV theory is a (non-conformal) super Yang--Mills theory with Chern--Simons interactions, see \cite{Faedo:2017fbv}. Still, the infrared develops a mass scale and the c-functions monotonically decrease to zero.
	
	Notably, the solutions introduced in the previous section allow us to construct theories that effectively become three-dimensional in the deep IR. However, when analyzing these theories from an IR perspective and tracing the RG flow toward higher energies, we encounter a puzzling scenario: as an additional dimension emerges and grows, the conventional expectations for c-functions come into question. In the following, we examine the implications of this dimensional transition and its impact on entanglement measures.
	
	\subsection{Top-down holographic example}\label{subsec:c_function_compactification}
	
	We consider a very specific class of Type II backgrounds dual to confining (gapped) field theories. The field theories we work with can be thought of as the deformation of a UV CFT via a VEV for a dimension-three operator. This is achieved by compactifying the UV-CFT on a circle with antiperiodic boundary conditions for fermions and periodic ones for bosons. This compactification breaks a fraction of the supersymmetry of the UV-CFT (if any SUSY was present). The above compactification is accompanied by a twist, a mixing between the compact circle and the R-symmetry (if any) of the CFT. This procedure is explained in detail for the case of ${\cal N}=4$ SYM in \cite{Kumar:2024pcz,Castellani:2024ial}. The procedure triggers an RG-flow from the UV-CFT$_d$ to a gapped IR QFT$_{d-1}$.
	
	The gravity background dual to this flow is inspired by solutions found in gauged supergravity by Anabal\'on and Ross \cite{Anabalon:2021tua,Anabalon:2022aig}. These solutions can be embedded in various backgrounds of Type II supergravity, see for example \cite{Nunez:2023xgl, Nunez:2023nnl,Fatemiabhari:2024aua, Fatemiabhari:2024lct, Chatzis:2024kdu, Chatzis:2024top,Anabalon:2024che,Anabalon:2024qhf}. Our prime example (that we generalize in Appendix \ref{app:confining}) is the one of a family of three-dimensional QFTs that follow from the twisted compactification of a family of CFT$_4$ on a twisted circle. 
	The family of dual backgrounds to this system
	is written in Sec.~2.3 of \cite{Chatzis:2024top}
	and with more detail in Sec.~4.2.1 of \cite{Chatzis:2024kdu}. Details of the construction are relegated to Appendix \ref{app:quiver-geometry}.
	
	In a nutshell, the field theory can be thought of as follows. We start from an infinite family  of six-dimensional ${\cal N}=(1,0)$ linear quiver SCFTs. These are compactified on a hyperbolic space $H_2$
	leading to an infinite family of four-dimensional SCFTs, studied in \cite{Merrikin:2022yho}. These four-dimensional SCFTs are then further twist-compactified on a circle, with a VEV that triggers an RG flow to a three-dimensional family of QFT$_3$ at low energies. There is a precise background describing these flows, for a detailed description see Appendix~\ref{app:quiver-geometry}. The metric and the dilaton (for the accompanying RR and NS fields, see \cite{Chatzis:2024kdu,Chatzis:2024top}) read
	\begin{eqnarray}\label{eq:metricOtherConfiningBackgrounds}
		\dd s_{10, \text{st}}^2 & = & 3\sqrt{6} \pi \sqrt{-\frac{\alpha(z)}{\alpha''(z)}}  \Big[r^2 (-\dd t^2+ \dd \rho^2+\rho^2 \dd\beta^2+ f(r) \dd\phi^2) +\frac{\dd r^2}{r^2 f(r)} + \frac{4(\dd x_1^2+\dd x_2^2)}{3(x_1^2+x_2^2-1)^2}  \nonumber\\
		& & \qquad\qquad\qquad - \frac{\alpha''(z)}{6\alpha(z)}\dd z^2 -\frac{\alpha(z)\alpha''(z)}{(6\alpha'(z)^2- 9 \alpha(z) \alpha''(z))} (\dd\chi^2+\sin^2\chi (\dd\xi +A_g+A_\phi)^2)\Big] \nonumber\\
		e^{-4\Phi} & = & \frac{2^5 3^3}{(18\pi)^{10}}\left( -\frac{\alpha''(z)}{\alpha(z)}\right)^3 (2\alpha'(z)^2-3\alpha(z)\alpha''(z))^2 \label{eq:6d4d}\\
		f(r) & = & 1-\frac{\mu}{r^4} -\frac{q^2}{r^6} \ ,\quad A_\phi= 3q\bigg(\frac{1}{r^2} -\frac{1}{\Rconf^2} \bigg) \dd\phi \ ,\quad A_g=\frac{1}{1-x_1^2-x_2^2}(x_1 \dd x_2-x_2 \dd x_1) \ .\nonumber
	\end{eqnarray}
	Here $\mu$ and $q$ are two free parameters. The parameter $\mu$ typically breaks SUSY and is associated with VEVs for $T_{\mu\nu}$ along the field theory directions. On the other hand, the parameter $q$ performs a twist, mixing the field theory direction denoted by $\phi$ with the $U(1)_\xi$ R-symmetry.  
	Notice that for $r\to\infty$ we have $f(r)\approx 1$, while away from the boundary there is a value of $r\equiv \Rconf$ for which $f(\Rconf) = 0$.
	It will be useful to cast the parameter $q$ in terms of $\Rconf$,
	\begin{equation}
		q^2=    
		\Rconf^2\left(\Rconf^4 - \mu\right)\,.
	\end{equation}
	In general, there would be a conical singularity at $r = \Rconf$, but requiring that the circle smoothly closes off, we find a relation between the size of the circle $\phi\in(0,L_\phi)$ and $\Rconf$,
	\begin{equation}\label{eq:bconf_general-5}
		L_\phi = \frac{4\pi}{\Rconf^2 f'(\Rconf)} = \frac{2\pi \Rconf^ 3}{3\Rconf^ 4 - \mu}\,.
	\end{equation}
	The background \eqref{eq:metricOtherConfiningBackgrounds} is (for $r\to\infty$) holographically dual to a family of CFT$_4$ and the space is asymptotically AdS$_5\times H_2\times M_3$ where $M_3$ is parametrized by the three coordinates $(z,\chi,\xi)$. Moving to smaller values of the radial coordinate $r$, the isometries of AdS$_5$ are broken. This corresponds to a non-trivial RG flow leading to a gapped QFT in three dimensions at low energies. The corresponding family of geometries is parametrized by the function $\alpha(z)$ that satisfies $\alpha''(z) < 0$. The function $\alpha(z)$ is determined by the dual linear quiver field theory, but for our purposes, we do not need its explicit form. For completeness, see \cite{Chatzis:2024kdu,Chatzis:2024top, Merrikin:2022yho} and Appendix~\ref{app:quiver-geometry} for details.
	
	Let us now calculate the entanglement entropy of a disk or radius $R$ and its complement, which will allow us to examine the flow of information afterwards. As usual, we need to find a minimal eight-dimensional hypersurface $\Sigma_8$ that is anchored to the disk at the boundary of spacetime. The symmetries of the problem allow us to assume that the embedding of $\Sigma_8$ is specified by
	\begin{equation}
		t = \text{constant}\,,\qquad \rho = \rho(r)\in[0,R]\ ,
	\end{equation}
	such that
	\begin{equation}\label{eq:boundaryConditionUV}
		\lim_{r\to\infty} \rho(r) = R\ ,
	\end{equation}
	which ensures that the minimal surface is anchored on the boundary of the entangling region as indicated by the RT formula. In particular, the hypersurface $\Sigma_8$ is expanded by $\{\beta,\phi, r,x_1,x_2,z,\chi,\xi\}$, and wraps the internal space completely. Its embedding function $\rho$ depends only on $r$, and  the metric restricted to $\Sigma_8$ becomes
	\begin{eqnarray}\label{eq:metric_compact_direction}
		& &  \dd s_8^2=  3\sqrt{6} \pi \sqrt{-\frac{\alpha}{\alpha''}}  \Big[r^2 \rho^2 \dd\beta^2+ r^2 f(r) \dd\phi^2 +\dd r^2\bigg( \frac{1}{r^2 f(r)} + r^2\rho'^2 \bigg) + \frac{4(\dd x_1^2+\dd x_2^2)}{3(x_1^2+x_2^2-1)^2}\nonumber\\
		& &\qquad\qquad\qquad-\frac{\alpha''}{6\alpha}\dd z^2 -\frac{\alpha\alpha''}{(6\alpha'^2- 9 \alpha \alpha'')} (\dd\chi^2+\sin^2\chi (\dd\xi +A_g+A_\phi)^2)\Big] \ .
	\end{eqnarray}
	The entanglement entropy then takes a simple form
	\begin{equation}\label{eq:EES1}
		S = \frac{1}{4G_{10}} \int \dd^8 x e^{-2\Phi}\sqrt{\det[g_8]}= \tilde{\NN} \int \dd r ~r~ \rho(r) \sqrt{1+ r^4 f(r) \rho'^2}
		\, ,
	\end{equation}
	with \begin{equation}\label{eq:def:tildeN}
		\tilde{{\cal N}} = \frac{ \vol{(H_2)} \vol{(S^2)} \vol{(S^1)}}{2\pi\times 3^5\,G_{10}}L_\phi \int_0^P \dd z (-\alpha(z) \alpha(z)'')\,.
	\end{equation}
	Here, the volumes of the spheres $\vol{(S^2)}$ and $\vol{(S^1)}$ come from integration over $\{\xi,\chi\}$ and $\beta$ respectively, while  $\vol{(H_2)}=\int(x_1^2+x_2^2-1)^{-1} \dd x_1\dd x_2$ is the volume of the two-dimensional hyperbolic plane.
	
	\begin{figure}
		\centering
		\includegraphics[width=0.7\linewidth]{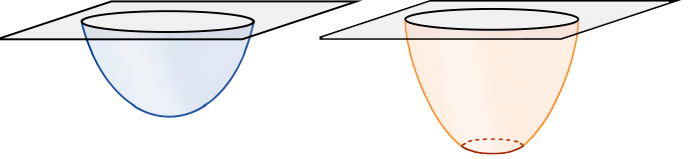}
		\put(-235,-12){(a)}
		\put(-95,-12){(b)}
		\caption{Illustration of two of the possible types of embeddings that solve \eqref{eq:eom3}. Cases (a) and (b) follow from setting \eqref{eq:bc1} and \eqref{eq:bc2} as a boundary condition, respectively. Figure adapted from~\cite{Jokela:2020wgs}.}
		\label{fig:picture_Disk_EE}
	\end{figure}
	
	To compute $S$, it is necessary to first find $\rho(r)$. Extremizing the  action functional \eqref{eq:EES1} leads to a second order non-linear differential equation for $\rho(r)$,
	\begin{equation}\label{eq:eom3}
		\rho'' +\left( \frac{r^4 f'(r)}{2} + 3 r^3 f(r)\right) \rho'^3 -\frac{\rho'^2}{\rho}  +\left( \frac{f'(r)}{f(r)} +\frac{5}{r}\right)\rho' -\frac{1}{r^4 f(r) \rho}=0 \ .
	\end{equation}
	This equation needs to be supplemented with appropriate boundary conditions for $\rho(r)$. The first condition is the UV boundary condition \eqref{eq:boundaryConditionUV} while the second is a regularity condition in the IR. 
	There are two possible qualitatively distinct ways the RT surface may cap off in a regular manner in the interior, which are illustrated in Fig.~\ref{fig:picture_Disk_EE} (a) and (b).
	The reason is that each constant-$r$ slice of the surface is topologically $S^{1}_{\beta}\times S^1_{\phi}\times M_{5}$, where $M_{5}$ is part of the compact manifold of the background, the first circle $S^1_{\beta}$ corresponds to a constant-$\rho$ circle of the entangling surface and the second circle $S^1_{\phi}$ is the compactified field theory spatial $\phi$-direction. In the first possibility represented in Fig.~\ref{fig:picture_Disk_EE} (a), $S^{1}_{\beta}$ shrinks to zero size and the surface caps off before reaching the bottom of the geometry, where $S^1_{\phi}$ shrinks.
	This corresponds to a turning point for the embedding $\rho(r)$ located at $r=r_*>\Rconf$, where the RT surface caps off smoothly if the embedding satisfies the regularity condition 
	\begin{equation}\label{eq:bc1}
		\rho(r_*) = 0,\qquad \rho' (r_*) = \infty\,.
	\end{equation}
	This boundary condition defines a one-parameter family of minimal surfaces parametrized by \mbox{$r_*\in (\Rconf,\infty)$} as in Fig.~\ref{fig:picture_Disk_EE} (a). 
	The other possibility, represented in Fig.~\ref{fig:picture_Disk_EE} (b), is that the embedding reaches the end-of-space at the bottom $r = \Rconf$ where $S^1_{\phi}$ shrinks to zero size. Because the internal manifold shrinks in a regular manner due to \eqref{eq:bconf_general-5}, the RT surface is automatically regular at $r = \Rconf$. The value of $\rho(r)$ remains free and finite at $r = \Rconf$ amounting to the boundary condition
	\begin{equation}\label{eq:bc2}
		\lim_{r\to \Rconf} \rho(r) = \rhoIR\,.
	\end{equation}
	This defines a one-parameter family of minimal surfaces parametrized by $\rhoIR\in (0,\infty)$ as in Fig.~\ref{fig:picture_Disk_EE}~(b).
	
	We have not been able to find analytical solutions in this case, and for this reason, we turn to numerics. For this it is useful to observe that the rescalings
	\begin{equation}\label{eq:scalings}
		\zeta = \frac{\Rconf}{r}\in(0,1]\,,\qquad
		\mut = \frac{\mu}{\Rconf^4}\,, \qquad \rt = \Rconf\, \rho\,,
	\end{equation}
	allow us to factor out the dependence on $\Rconf$, since 
	the equation for the embedding in (\ref{eq:eom3}) and the function $f(r)$ in (\ref{eq:6d4d}) become
	\begin{eqnarray}\label{eq:embeddingdimless-5}
		& & 	
		\ddot\rt(\zeta )+\dot\rt(\zeta )^3 \left(\frac{\dot  f(\zeta )}{2}-\frac{3 f(\zeta )}{\zeta
		}\right)-\frac{\dot\rt(\zeta )^2}{\rt(\zeta)}+\dot\rt(\zeta ) \left(\frac{\dot f(\zeta )}{f(\zeta
			)}-\frac{3}{\zeta }\right)-\frac{1}{f(\zeta )\rt(\zeta )}=0 \nonumber\\
		& &    f(\zeta) = -\mut \zeta ^{4}+(\mut-1) \zeta ^{6}+1\,,
	\end{eqnarray}
	where the dots 
	stand for derivatives with respect to $\zeta$. As promised, the dependence on $\Rconf$ has factored out and the only significant parameter left in the problem is $\mut$. 
	
	To solve this equation numerically, we use series expansions compatible with \eqref{eq:bc1}--\eqref{eq:bc2} to specify the boundary conditions of the numerical integrator. More precisely, solving the equation of motion \eqref{eq:embeddingdimless-5} using the boundary conditions in \eqref{eq:bc1} leads to a series expansion of the form
	\begin{equation}
		\varrho(\zeta) = \sum_{k=1}^\infty b_k (\zeta_*-\zeta)^{\frac{k}{2}}\,,
	\end{equation}
	where $\zeta_* = \Rconf/r_*$ corresponds to the position of the turning point in the new coordinate and all the coefficients $b_k$ are determined in terms of it.
	Similarly, \eqref{eq:bc2} leads to
	\begin{equation}
		\varrho(\zeta) = \varrho_* + \sum_{k=1}^\infty c_k (1-\zeta)^k\,,
	\end{equation}
	with $\varrho_* = \Rconf~ \rhoIR$ and all the coefficients $c_k$ determined in terms of it. Knowing these expansions, we can seed a numerical integrator such as Mathematica's \verb|NDSolve| and integrate up to a value close to the boundary. 
	Near the boundary ($\zeta\simeq 0$), the embedding behaves as
	\begin{equation}\label{eq:expansion_rho_3d}
		\rt(\zeta) = \sum_{k=0}^{\infty} \sum_{l = 0}^{k} a_{k,l} \left(\log \zeta\right)^l\zeta^k = a_{0,0}-\frac{\zeta ^2}{4 a_{0,0}}+\frac{\zeta ^4 \left(32 a_{4,0}
			a_{0,0}^3+\log (\zeta )\right)}{32 a_{0,0}^3}+\mathcal{O}\left(\zeta ^6\right)\,.
	\end{equation}
	Note that this expression contains two free (dimensionless) integration constants $a_{0,0}$ and $a_{4,0}$, that are not fixed asymptotically by the equations, but fixed by the UV and IR boundary conditions. These can be read off from the numerical solution. Physically, $a_{0,0}$ is the radius of the disk \mbox{$R = \Rconf^{-1} a_{0,0}$}, while we understand from the way the solutions are constructed that $a_{4,0}$ is fixed by regularity of the embedding in the bulk. This $a_{4,0}$ plays no role in what follows (though calculating it allows more precision in the results).
	
	\begin{figure}
		\centering
		\includegraphics[width=0.7\linewidth]{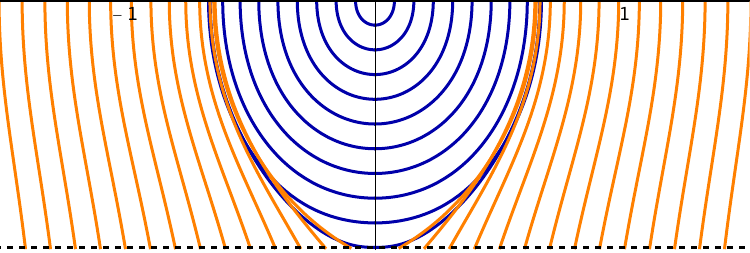}
		\put(10,0){$\zeta = 1$}
		\put(10,95){$\zeta = 0$}
		\caption{Numerical results for the small radius (blue) and large  radius (orange) embeddings $\rt(\zeta)$. We will keep this color coding in what follows. The dashed line is not a real boundary, but the surface where the $S^1$ of the background shrinks to zero size.}\label{fig:embeddings_compact}
	\end{figure}
	The numerical results for several embeddings are shown in Fig.~\ref{fig:embeddings_compact}, where we fixed $\mut = 0$, to preserve supersymmetry. Note that the embeddings that do not reach the end-of-space, as in Fig.~\ref{fig:picture_Disk_EE}~(a), exist for disks whose radius $R$ is smaller than a critical value $R_1\simeq 0.681\Rconf^{-1}$, {\emph{i.e.}}, $R\in(0,R_1)$. Thus, we refer to them as \textit{small radius embeddings}. In contrast, the second type of embeddings portrayed in Fig.~\ref{fig:picture_Disk_EE}~(b), appear for radii above $R_2\simeq0.639\Rconf^{-1}$, and we refer to them as \textit{large radius embeddings}. Crucially, $R_1>R_2$, which means that in between several embeddings exist for the same radius. 
	
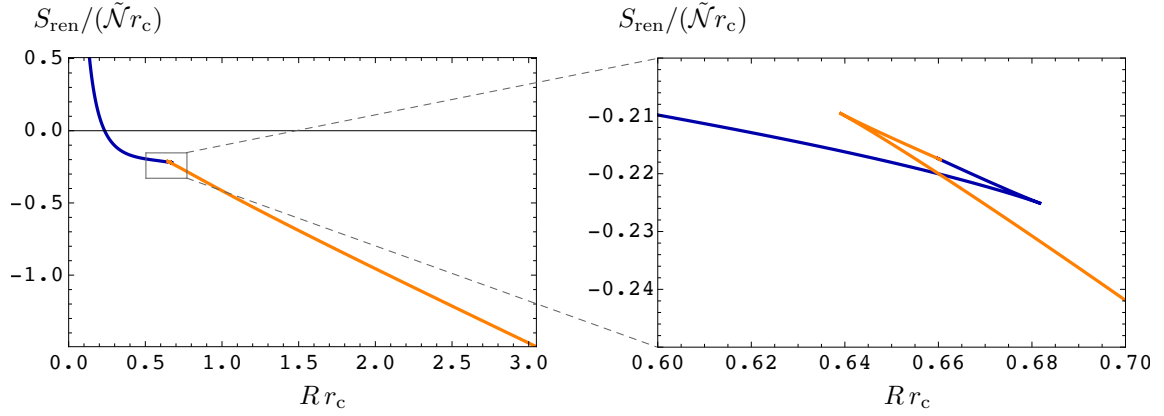
\begin{figure}[t]\centering
		\includegraphics[width=\linewidth]{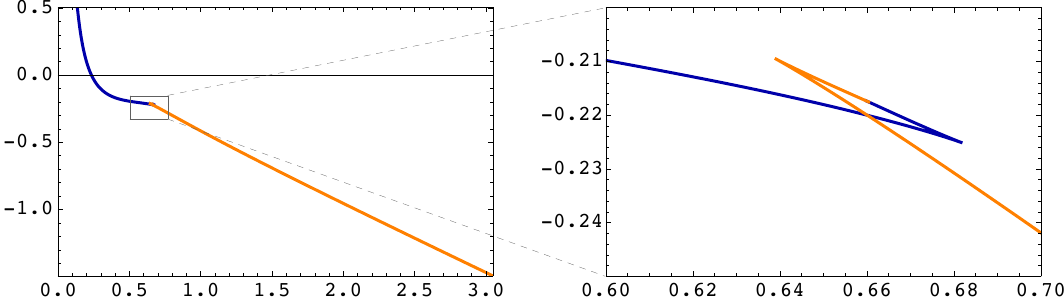}\put(-100,-12){$R\, \Rconf$}
		\put(-200,130){$S_{\text{ren}}/(\tilde\NN \Rconf)$}
		\put(-320,-12){$R\, \Rconf$}
		\put(-420,130){$S_{\text{ren}}/(\tilde\NN \Rconf)$}
		\caption{\small Renormalized entanglement entropy as a function of the radius with the zoomed in version focusing on the swallowtail on the right panel.
		}
		\label{fig:EE}
	\end{figure}

	Now that the embeddings are known, we can compute the EE. Using the redefinitions in \eqref{eq:scalings}, \eqref{eq:EES1} becomes
	\begin{equation} 
		S = -\tilde\NN \Rconf \int_{\zeta_*}^{\zUV}\dd\zeta\,  \frac{\rt}{\zeta ^3} \sqrt{\left(\zeta ^6 (\mut-1)-\zeta ^4 \mut+1\right) \dot\rt^2+1} \equiv  \int_{\zeta_*}^{\zUV} \dd \zeta \, {\cal L}(\rt,\dot\rt)
		\,,\label{dibumartinez}
	\end{equation}
	where we have regulated the divergent integral with a cut-off $\zeta = \zUV$. For the embeddings that reach the end-of-space, $r_*=\Rconf$ and $\zeta_*=1$. We will again focus on the SUSY preserving case $\tilde \mu = 0$. Near the boundary, the integrand possesses two divergent terms, 
	\be\label{eq:lagrEE}	
	{\cal L} = \tilde\NN \Rconf\left(-\frac{a_{0,0}}{\zeta^3} + \frac{1}{8 a_{0,0}\zeta} + \mathcal{O}(\zeta^0)\right) 
	\ee
	which implies that its indefinite integral diverges near the boundary as
	\begin{equation}\label{eq:SEEatUV}
		\int\dd\zeta \, (-{\cal L})= \tilde\NN \Rconf\left(-\frac{a_{0,0}}{2\zeta^2} 
		- \frac{\log(\zeta)}{8 a_{0,0}}+\mathcal{O}(\zeta^0) \right)\,. 
	\end{equation}
	Note that we added an extra minus sign since in \eqref{dibumartinez} the limits of integration are inverted. From this we learn which counterterms are needed at $\zeta = \zUV$ near the boundary before taking $\zUV\to 0$, to render the EE finite,
	\begin{equation}
		\label{eq:Sct}
		\Sct(\zUV) =  \tilde\NN \Rconf\left[\frac{a_{0,0}}{2\zUV^2} + \frac{\log(\zUV)}{8 a_{0,0}}\right]\,. 
	\end{equation}
	Indeed, defining $\Lct$ so that 
	\begin{equation}\label{eq:Lct}
		\int_{\zs}^{\zUV}\dd \zeta\, {\Lct}\equiv
		\tilde\NN \Rconf \int_{\zs}^{\zUV}\dd \zeta\, \parent{-\frac{a_{0,0}}{\zeta^3} + \frac{1}{8 a_{0,0}\zeta} }  = \Sct(\zUV) - \Sct(\zs)\,,
	\end{equation}
	we can define a finite renormalized entanglement entropy as
	\begin{equation}\label{eq:reg-EE}
		S_{\text{ren}} = \lim_{\zUV\to 0} \parent{S -\Sct(\zUV) } = \int_{\zeta_*}^0\dd \zeta\, ({\cal L}-{\Lct}) - \Sct (\zs)\,,
	\end{equation}
	The result is shown in Fig.~\ref{fig:EE}. Note that, as anticipated, there is a range of radii $R\in(R_2,R_1)$ where there are multiple extremal surfaces. The entanglement entropy is the one that minimizes $S$. For this reason, there is a phase transition at $R_{c} \simeq 0.660 r_c^{-1}$, when the curves of the two different branches cross.
	
	\subsection{C-functions upon compactification}\label{eq:c_compactification}
	
	In this section we aim to construct a c-function sensitive to the IR features of the theory. For a low energy observer, as we argued previously, the physics look three-dimensional at energies below the confining scale, dictated by $r_c$. The entangling regions considered in the previous section are four-dimensional cylinders,\footnote{Actually, they are solid tori, since the extra dimension S$^1_\Lrec$ is a circle. Nevertheless, we refer to them as cylinders since the periodicity of this particular direction does not play a significant role here.} but they appear as disks (balls) in the three-dimensional theory. For this reason, it is a natural expectation that if we apply the three-dimensional LM c-function it will have the desired properties of (i) being monotonic as a function of the radius of the disk, (ii) vanishing for large radii and (iii) blowing up towards plus infinity for small radii, due to the presence of an infinite number of Kaluza--Klein (KK) modes. These expectations, however, are not met for the reasons we explain next.
	
	Let us translate the results of the previous section into the notation of Sec.~\ref{sec:liu-mezei}. In four dimensions, the entangling regions are of the form $B_{2}(R)\times S^1_{\Lrec} \subset \mathbb{R}^{2}\times S^1_{\Lrec}$ where the circumference of the compact circle $\Lrec \equiv L_{\phi}$, see \eqref{eq:bconf_general-5}.
	The fact that the region wraps the extra direction completely and does not depend on this direction avoids additional complications.\footnote{See \cite{GonzalezLezcano:2022mcd} where also $(d-1)$-dimensional balls $B_{d-1}(R)$ are considered. Unlike cylinders, they self-intersect when the diameter of the ball is equal to the circumference of the compact circle $R = \Lrec\slash 2$.} The cut-off $\varepsilon$ used in Sec.~\ref{sec:liu-mezei} corresponds holographically to a cut-off in the radial coordinate $r$ \cite{Ryu:2006ef,Ryu:2006bv}. Therefore the cut-off in the $\zeta$-coordinate \eqref{eq:scalings} is given by
	\begin{equation}
		\zUV = \Rconf/r_{\text{\tiny UV}} = \Rconf \varepsilon \propto \varepsilon/\Lrec\,.
	\end{equation}
	The radius of the disk is given by $a_{0,0} = R r_c$. Therefore, from \eqref{eq:SEEatUV} we see that the entanglement entropy \eqref{dibumartinez} of the region $B_{2}(R)\times S^1_{\Lrec}$ has the divergence structure
	\begin{equation}
		S^{\text{cyl}}(R)\equiv S(R) = 
		\widetilde{p}_{4}\frac{\Lrec R}{\varepsilon^{2}} -b_{\text{UV}}\,\frac{\Lrec}{R}\log{\frac{R}{\varepsilon}}+\widetilde{f}(R)+\mathcal{O}(\varepsilon^2)\,,
		\label{eq:cylinder_entropy_d_4}
	\end{equation}
	which indeed coincides with that of $d=4$ cylinders, see \eqref{eq:cylinder_divergence_structure}. Here we identify
	$b_{\text{UV}}$ as the type~B Weyl anomaly coefficient, $\widetilde{p}_{4}$ is independent of $\Lrec,R$, and $\widetilde{f}(R)$ depends on $R$ only through the dimensionless combination $R\slash \Lrec$. The finite term vanishes $\widetilde{f}(0) = 0$ at the UV fixed point \mbox{$R\slash \Lrec = 0$} so that \eqref{eq:cylinder_entropy_d_4} coincides with \eqref{eq:UV_cylinder_entropy} in this case.
	
	The finite term $\widetilde{f}$ is renormalization scheme dependent: under a rescaling $\varepsilon\rightarrow \lambda\varepsilon$ of the UV cutoff, they transform as $\widetilde{p}_4\rightarrow \lambda^{-2}\,\widetilde{p}_4$ and
	\begin{equation}
		\widetilde{f}(R)\rightarrow \widetilde{f}(R)+b_{\text{UV}}\,\frac{\Lrec}{R}\log{\lambda}\,.
		\label{eq:scale_ambiguity}
	\end{equation}
	Now we can immediately see how problematic applying the $d=3$ LM operator to an entanglement entropy with cylinder divergent structure is. If we insist on doing so, we get
	\begin{equation}
		(R\,\partial_R -1)\,S^{\text{cyl}}(R) = \mathcal{F}^{\text{cyl}}_{\text{fin}}(R)-2b_{\text{UV}}\,\frac{\Lrec}{R}\,\biggl(\frac{1}{2}-\log{\frac{R}{\varepsilon}}\biggr)+\mathcal{O}(\varepsilon^2)\,,
		\label{eq:Scyl_3D}
	\end{equation}
	where we have defined the finite part $\mathcal{F}^{\text{cyl}}_{\text{fin}}(R) \equiv (R\,\partial_R -1)\,\widetilde{f}(R)$.
	Equation \eqref{eq:Scyl_3D} is UV divergent, because the $d = 3$ LM operator does not remove the logarithmic divergence. However, we can define a finite quantity $ \mathcal{F}^{\text{cyl}}(R)$ by multiplying \eqref{eq:Scyl_3D} by $R$, subtracting its value at a reference scale $R = R_{\text{\tiny{ref}}}$ and finally dividing by $R$. The result is
	\begin{equation}
		\mathcal{F}^{\text{cyl}}(R) \equiv \mathcal{F}^{\text{cyl}}_{\text{fin}}(R)-\frac{R_{\text{\tiny{ref}}}}{R}\,\mathcal{F}^{\text{cyl}}_{\text{fin}}(R_{\text{\tiny{ref}}})+2b_{\text{UV}}\,\frac{\Lrec}{R}\log{\frac{R}{R_{\text{\tiny{ref}}}}} \ .   
		\label{eq:subtracted_F_cyl} 
	\end{equation}
	We propose that \eqref{eq:subtracted_F_cyl} is a measure of number of degrees of freedom at scales larger than the reference (renormalization) scale $R\gtrsim R_{\text{\tiny{ref}}}$; the UV scale $\Lambda_{\text{UV}}$ corresponds to a reference length scale $R_{\text{ref}} = \Lambda_{\text{UV}}^{-1}$, and hence our c-function is defined at low energies, {\emph{i.e.}}, large $R$. In the IR, $R\rightarrow \infty$ limit with $R_{\text{\tiny{ref}}}$ fixed, we obtain
	\begin{equation}\label{eq:definition_of_tilde_F}
		\lim_{R\rightarrow \infty}\mathcal{F}^{\text{cyl}}(R) = \lim_{R\rightarrow \infty}\mathcal{F}^{\text{cyl}}_{\text{fin}}(R)\,,
	\end{equation}
	which can be interpreted as a topological entropy \cite{Kitaev:2005dm,Levin:2006zz} of a disk in the $d = 3$ IR theory arising from dimensional reduction \cite{Pakman:2008ui} when the $R\rightarrow \infty$ limit exists. The reason for this interpretation is that for large $R$, the solid torus in the four-dimensional UV theory becomes effectively a disk in the three-dimensional IR theory, and as briefly mentioned below equation \eqref{eq:ball_divergence_structure}, the finite $R$-independent constant in the entropy of a disk in a gapped three-dimensional theory is the topological entropy. Here it appears as the constant $\lim_{R\rightarrow \infty}\widetilde{f}(R)$ which is picked up in \eqref{eq:definition_of_tilde_F}.
	
	\begin{figure}[t]\centering
		\includegraphics[width=.49\textwidth]{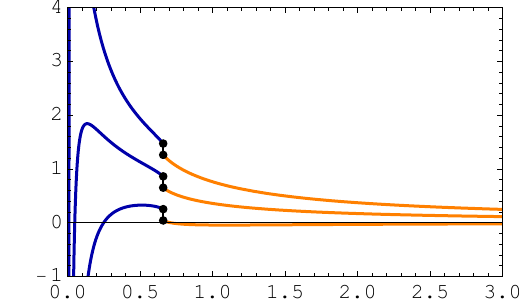}
		\put(-210,130){\small$\mathcal{F}^{\text{cyl}}(R)/(\tilde\NN \Rconf)$}
		\put(-120,-12){\small$R\, \Rconf$}
		\hfill
		\includegraphics[width=.49\textwidth]{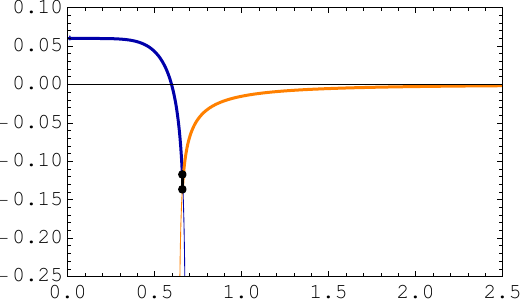}
		\put(-120,15){\includegraphics[width=.25\textwidth]{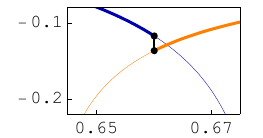}}
		\put(-120,-12){\small$R\, \Rconf$}
		\put(-210,130){\small$\displaystyle \mathcal{C}_{\text{ILN}}(R)/(\tilde\NN \Rconf)$}
		\caption{\small (Left) Proposed measure of the effective number of degrees of freedom from the lower dimensional perspective, \eqref{eq:subtracted_F_cyl}, for different choices of $R_{\text{\tiny{ref}}}\Rconf = 0.01$, $0.05$, and $0.25$. The different curves cross the horizontal axes precisely at these same $R=R_{\text{\tiny{ref}}}$. The ``desired'' properties for a c-function are recovered as $R_{\text{\tiny{ref}}}$ becomes small. The discontinuity, depicted by the vertical black straight line, corresponds to the phase transition of the entanglement entropy, see Fig.~\ref{fig:EE}. 
			(Right) Cylinder c-function as given in \eqref{eq:Ccyl} is depicted as a function of the base radius $R$.}\label{fig:cfunction}
	\end{figure}
	We will now evaluate \eqref{eq:subtracted_F_cyl} using the renormalized entropy computed numerically in the previous section. First, we find that \eqref{dibumartinez} has the divergence structure \eqref{eq:cylinder_entropy_d_4} with
	\begin{equation}\label{eq:terms_at_UV_cyl}
		\widetilde{p}_4 = \frac{\tilde{\mathcal{N}}}{2 \Lrec}\,,\quad b_{\text{UV}} = \frac{\tilde{\mathcal{N}}}{8 \Lrec}\,,\quad \widetilde{f}(R) = (R\,\partial_R - 1)\,S_{\text{ren}}(R)+\frac{\tilde{\mathcal{N}}}{8R}\,(1-2\log{(R r_c)})\,,
	\end{equation}
	where the renormalized entropy $S_{\text{ren}}$ is defined in \eqref{eq:reg-EE}. By substituting to \eqref{eq:subtracted_F_cyl}, we obtain
	\begin{equation}
		\mathcal{F}^{\text{cyl}}(R) = (R\,S_{\text{ren}}'(R)-S_{\text{ren}}(R))-\frac{R_{\text{\tiny{ref}}}}{R}\,(R_{\text{\tiny{ref}}}\,S_{\text{ren}}'(R_{\text{\tiny{ref}}})-S_{\text{ren}}(R_{\text{\tiny{ref}}}))\,.
	\end{equation}
	The numerical result for $\mathcal{F}^{\text{cyl}}(R)$ is shown in Fig.~\ref{fig:cfunction}~(left) for different representative choices of $R_{\text{\tiny{ref}}}$. Let us make some observations. First, $\mathcal{F}^{\text{cyl}}(R_{\text{\tiny{ref}}}) = 0$, which follows directly from its definition \eqref{eq:definition_of_tilde_F}. In particular, at $R=R_{\text{\tiny{ref}}}$ it is increasing from negative values for $R<R_{\text{\tiny{ref}}}$ to positive ones for $R>R_{\text{\tiny{ref}}}$. Second, the function has a global maximum at $R = R_{\text{\tiny max}}\simeq \mathcal{O}(1)\cdot R_{\text{\tiny{ref}}}$. Finally, for values above $R_{\text{\tiny max}}$ the function decreases monotonically towards zero as $R$ increases.
	
	The c-function \eqref{eq:subtracted_F_cyl} is renormalization scheme dependent (encoded in the arbitrary scale $R_{\text{\tiny{ref}}}$), because \eqref{eq:Scyl_3D} depends on the choice of cut-off $\varepsilon$ of the four-dimensional theory. However, this does not mean that \eqref{eq:subtracted_F_cyl} is meaningless: if the UV cut-off $\varepsilon$ of the four-dimensional theory is physical (as would be the case if the UV completion is a lattice theory with lattice spacing $\varepsilon$) then \eqref{eq:subtracted_F_cyl} is a well defined quantity given this fixed cut-off. Similar situation arises in two dimensions where the entanglement entropy \eqref{eq:ball_EE_2D} of an interval contains finite terms that are scheme dependent. In a lattice theory where $\varepsilon$ corresponds to the lattice spacing, these finite terms have well-defined meaning as boundary entropy (see the paragraph below \eqref{eq:ball_EE_2D}). 
	
	A more conservative approach leading to a renormalization scheme independent quantity is to use the ILN c-function \eqref{eq:Ccyl} adapted to cylinders as done in \cite{Ishihara:2012jg}. Applied to our case it gives
	\begin{equation}
		\mathcal{C}_{\text{ILN}}(R) =\frac{R}{2\Lrec}\,(R\,\partial_R-1)(R\,\partial_R+1)\, S(R) = \frac{R}{2\Lrec}\,(R\,\partial_R-1)(R\,\partial_R+1)\,S_{\text{ren}}(R)\,.
		\label{eq:Ccyl2}
	\end{equation}
	Note that this is now finite when the cut-off is taken to zero. 
	Moreover, applying it directly to \eqref{eq:cylinder_entropy_d_4} we discover that 
	\begin{equation}\label{eq:def_Ccyl}
		\mathcal{C}_{\text{ILN}}(R) = b_{\text{UV}}+\frac{R}{2\Lrec}\,(R\,\partial_R-1)(R\,\partial_R+1)\,\widetilde{f}(R)\,,
	\end{equation}
	from  which we conclude that the renormalization scale ambiguity \eqref{eq:scale_ambiguity} is now absent. We plot this quantity in Fig.~\ref{fig:cfunction}~(right). At the UV (small $R$), it goes to a constant, signaling that it is sensitive to the UV physics. While it is not monotonic, it approaches zero at the IR (large $R$). Our conjectured inequality \eqref{eq:Ccyl_inequality_conjecture} implies that $\mathcal{C}_{\text{ILN}}$ can only diverge to $-\infty$ and not to $+\infty$. This is indeed what we observe so that our conjecture holds in this case. More precisely, these divergences are a consequence of the swallowtail-like behavior of the entanglement entropy, characteristic of first-order phase transitions, see Fig.~\ref{fig:EE}~(right). At the cusps of the swallowtail, the slope of the derivative becomes infinite, which implies a divergence in the second derivative. This explains why the divergence appears in $\mathcal{C}_{\text{ILN}}(R)$ but not in $\mathcal{F}^{\text{cyl}}(R)$. Note, however, that this divergence would only occur along an unstable branch of the entanglement entropy curve; when restricting to stable configurations, $\mathcal{C}_{\text{ILN}}(R)$ remains finite.
	
	It is interesting to discuss a generalization of the holographic c-function in Sec.~\ref{sec:cholosubsec}, see (\ref{eq:genericmetriccc})--(\ref{eq:chol}), to a system that exhibits a flow in between different dimensions, such as in the background of (\ref{eq:metricOtherConfiningBackgrounds}). As we explain below, a simple adaptation of (\ref{eq:genericmetriccc}) displays the expected monotonic behavior.
	
	\subsection{A monotonic c-function along the flow}
	
	We briefly discuss a generalization of the holographic c-function \eqref{eq:chol} to a class of bulk metrics larger than \eqref{eq:genericmetriccc}. We use formulas developed in \cite{Bea:2015fja} and summarized in \cite{Merrikin:2022yho}.
	
	Consider a anisotropic field theory in dimension $d$. The dual string frame background of the (anisotropic) form is,
	\begin{equation}\label{eq:cholo_metric}
		\dd s^2 = -a_0 \dd t^2 + \sum_{\alpha=1}^{d-1} a_\alpha (\dd x^\alpha)^2+\prod_{\alpha=1}^{d-1}~(a_\alpha)^{\frac{1}{d-1}}~ \tilde{b}(r)~ \dd r^2+ \sum_{i,j=1}^{9-d}g_{ij}\,(\dd y^i-A^i)(\dd y^j-A^j)\,,
	\end{equation}
	where $a_{\alpha} = a_{\alpha}(r,y^i)
	$ and the system is also accompanied by a dilaton $\Phi(r,y^i)$.
	
	We consider the following integral 
	\begin{eqnarray}\label{eq:cholo_deerminant}
		\sqrt{\mathcal{H}(r)}\equiv \int_{M_{d-1}} \dd^{d-1}x\int_{M_{\text{int}}} \dd^{9-d}y\,e^{-2\Phi}\,\sqrt{\det[g_{8}]}
	\end{eqnarray}
	over the eight-dimensional constant-$t,r$ surface which is equipped with the metric $g_{8}$ (obtained by setting $t,r$ to constants in \eqref{eq:cholo_metric}) and parametrized by coordinates $[x^\alpha, y^i]$. The generalization of $c_{\text{hol}}(r)$ \eqref{eq:chol} to bulk metrics of the form \eqref{eq:cholo_metric} is defined as
	\begin{equation}
		\frac{\tilde{c}_{\text{hol}}(r)}{\vol{(M_{d-1})}}= \frac{(d-1)^{d-1}}{G_{10}}\, \tilde{b}(r)^{\frac{d-1}{2}}\,\frac{\mathcal{H}(r)^{\frac{2d-1}{2}}}{\mathcal{H}'(r)^{d-1}}\,,\label{cflowdiego}  
	\end{equation}
	where $\vol{(M_{d-1})}\equiv \int_{M_{d-1}}\dd^{d-1}x $. We call $\tilde{c}_{\text{hol}}(r)$ the anisotropic holographic c-function, because the metric \eqref{eq:cholo_metric} breaks Lorentz invariance in the field theory directions $x^\alpha$. Consider the isotropic limit in which $a_{\alpha}(r,y^i) = a(r,y^i)$, for $\alpha = 0,\ldots,d-1$, and in which all field theory directions are non-compact $M_{d-1} = \mathbb{R}^{d-1}$. In this case, the metric \eqref{eq:cholo_metric} reduces to \eqref{eq:genericmetriccc} and we see that
	\begin{eqnarray}
		\tilde{b}(r) = b(r)\,,\quad \sqrt{\mathcal{H}(r)} = \vol{(\mathbb{R}^{d-1})}\,\sqrt{H(r)}\ ,
	\end{eqnarray}
	where $b(r)$ and $H(r)$ are defined in \eqref{eq:genericmetriccc} and \eqref{eq:H_integral} respectively and $\vol{(\mathbb{R}^{d-1})}$ is the volume of the field theory directions (a constant). Therefore $\tilde{c}_{\text{hol}}(r) = c_{\text{hol}}(r)$ in the isotropic limit. When computed at conformal points (for isotropic field theories), $\tilde{c}_{\text{hol}}$ reproduces the results of \eqref{eq:chol}.
	
	\begin{figure}[t]\centering
		\includegraphics[width=.60\textwidth]{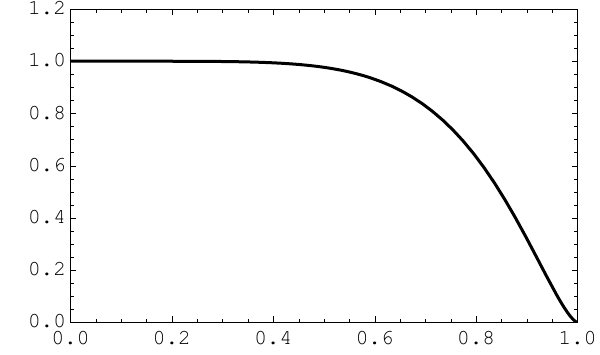}
		\put(-240,160){\small $4\pi\, \tilde{c}_{\text{hol}}/(\tilde\NN\vol(\mathbb{R}^2))$}
		\put(-120,-12){\small$\zeta$}
		\caption{\small Plot of the anisotropic holographic c-function \eqref{cflowdiego} in the $d=4$ backgrounds \eqref{eq:metricOtherConfiningBackgrounds}.}\label{fig:cfunctionHolo_sec4}
	\end{figure}
	
	We can now apply $\tilde{c}_{\text{hol}}$ to our family of  $d=4$ backgrounds \eqref{eq:metricOtherConfiningBackgrounds} for which $c_{\text{hol}}$ is not applicable. The relevant quantities appearing in \eqref{eq:cholo_metric}, \eqref{eq:cholo_deerminant}, and \eqref{cflowdiego} are
	\begin{gather}
		a_0=-a_1=-a_2=- \frac{a_3}{f(r)} = -3\sqrt{6}\pi \sqrt{-\frac{\alpha}{\alpha''}} r^2\,, \qquad \tilde{b}(r)= \frac{1}{r^4 f(r)^{\frac{4}{3}}}\,,\\
		\mathcal{H}=\mathcal{H}_0^2\, r^6 f(r)\,\text{ with }\, \mathcal{H}_0 = \frac{2}{3^5 \pi} \vol{(S^2)}\vol{(H_2)} \vol{(\mathbb{R}^2)}\,  L_\phi\int_0^P \dd z\,(-\alpha \alpha'') = \tilde{\mathcal{N}}\,\frac{2G_{10}\vol{(\mathbb{R}^2)}}{\pi}\,,\nonumber
	\end{gather}
	where we have used \eqref{eq:def:tildeN} and $\vol{(M_{d-1})} = L_\phi\vol{(\mathbb{R}^2)}$ where one of the field theory directions is a circle of circumference $L_\phi$. We obtain
	\begin{equation}
		\frac{\tilde{c}_{\text{hol}}(r)}{\text{Vol}(\mathbb{R}^2)} =  
		\frac{\tilde \NN}{4\pi }\parent{\frac{\sqrt{f(r)}}{f(r) + r f' (r)/6}}^ 3 = \frac{\tilde \NN}{4\pi }\parent{1-\zeta^6}^{\frac{3}{2}}\,
	\end{equation}
	in terms the coordinate $\zeta=\Rconf/r$ defined in \eqref{eq:scalings}. We plot this quantity in Fig.~\ref{fig:cfunctionHolo_sec4}. We find that it detects the UV fixed point (reaching a constant value in the UV) and vanishes in the IR, indicating a gapped QFT.
	
	Had we computed the holographic central charge $c_{\text{hol}}(r)$ \eqref{eq:chol} applicable only to metrics isotropic in field theory directions \eqref{eq:genericmetriccc}, we would have obtained a quantity that vanishes in the IR and diverges in the UV. The anisotropic generalization $\tilde{c}_{\text{hol}}(r)$ of this section captures the presence of an infinite number of KK  modes (encoded in the function $\tilde{b}(r)$), revealing the fixed point value of the UV CFT.
	
	\section{Non-monotonic entanglement c-functions in four dimensions}\label{sec:4d-non-monotonic}
	
	In this section we will illustrate with two examples that the $\CLM$ ceases to be monotonic in four dimensions. We will start by reviewing this issue in the Girardello--Petrini--Porrati--Zaffaroni deformation as originally pointed out in~\cite{Liu:2012eea}, but whose ten-dimensional uplift was only provided recently~\cite{Petrini:2018pjk,Bobev:2018eer,Pilch:2000fu,Baguet:2015sma}. This allows us to sharpen the discussion. We will then continue to a better-behaved Klebanov--Strassler case, the discussion of a similar non-monotonic $\CLM$ thereof distancing IR singularities from the spotlight.
	
	\subsection{Girardello--Petrini--Porrati--Zaffaroni deformation}
	\label{sec:GPPZ}
	
	We consider $\mathcal{N} = 1$ SUSY preserving deformations of $\mathcal{N} = 4$ SU$(N)$ super-Yang-Mills theory in four dimensions.  This SUSY field theory is written in terms of a vector multiplet and three chiral multiplets. We denote these chiral multiplets as $(\Phi_1,\Phi_2,\Phi_3)$, not to be confused with the bulk dilaton field with a similar notation. The deformation of $\mathcal{N}=4$ SYM is written in terms of these chiral multiplets by an addition to the superpotential of the form,
	\begin{equation}
		\Delta\mathcal{W} = m_1\textbf{tr} (\Phi_1^2) +m_2\textbf{tr} (\Phi_2^2)+m_3\textbf{tr} (\Phi_3^2)\,.
	\end{equation}
	This deformation gives masses to the three chiral multiplets. When $m_1=m_2\neq 0$ but $m_3 = 0$ the supersymmetry is enhanced to $\mathcal{N} = 2$ and the corresponding flow is the Pilch--Warner flow \cite{Pilch:2000ue}. When only one chiral multiplet is non-zero the theory flows in the IR to the Leigh--Strassler conformal fixed point \cite{Leigh:1995ep}. When all three masses are non-zero, the theory is called $\mathcal{N} = 1^*$ and has a rich, well-studied structure of vacua. GPPZ is an early attempt to describe the $\mathcal{N} = 1^*$ theory using five-dimensional supergravity \cite{Girardello:1999bd}. It has not been until recently that the ten-dimensional uplift of GPPZ has been understood, see the relevant references~\cite{Petrini:2018pjk, Bobev:2018eer,Pilch:2000fu,Baguet:2015sma}.
	
	The background considered by GPPZ can be obtained from the truncated action
	\begin{equation}\label{eq:actionGH2}
		I\,=\,\frac{1}{4\pi G_5}\int \dd^5x\sqrt{-G}\left(\frac{R}{4}-\frac12 \partial_M\phi\partial^M\phi
		-V(\phi)\right)\,, 
	\end{equation} where $M=0,\ldots,4$, the scalar potential $V(\phi)$ is given in terms of a superpotential $W(\phi)$, 
	\begin{equation}
		\label{eq:pot.from.superpot}
		V(\phi)= -\frac{4}{3}W(\phi)^2 + \frac{1}{2}\left(\frac{\partial W(\phi)}{\partial \phi}\right)^2 \ ,
	\end{equation}
	and
	\begin{equation}
		W(\phi) = -\frac{3}{4\Lads}\left(1+\cosh{\left(\frac{2\phi}{\sqrt{3}}\right)}\right)\,.
	\end{equation}
	Moreover, the five-dimensional Newton's constant relates to the ten-dimensional one as in \eqref{eq:relation_newton_constants} after integrating over the compact manifold, which in this case is a five-sphere, \begin{equation}
		G_5=\frac{G_{10}}{\pi^ 3\Lads^ 5}= \frac{\pi \Lads^ 3}{2N^ 2}\,.
	\end{equation} 
	In the last equality we used the usual holographic dictionary to translate to the rank of the gauge group of the microscopic SYM theory, $\Lads^4/l_s^4 = 4\pi g_sN^2$.
	
	Because the potential is derived from a superpotential, the background is found by solving a system of first order differential equations. Indeed, assuming a domain wall ansatz for the metric,
	\begin{equation}
		\dd s_5^2 = e^{2A}(-\dd t^2 + \dd \vec{x}^2) + \dd \rho^2 \ ,
	\end{equation}
	the ground state of the system is given by the solution to the following BPS-like equations,
	\begin{equation}\label{eq:BPS}
		\partial_\rho \phi = \partial_\phi W\,,\quad \partial_\rho A = - \frac{2}{3} W\,.
	\end{equation}
	We are using that the functions $A(\rho)$ and $\phi(\rho)$ depend only on the radial coordinate. We find it useful to perform the change of radial coordinate $ z = \Lads \exp(\rho/\Lads)$, in terms of which the metric becomes
	\begin{equation}\label{eq:metric.coord.z}
		\dd s_5^2 = e^{2A}(-\dd t^2 + \dd \vec{x}^2) + \frac{\Lads^2}{z^2}\dd z^2\,.
	\end{equation}
	The system \eqref{eq:BPS} can be solved analytically. In terms of the new coordinate the solution reads
	\begin{equation}
		\phi(z) = \sqrt{3}\,\text{arcoth}\parent{\frac{z\Lambda}{\sqrt{3}}}\,, \qquad
		e^{2 A(z)} = \frac{\Lads^2}{z^2} -{\frac{\Lambda^ 2\Lads^ 2}{3}}\,.
	\end{equation}
	Expanding the scalar near the boundary, we discover that the integration constant $\Lambda$ corresponds to the source of the operator dual to $\phi$, which has dimension $\Delta=3$,
	\begin{equation}
		\phi(z)= \Lambda z + \frac{\Lambda^ 3z^ 3}{9}+\ldots\,.
	\end{equation}
	The other integration constant has been fixed so that near the boundary the warp factor diverges as $e^{2A} = \Lads^ 2/z^ 2 + \cdots$. Also, note that the radial coordinate takes values in $z\in\parent{0,z_0}$, with $z_0 = \sqrt{3}/\Lambda$. This introduces an IR scale in the geometry, related to the confining nature of GPPZ flow.
	
\begin{figure}
		\centering
		\includegraphics[width=0.7\linewidth]{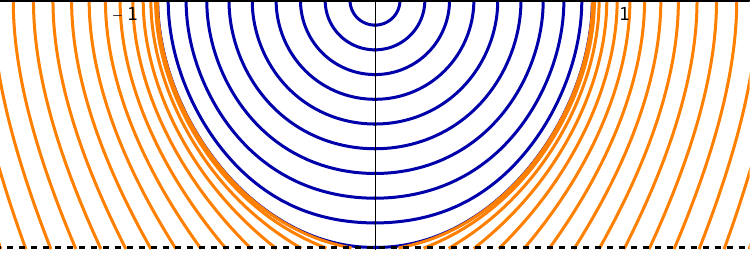}
		\put(-300,110){$\rho(z)/z_0$}
		\put(10,0){$z = z_0$}
		\put(10,95){$z = 0$}
		\caption{Numerical results for  the small radius (blue) and large  radius (orange) embeddings $\rho(z)$. We will keep this color coding in what follows.}\label{fig:embeddings_GPPZ}
	\end{figure}

	Let us now consider entanglement of balls in the background \eqref{eq:metric.coord.z}. The computation is quite similar to that in Section~\ref{sec:flow}. We perform the calculation in five-dimensional supergravity. First, we write the euclidean three-space in spherical coordinates, $\dd\vec{x}^2= \dd \rho^2+\rho^2\dd \Omega_2$. The embedding of the RT surface is specified by $t=$ constant, $\rho = \rho(z)$ with $[z,\Omega_2]$ parametrizing the surface. The three-manifold  (codimension-two manifold in five dimensions) is
	\begin{equation}
		\dd s_3^2=e^{2A(z)}\rho(z)^2~\dd \Omega_2 + \dd z^2\left( \frac{\Lads^2}{z^2} + e^{2A(z)} \rho'(z)^2\right) \,,
	\end{equation}
	where the prime $[']$ stands for derivative with respect to $\rho$.
	The entanglement entropy in this case is
	\begin{equation}\label{eq_SEE_GPPZ}
		S = \NN \int \dd z \, e^{2A(z)}\rho(z)^2\sqrt{\frac{\Lads^2}{z^2} + e^{2 A(z)} \rho'(z)^2}\ ,
	\end{equation} 
	Recall that $\NN$ was given by \eqref{eq:def_curly_N}, which in this case is simply\begin{equation}
		\NN = \frac{\vol{(S^2)}}{4G_5} = \frac{\pi}{G_5} = \frac{2N^2}{\Lads^3}\,.
	\end{equation}
	The corresponding equation of motion for $\rho(z)$,
	\begin{equation}\label{eq:eomtobesolved}
		\frac{\rho '(z) }{\Lads^2 z}\left(z A'(z) \left(3 z^2 e^{2 A(z)} \rho '(z)^2+4 \Lads^2\right)+\Lads^2\right)-\frac{2}{\rho (z)}
		\left(\frac{\Lads^2 e^{-2 A(z)}}{z^2}+\rho '(z)^2\right)+\rho ''(z)=0 \ ,
	\end{equation}
	is subject to the condition $\rho(z = 0) = R$, indicating that the embedding is attached to the surface of the ball of radius $R$ at the boundary. We solve (\ref{eq:eomtobesolved}) numerically. For small values of $z$, the embedding admits an expansion similar to that of~\eqref{eq:expansion_rho_3d}
	\begin{equation}
		\rho(z) = R-\frac{z^2}{2 R}+\frac{z^4 \left(6 a_4 R-\Lambda ^2 \log (z)\right)}{6
			R}+{\cal O}\left(z^6\right)\,,
	\end{equation}
	with $R$ and $a_4$ undetermined by the equations. This is compatible with situations (a) and (b) in Fig.~\ref{fig:picture_Disk_EE}, which correspond to
	\begin{equation}
		\rho(z) = \sqrt{z_*-z}\sum_{k=0}^\infty b_k (z_*-z)^{k}\,,\qquad \rho(z) = \rho_* +\sum_{k=0}^\infty c_k (z_0-z)^{k}\,,
	\end{equation}
	respectively. After constructing the numerical solutions, we can see that indeed these two distinct configurations are realized, as depicted in  Fig.~\ref{fig:embeddings_GPPZ}.
	
\begin{figure}[t]
		\centering
		\includegraphics[width=0.49\linewidth]{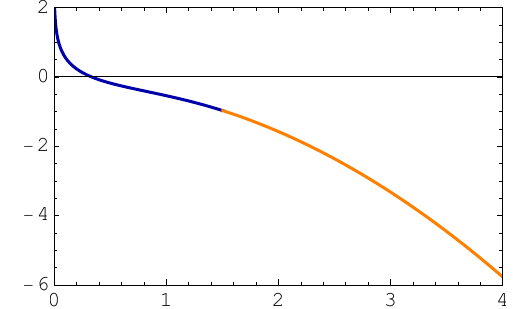}
		\put(-180,130){$S_{\mt{ren}}(R)/(2N^2)$}
		\put(-30,-10){$R \Lambda$}
		\hfill 
		\includegraphics[width=0.49\linewidth]{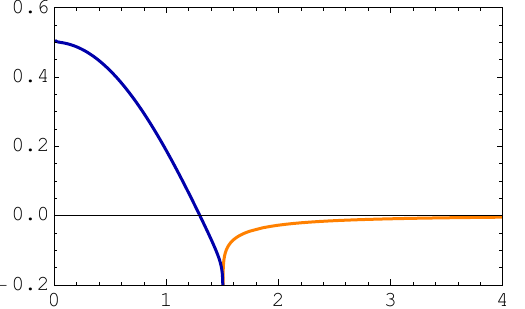}
		\put(-180,130){$\CLM/(2N^2)$}
		\put(-30,-10){$R \Lambda$}
		\caption{(Left) Entanglement entropy and (right) c-function as a function of ball radius $R$. The c-function is non-monotonic but assumes constant values at the fixed points and appears to be stationary to the precision of our numerics.}
		\label{fig:entanglement_GPPZ}
	\end{figure}

	Now, to compute the entanglement entropy, we need to regularize it by adding counterterms. Following the same procedure as in Section~\ref{subsec:c_function_compactification}, we write
	\begin{equation}
		S_{\text{ren}} = \lim_{{z_{\text{\tiny UV}}}\to 0} \parent{S -\Sct(z_{\text{\tiny UV}}) } = \int_{z_*}^0  ({\cal L}-{\Lct})\,\dd z- \Sct (z_*)\,,
	\end{equation}
	with
	\begin{equation}
		\Lct = 2N^2\left(\frac{ R^2}{z^3} - \frac{(3 + 2 R^2 \Lambda^2)}{6z}\right)
		\,,\quad 
		\Sct = -\frac{N^2}{3} \left( \frac{3 R^2}{z^2} + \left( 3 + 2 R^2 \Lambda^2 \right) \log z \right).
	\end{equation}
	The result is shown in Fig.~\ref{fig:entanglement_GPPZ}~(left). It does not coincide with that of \cite{Liu:2012eea} (see figure 10 there) since we used a different renormalization scheme. Nonetheless, we also find a transition between the small and the big radius embeddings at $R = R_c \simeq 0.870 z_0 = 1.507 \Lambda^{-1}$. Within our precision, the transition seems smooth in this case ({\emph{i.e.}}, it appears that the curve of $S_{\mt{reg}}$ as a function of $R$ is continuous, in contrast to Fig.~\ref{fig:EE}).
	
	Finally, we compute the c-function. Recalling its expression for $d=4$ from \eqref{eq:LM-c-function_bis}
	we plot it in Fig.~\ref{fig:entanglement_GPPZ}~(right). This is renormalization scheme-independent, thus the fact that it agrees with Ref.~\cite{Liu:2012eea} constitutes a nice check of our numerical approach. From our numerical results, we have checked that the first derivative of the entanglement entropy is smooth and continuous. In particular, this implies that its second derivative (and as a consequence, $\CLM(R)$) is finite. Due to limited numerical accuracy, we have not been able to assess whether these are continuous functions. We note that the bound in \eqref{eq:CLM_ball_inequality} is satisfied, however.
	
	As it was already pointed out in \cite{Liu:2012eea}, we obtain a c-function that is neither monotonic nor positive definite, even though it has the desired property of being a constant at $R=0$ 
	(corresponding to the UV CFT) and approaching zero as $R\to\infty$, when the confining IR is explored. One reason adduced for the non-monotonicity is the irregularity of the flow in the IR. Indeed, even the ten-dimensional uplift of the GPPZ suffers from an IR singularity, see for instance ref.~\cite{Petrini:2018pjk}. To understand whether regularity can cure this unsatisfactory behavior, in the next section we study the same quantity in a top-down model with a perfectly regular IR: the KS background \cite{Klebanov:2000hb}.
	
\begin{figure}[t]\centering
		\includegraphics[width=.60\textwidth]{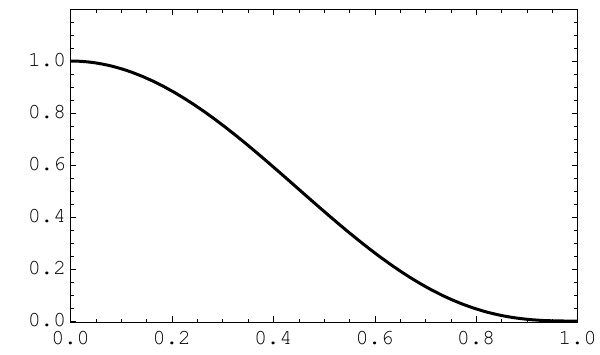}
		\put(-240,160){\small $4\pi\, c_{\text{hol}}/(N^2 {\vol(\mathbb{R}^3)})$}
		\put(-120,-12){\small$z/z_0$}
		\caption{\small The isotropic holographic c-function \eqref{eq:chol_GPPZ} in the GPPZ background.}\label{fig:cfunctionHolo_GPPZ}
	\end{figure}

	Finally, we compute the (isotropic) holographic central charge $c_{\text{hol}}$ defined in \eqref{eq:chol} for the case of the GPPZ background. In this case, it coincides with the anisotropic c-function $\tilde{c}_{\text{hol}}$ defined in \eqref{cflowdiego}, and is given by
	\begin{equation}\label{eq:chol_GPPZ}
		\frac{c_{\text{hol}}}{\vol(\mathbb{R}^3)}=\frac{\tilde{c}_{\text{hol}}}{\vol(\mathbb{R}^3)}= \frac{1}{8 G_5} \frac{\Lads^3}{z^3 A'^3}= \frac{\Lads^3}{8 G_5} \parent{1-  \frac{z^2 \Lambda^2}{3}}^3
		= \frac{N^2}{4\pi}  \parent{1-  \frac{z^2 }{z_0^2}}^3\,.
	\end{equation}
	We plot this in Fig.~\ref{fig:cfunctionHolo_GPPZ} and note that it behaves nicely, starting as a constant (with zero derivative) in the UV, smoothly decreasing towards the IR, eventually vanishing.
	
	\subsection{Klebanov--Strassler background}
	
	In the previous section we reviewed how the c-function obtained from spheres in the GPPZ background is not monotonic. As stated in Ref.~\cite{Liu:2012eea} the reason could be that the GPPZ background is singular in the IR. For this reason we turn our attention to the KS solution \cite{Klebanov:2000hb}, which is a regular solution to Type IIB supergravity equations. 
	
	The KS solution is sourced by $N$ D3-branes at the tip of the conifold, in addition to $M$ fractional three-branes sourcing a magnetic three-form flux through a three-sphere inside the conifold. This is dual to a cascading, confining SU$(N+M) \ \times \ $SU$(N)$ supersymmetric gauge theory, provided $N$ is a multiple of $M$. We refer to the original work for the details. The metric can be written as
	\begin{equation}\label{eq:KS_background}
		\dd s_{10}^2 = h^{-1/2}(\tau) \, \parent{-\dd t^ 2 + \dd \rho^ 2 + \rho^ 2 \dd \Omega^ 2_2} + h^{1/2}(\tau) \, \dd s_6^2 \ ,
	\end{equation}
	where $\tau\in (0,\infty)$ is the radial coordinate, $\dd \Omega^ 2_2$ is the line element of a two sphere,
	\begin{equation}
		\dd s_6^2 = \frac{1}{2} \eKS^{4/3} K(\tau) \left[ \frac{1}{3K^3(\tau)} \left( \dd \tau^2 + (g^5)^2 \right) + \cosh^2 \left( \frac{\tau}{2} \right) \left[ (g^3)^2 + (g^4)^2 \right] + \sinh^2 \left( \frac{\tau}{2} \right) \left[ (g^1)^2 + (g^2)^2 \right] \right]
	\end{equation}
	is the metric of the deformed conifold with deformation parameter $\eKS$, and 
	\begin{equation}
		K(\tau) = \frac{\parent{\sinh(2\tau) - 2\tau}^{1/3}}{2^{1/3}\sinh(\tau)},\quad 
		h(\tau) =
		\frac{(g_s l_s^2 M)^2 2^{2/3}}{\eKS^{8/3}}
		\int_\tau^\infty \dd x \frac{x\coth x-1}{\sinh^2 x} (\sinh (2x) - 2x)^{1/3}\,.
	\end{equation}
	The dilaton is constant and set to zero. The details concerning the angular forms $g_i$ can be found in refs.~\cite{Klebanov:2000hb,Herzog:2001xk}.
	
	This geometry has a perfectly regular IR at $\tau = 0$. In this region we find two meaningful energy scales,
	\begin{equation}
		\Lambda_s = \frac{\eKS^{2/3}}{l_s^2\sqrt{g_s M}}\,,\qquad \Lambda_{\text{\tiny KK}}  = \frac{\eKS^{2/3}}{l_s^2{g_s M}}\,,
	\end{equation}
	related to the confining string tension and the masses of the glueball and KK states, respectively.
	
	To study entanglement entropy of spheres in this background we consider an eight-dimensional hypersurface that wraps completely the compact space spanned by the forms $g^i$ and is attached to the sphere on the boundary at a constant time slice. Similar to the previous cases, the embedding is given in terms of an embedding function  $\rho(\tau)$, that depends only on the radial coordinate. More precisely, the induced metric on $\Sigma_8$ becomes
	\begin{align}
		\dd s_{8}^2 = &\Bigg[h^{-1/2}\rho'(\tau)^2 + \frac{h^ {1/2}\eKS^{4/3}}{6K^ 2}\Bigg]\dd\tau^2 +h^{-1/2} \rho^ 2 \dd \Omega^ 2_2\nonumber \\
		&\qquad+ \frac{h^ {1/2}\eKS^{4/3}}{6K^ 2}g_5^2 + \frac{1}{2}h^{1/2}\eKS^{4/3}K\parent{\cosh\parent{\frac{\tau}{2}}^2(g_3^ 2+g_4^ 2) +\sinh\parent{\frac{\tau}{2}}^2(g_1^ 2+g_2^ 2) } \ ,
	\end{align}
	which leads to
	\begin{equation}\label{eq:entanglement_KS}
		S = \frac{1}{4G_{10}}\frac{\eKS^{10/3}}{96}\int\dd\Omega_2\int g_1\wedge\cdots \wedge g_5\int\dd \tau \sqrt{h(\tau )} K(\tau ) \rho(\tau )^2 \sinh ^2(\tau ) \sqrt{\frac{\eKS ^{4/3} h(\tau )}{ K(\tau
				)^2}+6\rho'(\tau )^2}  \ . 
	\end{equation}
	We decide to work in terms of the dimensionless variables 
	\begin{equation}
		\rt(\tau) = \Lambda_{\text{\tiny KK}}\rho(\tau)/2^{1/3} \ , \quad \htt(\tau) = \eKS^{4/3}\Lambda^{2}_{\text{\tiny KK}} h(\tau)/2^{2/3}
	\end{equation}
	in terms of which \eqref{eq:entanglement_KS} becomes
	\begin{align}\label{eq:entanglement_KS_dimless}
		S &= g_s^2 M^4 \frac{16\cdot 2^{1/3}}{\pi^2}\int\dd \tau \sqrt{\htt(\tau )} K(\tau ) \rt(\tau )^2 \sinh ^2(\tau ) \sqrt{\frac{\htt(\tau )}{ K(\tau
				)^2}+6\rt'(\tau )^2} \\
		&\equiv g_s^2 M^4\int\dd\tau\, \LL(\rt,\rt') \ ,
	\end{align}
	where we have used that $\int g_1\wedge\cdots \wedge g_5 = 64\pi^3$.
	
	\begin{figure}
		\centering
		\includegraphics[width=0.7\linewidth]{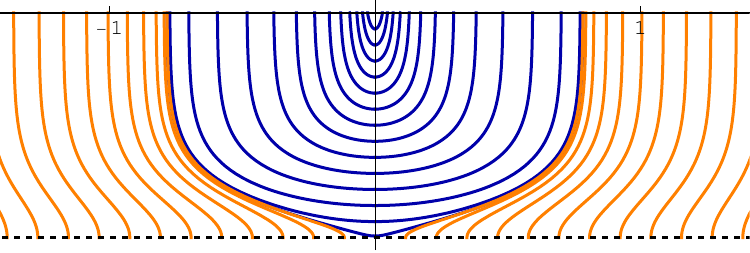}
		\put(-300,110){$\rt(\tau)$}
		\put(10,0){$\tau = 0$}
		\put(10,95){$\tau=\tau_c$}
		\caption{Embedding functions as a function of the radial coordinate. Here $\tau =0$ corresponds to the IR and the boundary is at $\tau\to\infty$. Blue curves stand for ``disconnected'' configurations, while orange curves correspond to ``connected'' configurations.}\label{fig:embeddings_KS}
	\end{figure}
	The equation of motion for $\rho(\tau)$ is again obtained in the usual way,
	\begin{equation}
		\label{eom.spheres.KS}
		\frac{\partial\LL}{\partial \rt(\tau)} - \partial_\tau \frac{\partial\LL}{\partial \rt'(\tau)}=0\,.
	\end{equation}
	The explicit expression in this case is quite involved and not particularly insightful, so we chose not to write it out explicitly. It has to be supplemented with the boundary condition $\rt(\tau\to\infty) = a_0 = \Lambda_{\text{\tiny KK}}R/2^{1/3}$, with $R$ the radius of the corresponding sphere. It turns out that the perturbative expansion about the UV that fulfills this requirement is richer than in the previous sections. This is inherited from the lack of conformality at the UV. Indeed, the embeddings approach the boundary as
	\begin{equation}\label{eq:UV:sphere:KS}
		\rt(\tau) = a_0 + f_1(\tau)e^{-2\tau/3}+ f_2(\tau)e^{-4\tau/3}+ f_3(\tau)e^{-2\tau}+\ldots\,.
	\end{equation}
	When this ansatz is substituted into the equation of motion \eqref{eom.spheres.KS}, at every order in $e^{-2\tau/3}$ we encounter a differential equation for the coefficients $f_i(\tau)$. The two first coefficients can be solved analytically, and we find,
	\begin{equation}\label{eq:coefficientsUV}
		\begin{aligned}
			f_1(\tau) &= \frac{9-36\tau}{32a_0}\,, \\
			f_2(\tau) &=  a_4 + \frac{3}{256 a_0^3} \left(216 \tau - 108 \tau^2 - 243 e^{(4\tau-1)/3} 
			\text{Ei}\left(\frac{1 - 4 \tau}{3}\right)\right) \\
			&=
			-\frac{81 \tau^{2}}{64 a_0^3} 
			+ \frac{81 \tau}{32 a_0^3} 
			+ a_4
			+ \frac{2187}{1024 a_0^3 \tau} 
			- \frac{2187}{2048 a_0^3 \tau^2} 
			+ \frac{28431}{16384 a_0^3 \tau^3} 
			+ \ldots\,,
		\end{aligned}
	\end{equation}
	where Ei$(x)$ is the exponential integral function of $x$.
	Higher-order coefficients $f_i(\tau)$ can be found in series expansion of $\tau$, but we will not need them. At the end of the day, we are left with two undetermined coefficients, $a_0$ and $a_4$. 
	\begin{figure}[t]
		\centering
		\includegraphics[width=0.49\linewidth]{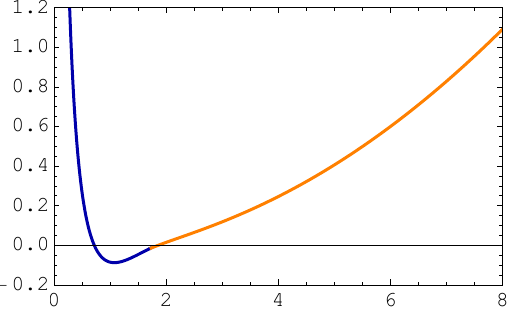}
		\put(-190,130){$S_{\mt{ren}}(R)/(g_s^2M^4)$}
		\put(-40,-10){$R \Lambda_{\text{\tiny KK}}$}
		\includegraphics[width=0.49\linewidth]{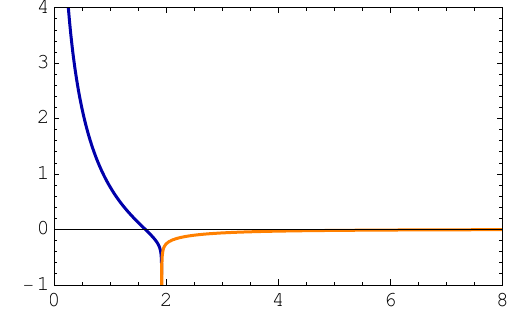}
		\put(-190,130){$\CLM(R)/(g_s^2M^4)$}
		\put(-40,-10){$ R \Lambda_{\text{\tiny KK}}$}
		\hfill
		\caption{(Left) Sphere entanglement entropy as a function of the radius in KS geometry. (Right) Corresponding c-function.    }
		\label{fig:entropy.and.c_KS}
	\end{figure}
	
	In the KS background we also find small and large radius embeddings, see Fig.~\ref{fig:picture_Disk_EE} (a) and (b). The corresponding boundary conditions that these fulfill in the bulk are, respectively
	\begin{equation}
		\rt(\tau) = \sqrt{\tau-\tau_*} \sum_{k=0}^\infty b_k (\tau-\tau_*)^k\,,\qquad \rt(\tau) = \rt_* + \sum_{k=1}^\infty b_k \tau^{2k}\,.
	\end{equation}
	These expansions are used to seed the integrator and find the solution numerically, up to a large value $\tau = \tau_c$  sufficiently close to the boundary. Due to the exponential dependence on $\tau$, it turns out that a choice $\tau_c = 14$ suffices for the numerical precision. Some representative solutions are shown in Fig.~\ref{fig:embeddings_KS}.  We can then evaluate each numerical solution at $\tau =\tau_c$ to find the corresponding values of $a_0$ and $a_4$, by equating the result to \eqref{eq:UV:sphere:KS}--\eqref{eq:coefficientsUV}.
	Furthermore, the UV expansion can be used to find appropriate counterterms to regulate the entanglement entropy. Indeed, near the UV
	\begin{equation}\label{eq:lagrangian_KS_UV}
		\LL = a_0^2\frac{  (4 \tau -1)}{32 \pi ^2}e^{2 \tau /3}-\frac{9 \left(16 \tau ^2+40 \tau -47\right)}{1024 \pi ^2} + \mathcal{O}(e^{-2\tau/3})\,,
	\end{equation}
	and so, 
	\begin{equation}\begin{aligned}
			\Lct(\tau) &= a_0^2\frac{  (4 \tau -1)}{32 \pi ^2}e^{2 \tau /3}-\frac{9 \left(16 \tau ^2+40 \tau -47\right)}{1024 \pi ^2}\,, \\
			\Sct(\tau) &= g_s^2M^4\left( a_0^2\, \frac{3  (4 \tau -7)}{64 \pi ^2}e^{2 \tau /3} -\frac{3 \tau  \left(16 \tau ^2+60 \tau -141\right)}{1024
				\pi ^2}\right)\,,
		\end{aligned}
	\end{equation}
	can be used to regularize the integral. It is worth noticing that this has two parts. The first term is proportional to $a_0^2= \Lambda_{\text{\tiny KK}}^2R^2/2^{2/3}$, and it thus corresponds to the usual area-law term. The second term, on the other hand, is independent of $a_0$ and will be unimportant when we take derivatives with respect to the radius (for example, when computing the c-function). 
	
	The renormalized entanglement entropy now becomes
	\begin{equation}\label{eq:SEE_reg_KS}
		S_{\text{ren}} = \lim_{\tau_{\text{\tiny UV}}\to\infty} (S-\Sct (\tau_{\mt{UV}})) = g_s^2M^4\int_{\tau_*}^\infty({\cal L} - \Lct)\,\dd\tau  - \Sct (\tau_*)\,,
	\end{equation}
	which is a regular quantity.\footnote{In practice, we cannot perform the integral all the way up to $\tau=\infty$, but up to $\tau=\tau_c$. To minimize the error made by cutting off the integral in this way, we actually added the integral of the order $e^{-2\tau/3}$ in $\LL$ ({\emph{i.e.}}, the first order that is not shown in \eqref{eq:lagrangian_KS_UV}), between $\tau_c$ and $\infty$, which can be given analytically in terms of $a_0$ and $a_4$.} For large radius embeddings,  $\tau_*  = 0$.
	
	The results for the entanglement entropy are shown in Fig.~\ref{fig:entropy.and.c_KS}~(left). Like in the previous cases, we see there is a phase transition between the two possible configurations. Now we can compute the $d=4$ c-function in this background, \eqref{eq:LM-c-function_bis},
	\begin{equation}\label{eq:cfuncKS}
		\CLM(R)=\frac{1}{2}(R\partial_{R})(R\partial_{R}S - 2S) = \frac{1}{2}(R\partial_{R})(R\partial_{R}S_{\mt{ren}} - 2S_{\mt{ren}})\,.
	\end{equation}
	In any case, we believe that the sudden change originates from the higher derivatives in \eqref{eq:LM-c-function_bis}, which make the c-function sensitive to the curvature of the entanglement entropy. Finally, we see that the main (and actually, unique) qualitatively distinct element that differentiates GPPZ from KS is that in the latter the c-function diverges in the UV ({\emph{i.e.}}, for small values of $R$), while in the former it  reaches a constant value. This is nonetheless expected, given that the UV of KS is dual to a cascading theory whose number of degrees of freedom is growing, while GPPZ flows from a conformal theory.
	
	\begin{figure}[t]\centering
		\includegraphics[width=.5\textwidth]{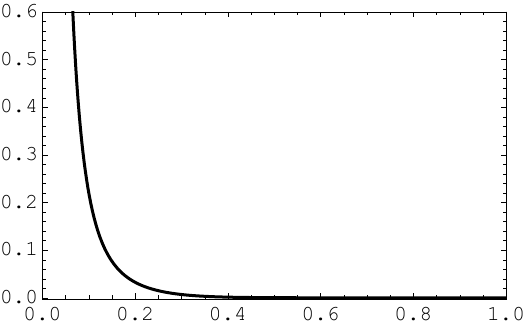}
		\put(-240,140){\small ${c_{\text{hol}}}/(g_s^2M^4 \vol(\mathbb{R}^3))$}
		\put(-120,-12){\small$\mathbf{x}=(1+\tau)^{-1}$}
		\caption{\small The isotropic holographic c-function in the KS background as a function of $\mathbf{x}=(1+\tau)^{-1}$. The UV is on the left of the plot ($\mathbf{x}=0$) and the IR on the right ($\mathbf{x}=1$), conforming with the conventions in this paper.
		}\label{fig:cfunctionFlow_KS}
	\end{figure}
	
	It is interesting to compare our result with the isotropic holographic c-function \eqref{eq:chol} in the KS background. In this case with \eqref{eq:chol}, \eqref{eq:H_integral} become
	\begin{equation}\label{eq:chol_KS}
		\begin{aligned}
			\sqrt{H(r)} &= \vol(\mathbb{R}^3) \frac{2\cdot 2^{5/6}\pi ^3 }{3^{1/2}} (l_s^2g_s M \epsilon ^2)\times \sqrt{\htt(\tau )} K(\tau ) \sinh ^2(\tau ) \\
			\frac{c_{\text{hol}}}{\vol(\mathbb{R}^3)}&=g_s^2M^4 \frac{3}{2 \cdot 2^{2/3}\pi^3}\frac{\sinh^2(\tau)\, \htt(\tau)^2}{\displaystyle K(\tau)^2 \left( \frac{2 K'(\tau)}{K(\tau)} + 4 \coth(\tau) + \frac{\htt'(\tau)}{\htt(\tau)} \right)^3}\ ,
			\,
		\end{aligned}  
	\end{equation}
	respectively. The result is plotted in Fig.~\ref{fig:cfunctionFlow_KS}. It shares features of the LM c-function at the end-points of the flow: it diverges in the UV and approaches zero in the IR. However, along the flow, it interpolates between the two values monotonically.

	\section{Conclusions and discussion}\label{sec:discussion}
	
	In this work, we explored the effectiveness of (entanglement) c-functions in identifying the number of degrees of freedom across various quantum field theories. Utilizing the principles of holography, we were able to conduct this investigation within several strongly coupled theories. Our primary focus was on scenarios where the geometric framework is clearly delineated through the top-down approach of brane construction.
	
	We carried out a systematic analysis in ${\cal N}=2$ superconformal field theories in generic number of spacetime dimensions $d+1=2,3,\ldots,7$ constructed in~\cite{Lozano:2020txg, Lozano:2020bxo, Assel:2011xz, Lozano:2019zvg, Akhond:2021ffz, DHoker:2016ysh, Legramandi:2021uds, Nunez:2019gbg, Gaiotto:2009tk, Apruzzi:2015wna}. In these cases the dual geometry had an explicit AdS metric factor, taking the general form written in \eqref{eq:general_metric}. The corresponding LM c-functions were determined to be constants and be related to the type A Weyl anomaly coefficient for the different CFTs, summarized in Table~\ref{table:differentQs}. 
	
	After recognizing that LM and ILN c-functions provide potential measures for the degrees of freedom in CFTs, we extended our analysis to QFTs with intrinsic energy scales, especially those pertinent to confining physics. Our initial focus was on the deformation of AdS$_5$ backgrounds~\cite{Chatzis:2024kdu,Chatzis:2024top} given in \eqref{eq:metricOtherConfiningBackgrounds}, which preserve parts of the supersymmetry according to the framework established by Anabal\'on and Ross \cite{Anabalon:2021tua,Anabalon:2022aig}. 
	On the field theory side one finds a family of  4d ${\cal N}=1$ SCFT compactified on a circle.
	The corresponding dual QFTs are effectively $(2+1)$-dimensional at low energies, growing one dimension as we move towards the UV. In this case, like in the \mbox{$(3+1)$}-dimensional bottom-up AdS soliton models studied in \cite{Ishihara:2012jg,Fujita:2023bdk}, the ILN c-function reaches large
	negative values associated with a transition in the RT surface. This indicates that non-monotonicity is not an artifact of the bottom-up models, but also persists in IR-regular top-down constructions. 
	
	Because the theories become effectively three-dimensional in the IR, we attempted to apply the three-dimensional LM c-function to these geometries. Ultimately, this was motivated by the existence of a monotonicity theorem in three dimensions \cite{Casini:2012ei,Casini:2015woa,Casini:2017vbe}. Unfortunately, the entanglement entropy of these regions exhibits the divergence structure of cylinders at small scales, which the three-dimensional LM c-function is not designed to address. This has dramatic consequences: if we persist on using this c-function, the outcome becomes dependent on the renormalization scheme. Despite this, we managed to  construct an IR-adapted effective c-function that retains desired properties below a renormalization energy scale.
	
	We also investigated four-dimensional confining theories that do not rely on the shrinking of one of the field theory directions. We began by revisiting the GPPZ model, originally analyzed in \cite{Liu:2012eea}. As pointed out there, the four-dimensional LM c-function is neither monotonically decreasing nor positive definite in this setup. The problematic singular nature of the GPPZ flow at the IR was cited as the culprit for this inadequacy. In our work, we extended this analysis to the KS background, which is perfectly regular in the IR. Despite this, we observed similar behavior, with the c-function failing to exhibit positivity or monotonicity. Therefore IR singularities appear not to be connected with non-monotonicity.
	
	Comparing the two scenarios studied, we observe that the LM and ILN c-functions behave almost identically: both are non-monotonic and exhibit a sudden change of behavior at some negative value. This is notable given that the LM c-function matches the type A Weyl anomaly coefficient at the UV fixed point, while the ILN c-function is sensitive to the type B coefficient. In \cite{Ishihara:2012jg}, the non-monotonicity of the ILN c-function is attributed to the type B coefficient being an unreliable measure of degrees of freedom.\footnote{The type B coefficient need not decrease along RG flows \cite{Anselmi:1997am,Anselmi:1997ys} and may vary under marginal deformations in holographic settings \cite{Nakayama:2017oye}.} However, we find this unsatisfactory, since the LM c-function, tied to the monotonic type A coefficient \cite{Komargodski:2011vj}, behaves non-monotonically in the same way. This view is supported by the fact that, in holographic theories dual to Einstein gravity, the type A and B coefficients coincide \cite{Henningson:1998gx,Henningson:1998ey}. We verified this explicitly for 4d ${\cal N}=2$ SCFTs dual to our top-down supergravity backgrounds in Section~\ref{sec:3d-non-monotonic}. These linear quiver theories require, for the holographic regime, long quivers ($P \to \infty$), large-rank gauge nodes ($N_i \to \infty$), and sparse flavor groups. Under these conditions, the type A and B coefficients agree, and the ILN c-function captures the type A coefficient of the UV theory in our setup.
	
	As alluded to in the introduction, we believe that the underlying reason for the non-monotonic behavior of both the LM and ILN c-functions is a phase transition in the RT surface, which is characteristic of confining models at large-$N$. This transition appears to be at least second order, allowing the second derivatives of the entanglement entropy and consequently the c-functions to switch from decreasing to increasing behavior, thereby violating monotonicity. We conclude that defining a fully monotonic c-function from the entanglement entropy of compact regions in gapped theories, which satisfies both monotonicity and fixed point criteria, remains an open challenge.
	
	We also considered an alternative c-function which may be constructed in holographic setups following \cite{Macpherson:2014eza,Bea:2015fja}. For the backgrounds we studied this has the desired properties: it interpolates monotonically between constants in the UV and the IR. This applies both to flows induced by circle compactifications and by deformations. It may be interesting in future work to further study our c-functions. For example, it is natural to ask if there is a more covariant way of writing the expressions in equations \eqref{eq:chol} and \eqref{cflowdiego}. Along the same line, it is known that certain non-singular supergravity backgrounds  display non-monotonic c-functions \cite{Hoyos:2021vhl}. It would be nice to understand the mechanism and dynamical reasons for this. This might teach us something about the proposed c-function, or about these special backgrounds, or both. We leave this for future work.
	
	\begin{acknowledgments}
		We thank C\'esar Ag\'on and Leopoldo Pando Zayas for useful discussions.  N.~J. has been supported in part by the Research Council of Finland grant no.~13545331. J.~K. is supported by the Deutsche Forschungsgemeinschaft (DFG, German Research Foundation) through the German-Israeli Project Cooperation (DIP) grant ‘Holography and the Swampland’, as well as under Germany’s Excellence Strategy through the W\"{u}rzburg-Dresden Cluster of Excellence on Complexity and Topology in Quantum Matter - ct.qmat (EXC 2147, project-id 390858490). C.~ N. is supported by STFC's grants ST/Y509644-1, ST/X000648/1 and ST/T000813/1. H.~R. has been supported in part by the Finnish Cultural Foundation. 
	\end{acknowledgments}
	
	\appendix
	
	\section{Entanglement inequalities for slabs}\label{app:strips}
	
	An infinite slab of width $2R$ is the subregion $(-L,L)^{d-2}\times (-R,R)\subset \mathbb{R}^{d-1}$ where $L\rightarrow \infty$. In $d = 3$, the $L\rightarrow \infty$ limit is known as a (infinite) strip while in $d\geq 4$ it is known as a (infinite) slab. At the UV fixed point, its entanglement entropy in $d\geq 3$ dimensions behaves as \cite{Ryu:2006bv,Ryu:2006ef}
	\begin{equation}
		S_0(R,L) = d_1\,\frac{L^{d-2}}{\varepsilon^{d-2}} - s_1\,\frac{L^{d-2}}{R^{d-2}}\,.
		\label{eq:S0_slab}
	\end{equation}
	In $d = 1$, the slab reduces to an interval of length $2R$ whose entropy coincides with \eqref{eq:ball_EE_2D}.
	
	Now the derivative of the entanglement entropy with respect to the width of the slab $\partial_RS(R,L)$ is a UV finite quantity, at least at the UV fixed point. Strong subadditivity \eqref{eq:SSA} implies for $d\geq 2$ \cite{Hirata:2006jx,Myers:2012ed}
	\begin{equation}
		R\,\partial_R^2\,S(R,L)+\partial_R S(R,L) \leq 0\,.
		\label{eq:slab_ineq}
	\end{equation}
	The explicit proof of this inequality can be found in \cite{Myers:2012ed}. The $d = 1$ case is an interval and satisfies $R\,S''(R)\leq 0$ as above. For slabs there is no need to use the UV subtracted entropy \eqref{eq:DeltaS_A} nor the Markov property in contrast to solid balls above, because there are no cusps in the boosted subregions.
	
	Defining the quantity
	\begin{equation}
		C(R,L) = \frac{1}{L^{d-2}}\,R\,\partial_R S(R,L)\,,
		\label{eq:slab_C_L_R}
	\end{equation}
	the inequality \eqref{eq:slab_ineq} takes the form
	\begin{equation}
		\partial_R C(R,L)\leq 0\,,
		\label{eq:slab_monotonicity}
	\end{equation}
	which implies $C(R,L)$ decreases monotonically as a function of $R$. At the UV fixed point, we get
	\begin{equation}
		C_0(R,L) = \frac{(d-2)\,s_1}{R^{d-2}}\,,
	\end{equation}
	which clearly obeys \eqref{eq:slab_monotonicity} \cite{Hirata:2006jx}.
	
	\paragraph{C-functions from slabs.}
	
	Using the entanglement entropy $S(R,L)$ of a slab, we define the quantity \cite{Myers:2012ed}
	\begin{equation}
		\mathcal{C}^{\text{slab}}(R,L) \equiv R^{d-2}\, C(R,L) =  \frac{R^{d-2}}{L^{d-2}}\,R\,\partial_R S(R,L)\,,
		\label{eq:C_slab}
	\end{equation}
	where $C(R,L)$ has been defined in \eqref{eq:slab_C_L_R}. It is defined such that at the UV fixed point, where the entropy takes the form \eqref{eq:S0_slab}, it is equal to an $R,L$-independent constant
	\begin{equation}
		\mathcal{C}^{\text{slab}}_0(R,L) = (d-2)\,s_1\,.
	\end{equation}
	The inequality \eqref{eq:slab_monotonicity} then implies that \cite{Myers:2012ed}
	\begin{equation}
		\partial_R\mathcal{C}^{\text{slab}}(R,L) \leq \frac{d-2}{R}\,\mathcal{C}^{\text{slab}}(R,L) \ .
	\end{equation}
	
	\section{Weyl anomaly and entanglement entropy}\label{app:Weyl_conventions}
	
	Let $W[g]$ be the effective action of a four-dimensional CFT in a Euclidean background metric $g$. The functional derivative of $W[g]$ defines the stress tensor one-point function
	\begin{equation}
		\langle T_{ab}\rangle \equiv -\frac{2}{\sqrt{g}}\frac{\delta W[g]}{\delta g^{ab}}\,,
	\end{equation}
	where $a,b=0,\ldots,3$. It exhibits a Weyl anomaly in a form fixed completely by the Wess--Zumino consistency conditions
	\begin{equation}
		\int \dd^4x\sqrt{g}\,g^{ab}\,\langle T_{ab}\rangle = -\frac{4A_{\text{UV}}}{\pi^2}\,\mathcal{E}_4 - \frac{4B_{\text{UV}}}{\pi^2}\,\mathcal{I}_4\,,
		\label{eq:Weyl_anomaly}
	\end{equation}
	where $A_{\text{UV}}$,$B_{\text{UV}}$ are constants in the conventions of \cite{Shapere:2008zf,Nunez:2023loo} and we have defined (in the conventions of \cite{Solodukhin:2008dh})
	\begin{align}
		\mathcal{E}_4 &\equiv \frac{1}{64}\int \dd^4x\sqrt{g}\,(R_{abcd}R^{abcd} - 4R_{ab}R^{ab} + R^2)\,, \\
		\mathcal{I}_4 &\equiv -\frac{1}{64}\int \dd^4x\sqrt{g}\, \biggl(R_{abcd}R^{abcd} - 2R_{ab}R^{ab} + \frac{1}{3}R^2\biggr)\,.
	\end{align}
	The constants $A_{\text{UV}}$ and $B_{\text{UV}}$ are known as type A and B Weyl anomaly coefficients, respectively.
	
	The Weyl anomaly determines the coefficients of logarithmic divergences appearing in entanglement entropies of ball- and cylinder-shaped subregions \cite{Solodukhin:2008dh} (see also \cite{Myers:2010tj,Hung:2011xb}). This is done using the replica trick where Rényi entropy of order $n$ is related to a Euclidean path integral over a replica manifold with $n$ sheets. The replica manifold is diffeomorphic to a single sheeted manifold with conical singularities along the entangling surface and the universal contribution to the entropy is captured by the effective action of this manifold. The singularities must be regulated by introducing a regulator $\varepsilon\rightarrow 0$ of dimension length for example by cutting a tube of radius $\varepsilon$ around the entangling surface or by smoothing out the singularities. The result is the regularized effective action $W[g;\varepsilon]$ from which the logarithmic divergences of the entropy can be extracted \cite{Solodukhin:2008dh}.
	
	The regularized effective action satisfies scale invariance $W[\lambda^{2} g_{ab};\lambda\varepsilon] = W[g_{ab};\varepsilon]$ where $\lambda$ is a dimensionless constant. The interpretation of this equation is that the length scales set by the metric $g_{ab}$ and by the cut-off $\varepsilon$ are not independent. Note that for $\varepsilon$ to preserve diffeomorphism invariance, it must be determined by the metric, for example by equating it to the proper radius of the cut-out tube. The scaling comes from the assumption that $g_{ab}$ has units of length squared (to ensure that the line element has units of length squared when the coordinates are dimensionless) while $\varepsilon$ has units of length. Setting $\lambda = 1+\delta\lambda$ and expanding to linear order in $\delta\lambda$, we obtain
	\begin{equation}
		\biggl(\varepsilon\frac{\partial}{\partial \varepsilon} - 2\int \dd^4 x\,g^{ab}\,\frac{\delta}{\delta g^{ab}}\biggr)\,W[g;\varepsilon] = 0\,,
	\end{equation}
	where we used that $g^{ab}$ is scaled as $\lambda^{-2} g^{ab}$. Substituting \eqref{eq:Weyl_anomaly} and integrating with respect to $\varepsilon$ gives \cite{Solodukhin:2008dh}
	\begin{equation}
		W[g;\varepsilon] = \biggl(\frac{4A_{\text{UV}}}{\pi^2}\,\mathcal{E}_4 + \frac{4B_{\text{UV}}}{\pi^2}\,\mathcal{I}_4\biggr)\log{\varepsilon} + \ldots\,.
	\end{equation}
	By the argument explained above involving conical singularities, the entanglement entropies of a ball and a cylinder are logarithmically divergent as \cite{Solodukhin:2008dh}
	\begin{equation}
		S_{0}(R) = a_{\text{UV}}\log{\varepsilon}+\ldots\,,\quad S_{0}^{\text{cyl}}(R) =b_{\text{UV}}\,\frac{L}{R}\log{\varepsilon}+\ldots\ ,
		\label{eq:entropy_log_divergences_app}
	\end{equation}
	where the coefficients are controlled by the coefficients in the effective action as
	\begin{equation}
		a_{\text{UV}} = 4A_{\text{UV}}\,,\quad b_{\text{UV}} = \frac{1}{2}\,B_{\text{UV}}\,.
	\end{equation}
	The divergences in \eqref{eq:entropy_log_divergences_app} coincide with equations \eqref{eq:d_4_EE} and \eqref{eq:UV_cylinder_entropy} in the main text.
	
	\section{A simple bottom-up model between CFTs}\label{app:cylinders}
	
	In this Appendix we study a simple, bottom-up holographic model that describes theories with an RG flow between two CFTs. On the gravity side, the background solutions are obtained from the same gravity action as in Section~\ref{sec:GPPZ}, given in \eqref{eq:actionGH2}. Here we choose a different superpotential, which will account for the physics we are interested in.
	
	In particular, let us now specify the model by choosing an appropriate superpotential $W$ which allow us to describe a flow between two CFTs. For this, we want it to have a maximum at $\phi=0$ and a minimum at $\phi = \pIR$, dual to the UV and the IR CFTs respectively. Imposing that the superpotential is also invariant under $\phi\to -\phi$, the simplest model we can consider is one for which the derivative of the superpotential reads
	\begin{equation}
		W'(\phi) = W_0 \phi (\phi + \pIR)(\phi - \pIR)\,,
	\end{equation}
	with $W_0$ a constant to be determined. Integrating this simple polynomial we obtain
	\begin{equation}
		W(\phi) = \frac{1}{4}W_0(\phi^4 -2\phi^2\pIR^2) + U_0\,,
	\end{equation}
	with $U_0$ an integration constant.
	Now, from \eqref{eq:pot.from.superpot} we can get the potential,
	\begin{equation}
		V(\phi) = -\frac{\phi^8 + 36 \pIR^4 + 18 \phi^2 \pIR^4 + 4 \phi^4 \pIR^4 - 2 \phi^6 (3 + 2 \pIR^2)}{12 \Lads^2 \pIR^4}\ .
	\end{equation}
	Note we have fixed the integration constant $U_0$ imposing that $V(0) = -3/\Lads^2$, in such a way that the asymptotic AdS radius is $\Lads$. Similarly, we chose $W_0 = 1/(\Lads\pIR^2)$, so that the dimension of the operator dual to $\phi$ at the UV is $\Delta = 3$. As a consequence, the only parameter that we are left with is $\pIR$, the position of the IR fixed point, related to the dimension of the operator dual to $\phi$ in the IR CFT.
	
	Using the same Ansatz for the metric as in \eqref{eq:metric.coord.z}, the ground state can be found analytically by solving \eqref{eq:BPS}. We obtain
	\begin{equation}
		\phi(z) = \frac{z \, \Lambda \, \pIR}{\sqrt{z^2 \, \Lambda^2 + \pIR^2}}\,,\quad
		e^{2 A(z)} = \frac{\Lads^2}{z^2} \exp\parent{-\frac{z^2 \, \Lambda^2 \, \pIR^2}{6 \, (z^2 \, \Lambda^2 + \pIR^2)}} \pIR^{{\pIR^2}/{3}} \, (z^2 \, \Lambda^2 + \pIR^2)^{-{\pIR^2}/{6}}\,,
	\end{equation}
	where $\Lambda$ corresponds to the source of the operator dual to $\phi$, since near the boundary
	\begin{equation}
		\phi(z)= \Lambda z + \ldots \ ,
	\end{equation}
	which breaks conformal invariance explicitly. We can therefore test the ideas from Section~\ref{sec:liu-mezei} in this non-conformal, non-confining model. We do this by studying entanglement entropies of solid balls and cylinders.
	
	\subsection{Balls}
	
	\begin{figure}[t]
		\centering
		\includegraphics[width=0.49\linewidth]{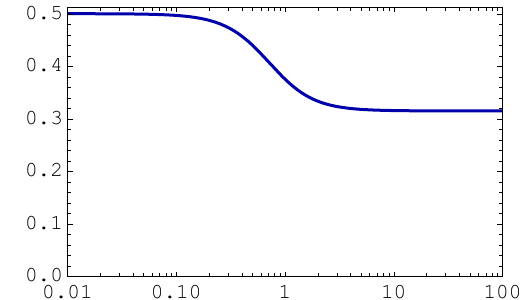}
		\put(-180,130){$\CLM(R)\,/\mathcal{N}$}
		\put(-100,-15){$R\Lambda$}
		\hfill
		\includegraphics[width=0.49\linewidth]{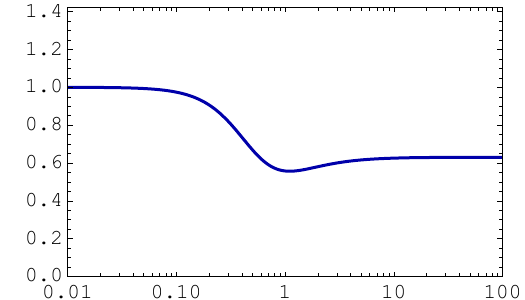}
		\put(-200,130){$\displaystyle  \mathcal{C}_{\text{ILN}}(R)\, 8 L/{\tilde{\mathcal{N}}}$}
		\put(-100,-15){$R\Lambda$}
		\caption{(Left) Entanglement entropy of balls as a function of the radius $R$. The c-function is monotonic and constant at the fixed points. (Right) Entanglement entropy for cylinders. Note that at the UV ($R\simeq 0$) we recover $b_{\text{UV}} = \tilde {\mathcal{N}}/(8L)$ as in \eqref{eq:terms_at_UV_cyl}.}
		\label{fig:appendix_EE}
	\end{figure}
	
	We are interested in understanding how the entanglement entropy of a ball with its complement varies as we change the radius of the ball, $R$. 
	The entanglement entropy in this case is as in \eqref{eq_SEE_GPPZ}, 
	\begin{equation}
		S = \NN \int \dd z \, e^{2A(z)}\rho(z)^2\sqrt{\frac{\Lads^2}{z^2} + e^{2 A(z)} \rho'(z)^2}\ ,
	\end{equation}
	with $ \NN ={\pi}/{G_5}$. Again, It is easy to solve the second order differential equation obtained from varying this action numerically, by imposing that the embedding has a tip at some particular value of the radial coordinate, $z=z_*$,
	\begin{equation}
		\rho(z) = \sqrt{z_*-z}\sum_{k=0}^\infty C^{(\rho)}_k (z-z_*)^{k}\,.
	\end{equation}
	All of the coefficients $C^{(\rho)}_k$ are determined once the position of the tip $z_*$ is specified. This corresponds to an embedding that is qualitatively the same as that of Fig.~\ref{fig:picture_Disk_EE} (left). Since in here the theory flows to an IR CFT, embeddings like Fig.~\ref{fig:picture_Disk_EE} (right) are not realized in this case.
	
	With this information, we evaluate the entanglement entropy using the appropriate counterterms. Note that the UV behavior of the backgrounds and embeddings is the same as in Section~\ref{sec:GPPZ}, where the details of the computation can be found. From the entanglement entropy we then compute the c-function~\eqref{eq:LM-c-function_bis}. The result is shown in Fig.~\ref{fig:appendix_EE} (left): it is a monotonically decreasing function between UV and IR fixed points.
	
	\subsection{Cylinders}
	
	Finally, we consider cylinders in this same theory. After writing $\vec{x}$ in \eqref{eq:metric.coord.z} in cylindrical coordinate, the embedding is specified by $t=$ constant, $r = r(z)$ and $\phi$, $x_3$ free. The entanglement entropy in this case is,
	\begin{equation}\label{eq:SEEcyl}
		S^{\text{cyl}} = \tilde{\mathcal{N}} \int \dd z \, e^{2A(z)}r(z)\sqrt{\frac{\Lads^2}{z^2} + e^{2 A(z)} \left( r' \right)^2}\,.
	\end{equation}
	with $\tilde{\mathcal{N}} = {\mathcal{N} L }/{2}$ and  $L$ the size of the cylinder.
	
	Like for spheres, here we need to find the embeddings numerically, imposing that at the tip of the embedding 
	\begin{equation}
		r(z) = \sqrt{z_*-z}\sum_{k=0}^\infty C^{(r)}_k (z-z_*)^{k}\,.
	\end{equation}
	All coefficients $C^{(r)}_k$ are again determined once the position of the tip $z_*$ is specified. With the numerical embeddings we can then compute the entanglement entropy after regulating \eqref{eq:SEEcyl} with the appropriate counterterms. With this information, we can compute $\mathcal{C}_{\text{ILN}}(R)$ from \eqref{eq:Ccyl}. The result is shown in Fig.~\ref{fig:appendix_EE} (right): unlike the LM c-function, it is non-monotonic in the intermediate region. However, we can see that our conjectured bound \eqref{eq:Ccyl_inequality_conjecture} is satisfied.
	
	\section{Confining backgrounds in general dimension}\label{app:confining}
	
	Let us generalize the family of confining holographic duals in \eqref{eq:6d4d} to a generic dimensionality. After that, we go over the expressions needed to study the numerical evaluation of the entanglement entropy. Generalizing \eqref{eq:6d4d}, we propose background string frame metric and dilaton (depending on internal coordinates) of the form
	\begin{eqnarray}
		\dd s_{10}^2 & = & f_1(\vec{y})\left[r^2(-\dd t^2+ \dd \rho^2 +\rho^2 \dd \Omega_{d-3} + f(r) \dd\phi^2) + \frac{\dd r^2}{r^2 f(r)}\right] +\dd s_{\text{int},9-d}^2\label{eq:metricaARgeneric}\\
		\dd s_{\text{int},9-d}^2 & = & g_{ij,8-d}(\vec{y}) \dd y^i \dd y^j+ f_2(\vec{y})\left(\dd\xi + A_i(\vec{y})\dd y^i +A_\phi(R)\dd\phi\right)^2 \ ,\qquad \Phi=\Phi(\vec{y}) \ ,\nonumber 
	\end{eqnarray}
	where the $\phi$-direction is fibered over a $U(1)$ subgroup of the R-symmetry of the original CFT, here represented by the Reeb vector $\partial_\xi$. The function $f(r)$ is given by
	\begin{equation}
		f(r)= 1- \frac{\mu}{r^{d}} -\frac{q^2}{r^{2(d-1)}} \ .\label{eq:genericfR}
	\end{equation}
	To avoid a conical singularity, we need to choose the periodicity of the $\phi$-coordinate to be \mbox{$\phi \sim \phi + L_\phi$} with period
	\begin{equation}
		L_\phi= \frac{4\pi \ell^2}{\Rconf^2f'(\Rconf)} \ .\label{eq:periodphi}  
	\end{equation}
	In our case the radius of curvature $\ell=1$ and $\Rconf$ is the value of the radial coordinate at which the $\phi$-direction pinches off, namely $f(\Rconf)=0$.
	
	The case $\mu=0$ preserves SUSY. In other words, the CFT$_{d}$ in the far UV is compactified according to 
	\begin{equation}
		\text{CFT}_{d}\to \text{QFT}_{d-1}\times S^1_\phi \ ,
	\end{equation}
	where we define the sphere $S^{d-3}$ inside  QFT$_{d-1}$. We parametrize the eight manifold by $\Sigma_8=[\Omega_{d-3}, \phi, r, \vec{y},\xi]$ and compute the holographic EE, resulting in
	\begin{align}
		\dd s_8^2 &=  f_1\left[r^2\rho^2 \dd \Omega_{d-3} + r^2 f(r) \dd\phi^2 + \frac{\dd r^2}{r^2 f(r)}(1+ r^4f(r) \rho'^2)\right] +\dd s_{\text{int}, 9-d}^2 \label{eq:ind8}\\
		S&=  \tilde{\NN}\int \dd r \left( \rho ~r\right)^{d-3}\sqrt{1+ r^4 f(r) \rho'^2}\label{eq:EEd3-r}\\
		\tilde{{\cal N}}&= \frac{1}{4G_{10}}\int \dd^{9-d} y ~\dd\phi e^{-2\Phi}\sqrt{ f_1^{d-1} \det[g_{ij,8-d}] f_2} \label{eq:def:tildeN-app}\ . 
	\end{align}
	The equation of motion derived from \eqref{eq:ind8} gives
	\begin{equation}
		\rho'' + \frac{r^3 \rho'^3}{2}\left(f(r)( 2(d-1)) + rf'(r) \right) +(3-d)\frac{\rho'^2}{\rho}+\left(\frac{d+1}{r}+\frac{f'}{f}\right)\rho' +\frac{(3-d)}{r^4 f(r) \rho} = 0 \ .\label{eq:eqmovconf}
	\end{equation}
	Let us discuss how to set the numerical analysis for generic $d$.
	
	\subsection{Numerical computation of the embeddings}
	
	Let us now discuss how to solve the embedding equation numerically. It is useful to perform some redefinitions. Confinement has an intrinsic energy scale, determined by the position where the function $f(r)$ vanishes, denoted by $\Rconf$. 
	Using this, the parameter $q$ in \eqref{eq:genericfR} is given by
	\begin{equation}
		q^2=    
		\Rconf^{d-3}\left(\Rconf^{d+1} - \Rconf~ \mu\right)\,.\label{eq:paramq}
	\end{equation}
	Furthermore, using the definitions of
	\begin{equation}\label{eq:zetadefinition}
		\zeta \equiv \frac{\Rconf}{r} \in(0,1] \ ,\qquad	\mut = \frac{\mu}{\Rconf^d}\,, \qquad \rt = \Rconf \rho\,,
	\end{equation}
	the equation for the embedding in \eqref{eq:eqmovconf} becomes 
	\begin{equation}
		\rt'(\zeta )^3 \left(\frac{f'(\zeta )}{2}-\frac{(d-1) f(\zeta )}{\zeta
		}\right)+\rt'(\zeta ) \left(\frac{f'(\zeta )}{f(\zeta
			)}+\frac{1-d}{\zeta }\right) +
		\parent{\frac{1}{f(\zeta )} + \text{$\rt
				$}'(\zeta )^2}
		\frac{3-d}{\rt(\zeta )}+\rt''(\zeta )=0\,,\label{eq:embeddingdimless}
	\end{equation}
	where the function $f(\zeta)$ in  \eqref{eq:genericfR} using the coordinates in (\ref{eq:zetadefinition}) reads
	\begin{equation}\label{eq:fzeta}
		f(\zeta) = -\mut \zeta ^{d}+(\mut-1) \zeta ^{2(d-1)}+1 \ .
	\end{equation}
	Note that the dependence on $\Rconf$ has factored out, and thus \eqref{eq:fzeta} only depends on $d$ and $\mut$. The avoidance of a conical singularity at $r = \Rconf$ implies, according to \eqref{eq:periodphi}, a relation between the size of the circle $\phi\in(0,L_\phi)$ and $\Rconf$:
	\begin{equation}\label{eq:bconf_general}
		L_\phi = \frac{4\pi}{\Rconf^2 f'(\Rconf)} = \frac{4\pi}{r_c\left(d\tilde \mu-2(d-1)(\tilde\mu-1)\right)} \ .
	\end{equation}
	After this, the numerical analysis of \eqref{eq:embeddingdimless} follow similarly as discussed for the case $d=4$ below \eqref{eq:eom3}.
	
	\subsection{The special case of \texorpdfstring{$\text{SCFT}_3 \to \text{QFT}_2$}{} }\label{app:Q4example}
	
	In the case of $d=3$, the functional of the holographic entanglement entropy and the equation of motion \eqref{eq:embeddingdimless} using the coordinates \eqref{eq:scalings} become
	\begin{align}
		S = \tilde{\NN}r_c\int \dd \zeta &\frac{1}{\zeta^2}\sqrt{f(\zeta)\varrho'(\zeta)^2+1} \label{eq:SEEQ=4} \\\varrho'(\zeta)^3\left(\frac{f'(\zeta)}{2}-\frac{2f(\zeta)}{\zeta}\right)+&\varrho'(\zeta)\left(\frac{f'(\zeta)}{f(\zeta)}-\frac{2}{\zeta}\right)+\varrho''(\zeta)=0 \ , 
	\end{align}
	where $f(\zeta)$ is given by \eqref{eq:fzeta} and $\tilde{\NN}$ is given by \eqref{eq:def:tildeN-app}. 
	Since the Lagrangian in (\ref{eq:SEEQ=4}) does not explicitly depend on $\varrho$, there exists a conserved quantity associated with translational symmetry of the entangling region. From the UV QFT perspective, the entangling region is a strip, and translations orthogonal to the strip leave the setup invariant. The associated conserved momentum arises naturally via the Hamiltonian formalism, reflecting this symmetry. We can therefore perform the first integral and find
	\begin{equation}\label{eq:rho2}
		\varrho'^2 = \frac{1}{f(\zeta)\left(\frac{\zeta_*^4f(\zeta)}{\zeta^4 f(\zeta_*)}-1\right)} \ ,
	\end{equation}
	where we identify the turning point as the value of $\zeta=\zeta_*$ at which the $\rho'(\zeta_*) \to \infty$. Integrating over \eqref{eq:rho2} yields the $R(\zeta_*)$ and plugging the $\varrho'^2$ into the area functional \eqref{eq:SEEQ=4} gives us the entanglement entropy $S(\zeta_*)$ and strip width $R(\zeta_*)$ as 
	\begin{align}
		S(\zeta_*) &= \tilde{\NN}r_c \int_0^{\zeta_*} \dd \zeta \,\frac{1}{\zeta^2}\frac{1}{\sqrt{1-\frac{\zeta^4f(\zeta_*)}{\zeta_*^4f(\zeta)}}} \label{eq:EEd3-zeta} \\
		R(\zeta_*) &= r_c^{-1}\sqrt{f(\zeta_*)}\int_0^{\zeta_*} \dd \zeta \,\frac{\zeta^2}{\zeta_*^2}\frac{1}{f(\zeta)\sqrt{1-\frac{\zeta^4f(\zeta_*)}{\zeta_*^4f(\zeta)}}} \ .\label{eq:Rzeta-d3}
	\end{align}
	We can then use the chain rule \cite{Bilson:2010ff,Jokela:2020auu} to obtain the relation
	\begin{equation}\label{eq:chain-rule}
		\frac{\dd S}{\dd R}=\frac{\dd S}{\dd \zeta_*}\frac{\dd \zeta_*}{\dd R}=\tilde{\NN}r_c^2\,\zeta_*^{-2}\sqrt{f(\zeta_*)} \ ,
	\end{equation}
	which can be used for slab-shaped entangling surfaces directly without performing the integrals and is hence void of scheme dependence or subtleties of regularization.
	\begin{figure}[t]
		\centering
		\includegraphics[width=0.47\linewidth]{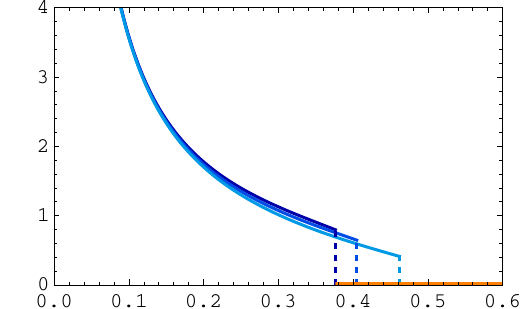}
		\put(-190,130){$\CLM(R)/(\tilde{\NN} r_c)$}
		\put(-100,-12){$R\,r_c$}
		\includegraphics[width=0.47\linewidth]{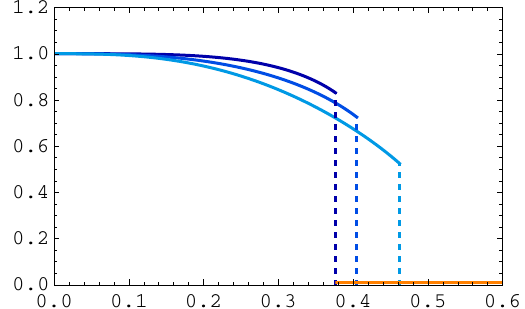}
		\put(-190,130){$\mathcal{C}^{\text{slab}}(R)/\left(\tilde{\NN}L_\phi^{-1}\pi\left(\frac{\Gamma\left(3/4\right)}{\Gamma\left(1/4\right)} \right)^2\right)$}
		\put(-100,-12){$R\,r_c$}
		\caption{(Left) The IR LM function as obtained through the use of the chain rule. The transition to the trivial, ``disconnected" configuration shown in orange, marks the point of finite correlation length. (Right) The UV LM c-function relevant for strips. The curves correspond to three different values of $\mu = \{ 0, \frac{3}{2}, 3\}$ from dark blue to light blue, respectively.}
		\label{fig:Cfuncd3}
	\end{figure}
	From the relation \eqref{eq:chain-rule}, we can directly compute the c-function. 
	
	At this point, we  have a choice for two different measures of degrees of freedom: lower-dimensional and higher-dimensional c-functions. In the IR, the theory can be effectively described using the lower-dimensional theory, {\emph{i.e.}}, the $(1+1)$-dimensional QFT. In this case, the entangling region is an interval and the corresponding LM c-function emerges 
	\begin{equation}\label{eq:cLM-app}    \CLM(R)=RS'(R)=\tilde{\NN}r_c\,\zeta_*^{-2}\sqrt{f(\zeta_*)}\,R(\zeta_*) \ ,
	\end{equation}
	where $R(\zeta_*)$ is given in \eqref{eq:Rzeta-d3} and so one needs to revert $\zeta_*=\zeta_*(R)$. This c-function in \eqref{eq:cLM-app} is plotted in Fig.~\ref{fig:Cfuncd3} (left) for three choices of $\mu$. The plot shows how in the UV region the $\CLM(R)$ grows without bound as $R^{-1}$ for small values of $R$, which correspond to small values of $\zeta_*$. This is signaling that the c-function no longer accounts for the correct number of degrees of freedom, as in this limit the flow ends in a theory in one dimension higher. Conversely, if $\zeta_*$ is large, we are exploring the deep IR, in which case we encounter the expected: there are no degrees of freedom visible for scales larger than the correlation length set by $r_c^{-1}$.
	
	However, looking the theory from UV perspective, we will use the strip c-function for $(2+1)$-dimensional field theory (see below equation \eqref{eq:C_slab})
	\begin{equation}\label{eq:cstrip}
		\mathcal{C}^{\text{slab}}(R) = \frac{R}{L_\phi}\,R\, S'(R) \ ,
	\end{equation}
	where $ L_\phi$ is given by \eqref{eq:bconf_general}. The c-function \eqref{eq:cstrip} captures the UV fixed point, and flows to a constant value of $\tilde{\NN} L_\phi^{-1}\pi\left(\frac{\Gamma\left(3/4\right)}{\Gamma\left(1/4\right)} \right)^2$   at the UV as shown in Fig.~\ref{fig:Cfuncd3} (right).
	In both cases, the c-function drops to zero at the critical point where the entanglement phase transition occurs. 
	
\begin{figure}
		\centering
		\includegraphics[width=0.5\linewidth]{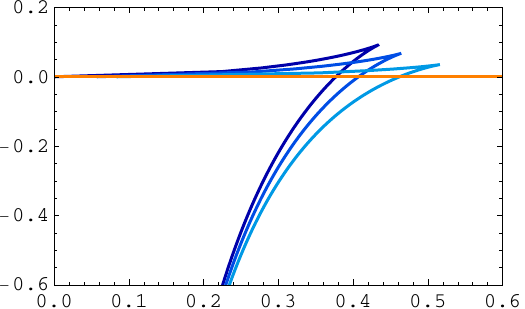}
		\put(-210,135){$S_{\text{ren}}(R)/(\tilde{\NN}r_c)$}
		\put(-100,-12){$R\,r_c$}
		\caption{Entanglement entropy for three different values of $\mu = \{0, \, \frac{3}{2},\, 3\}$ from dark to light blue, respectively. The phase transition from ``connected" configuration to ``disconnected" one occurs at the point where the blue curves (``connected") intersect with the orange one (``disconnected").}
		\label{fig:EEd3}
	\end{figure}

	If one is interested in the result for the entanglement entropy, then one needs to regularize it. One option is to subtract the trivial embedding, the ``disconnected" configuration $\sim \int \dd\zeta \frac{1}{\zeta^{2}}$ from \eqref{eq:EEd3-zeta}. One then expects a phase transition in the EE, following what is discussed in \cite{Kol:2014nqa}. The results for EE are shown in Fig.~\ref{fig:EEd3}, where the phase transition is evident.
	
	\paragraph{The $\mu = 0$ SUSY background}
	Curiously, for $\mu=0$ we have a SUSY-preserving background, and we can integrate the on-shell action giving us an analytical result in terms of hypergeometric functions. The entanglement entropy $S(\zeta_*)$ and strip width $R(\zeta)$ read
	\begin{align}
		S(\zeta_*) &= \tilde{\NN}r_c \left( \frac{25\pi^{3/2}\left(\zeta_*^4-1\right){}_2F_1\left(\frac{1}{2},\frac{3}{4};\frac{1}{4};\zeta_*^4\right)}{128\sqrt{2}\,\zeta_*\, \Gamma\left(\frac{9}{4}\right)^2}+1\right)\label{eq:Sd3-mu0} \\
		R(\zeta_*) &= r_c^{-1}\frac{\pi^{1/2}\sqrt{\zeta_*^{-4}-1}\,\zeta_*^3\,\Gamma\left(\frac{7}{4}\right){}_2F_1\left(\frac{1}{2},\frac{3}{4};\frac{5}{4};\zeta_*^4\right)}{3\Gamma\left(\frac{5}{4}\right)} \ .\label{eq:Rd3-mu0}
	\end{align}
	The entanglement entropy here has the same form as the one derived for the QFT on a strip. In Fig.~\ref{fig:EEd3} we show the $S(R)$ curves solved for three values of $\mu$, where the solutions are obtained by numerically integrating \eqref{eq:EEd3-zeta} and \eqref{eq:Rzeta-d3}. The analytical solution obtained for $\mu = 0$ coincides with the numerical solution, and matches with the darkest blue curve. The c-function curve for analytical solution similarly matches with the $\mu = 0$ curve in Fig. \ref{fig:EEd3} (dark blue) and it drops to zero at the critical point. However, this behavior disappears in the limit of $q\to 0$ in which the background is no longer confining. 
	
	\section{Details of the holographic setup of Section~\ref{sec:flow}}\label{app:quiver-geometry}
	
	In this appendix, we describe the holographic dual to a family of four-dimensional CFTs that gets compactified on a circle (with a twist). The construction starts with duals to six-dimensional SCFTs with ${\cal N}=(1,0)$ SUSY and ends with the family of backgrounds in \eqref{eq:6d4d}. The construction proceeds in three steps that we describe in Subsections~\ref{secthol},~\ref{appc2},~and~\ref{appc3} below. A different perspective on this can be found in \cite{Giliberti:2024eii}.
	
	Let us start with a summary of six-dimensional ${\cal N}=(1,0)$ conformal field theories and their holographic description.
	The relevant Hanany--Witten setups  \cite{Hanany:1996ie} were presented in \cite{Hanany:1997gh}. The associated field theories preserve eight Poincar\'e supercharges, have $SO(1,5)$ Lorentz and  $SU(2)$ R-symmetries. In more detail, the field theories with ${\cal N}=(1,0)$ SUSY are constructed in terms of the following multiplets:
	\begin{itemize}
		\item{Tensor multiplets with field content   $(B_{\mu\nu},\lambda_1,\lambda_2,\phi)$: a two-form with self-dual curvature $H_3=dB_2$, two fermions, and a real scalar.}
		\item{Vector multiplets with field content  $(A_\mu,\hat{\lambda}_1, \hat{\lambda}_2)$: a six-dimensional vector and two fermions.}
		\item{Hypermultiplets with field content  $(\varphi_1,\varphi_2,\psi_1,\psi_2)$: two scalars and two fermions.}
		\item{Linear multiplets with field content  $(\vec{\pi}, c,\tilde{\xi})$ an $SU(2)$: a triplet and a singlet, together with a fermion.}
	\end{itemize}
	The field theories have a `tensor branch' when the scalar $\phi$ acquires a non-zero VEV. In this case, the $SU(2)_R$ symmetry is preserved. On the other hand, when the scalars inside the hypermultiplet or the linear multiplet get VEVs, we explore the Higgs branch breaking the R-symmetry. In what follows we will be concerned with the tensor branch only.
	
	To reproduce the Lorentz and R-symmetry mentioned above, the authors of \cite{Hanany:1997gh} distributed D6-, NS5-, and D8-branes according to Table \ref{table:BraneSetup}.
	
	\begin{table}[ht]
		\centering
		\begin{tabular}{c||c c c c c c |c|c c c}
			& $t$ &  $x_1$ & $x_2$ & $x_3$ & $x_4$ & $x_5$ & $x_6$ & $x_7$ & $x_8$ & $x_9$ \\ [0.5ex] 
			\hline\hline 
			NS5 & $\bullet$ & $\bullet$ & $\bullet$ & $\bullet$ & $\bullet$ & $\bullet$ & $\cdot$ & $\cdot$ & $\cdot$ & $\cdot$ \\ 
			\hline\hline 
			D6 & $\bullet$ & $\bullet$ & $\bullet$ & $\bullet$ & $\bullet$ & $\bullet$ & $\bullet$ & $\cdot$ & $\cdot$ & $\cdot$ \\
			\hline
			D8 & $\bullet$ & $\bullet$ & $\bullet$ & $\bullet$ & $\bullet$ & $\bullet$ & $\cdot$ & $\bullet$ & $\bullet$ & $\bullet$ 
		\end{tabular}
		\caption{The generic brane set-ups. All the branes are extended on the Minkowski ${\mathbb{R}}^{1,5}$ directions. The D6-branes also extend over $x_6$ where they have finite size extension between NS5-branes. The D8-branes also extend along the $x_7,x_8$, and $x_9$ directions, preserving $SO(3)_R$ symmetry.}
		\label{table:BraneSetup}
	\end{table}
	When the NS5-branes get very close together, at the origin of the tensor branch, the system goes to strong coupling and reaches a UV fixed point. We describe holographically this fixed point, to begin with.
	
	\subsection{Holographic description of the 6d SCFTs}\label{secthol}
	
	Let us now discuss the holographic description of the CFTs that appear when we move to the origin of the tensor branch. This description 
	was developed in a set of papers, most notably \cite{Gaiotto:2014lca,Apruzzi:2014qva,Apruzzi:2015wna,Passias:2015gya,Apruzzi:2017nck,Macpherson:2016xwk,Bobev:2016phc,Cremonesi:2015bld}. We adopt the notation of \cite{Cremonesi:2015bld}. 
	The six-dimensional SCFTs have $SO(2,6)\times SU(2)_R$ bosonic symmetries, see for example \cite{Nahm:1977tg,Minwalla:1997ka}. They are realized as the isometries of a Massive Type IIA background of the form as
	\begin{eqnarray}\label{eq:TomasielloGeometryGeneral}
		& & \dd s^2=f_1(z) \dd s^2_{AdS_7}+f_2(z) \dd z^2+f_3(z)\dd \Omega_2^2(\chi, \xi) \nonumber\\
		& & B_2=f_4(z)  \mathrm{Vol}(\Omega_2),\quad F_2= f_5(z)  \,\mathrm{Vol}_{\Omega_2} \ , \quad e^{\phi}=f_6(z)\ ,\quad F_0= F_0(z) \ ,\label{eq:backgroundads7xm3}
	\end{eqnarray}
	where we have defined $d \Omega_2^2(\chi, \xi)=\dd \chi^2+ \sin^2 \chi\, \dd \xi^2$ and $\mathrm{Vol}_{\Omega_2}=\sin\chi\,\dd\chi\wedge \dd\xi$ is the volume form on $S^2$.
	
	If we impose that the ${\cal N}=(1,0)$ SUSY is preserved by the background, we need the functions $f_i(z)$ to satisfy some first-order and nonlinear differential equations.
	These BPS equations are solved if the functions $f_i(z)$ in \eqref{eq:TomasielloGeometryGeneral} are all defined in terms of a function $\alpha(z)$ and its derivatives---see \cite{Apruzzi:2014qva,Apruzzi:2015wna,Passias:2015gya,Apruzzi:2017nck,Macpherson:2016xwk,Bobev:2016phc,Cremonesi:2015bld} for the details,
	\begin{eqnarray}\label{eq:TomasielloGeometriesFunctions}
		& & f_1(z)= 8 \sqrt{2} \pi  \sqrt{-\frac{\alpha}{{\alpha''}}} \ ,\quad f_2(z)= \sqrt{2} \pi \sqrt{-\frac{{\alpha''}}{{\alpha}}},\nonumber \\
		& & f_3(z)= \sqrt{2} \pi \sqrt{-\frac{{\alpha''}}{{\alpha}}}\left( \frac{\alpha^2}{{\alpha'}^2-2 \alpha {\alpha''}}\right) \ , \quad 
		f_4(z)=\pi \left(-z +\frac{\alpha {\alpha'}} {{{\alpha'}}^2-2 \alpha {\alpha''}}\right)\ ,\qquad \\ & & f_5(z)=\left( \frac{{\alpha''}}{162 \pi^2}+ \frac{\pi F_0 \alpha {\alpha'}}{ {\alpha'}^2-2 \alpha {\alpha''}}  \right) , \ \quad f_6(z)=2^{5/4} \pi^{5/2}3^4 \frac{(-\alpha/ {\alpha''})^{3/4}}{\sqrt{{\alpha'}^2-2 \alpha {\alpha''}}} \ ,\nonumber
	\end{eqnarray}
	where $\alpha(z)$ is required to satisfy the differential equation 
	\begin{equation}\label{eq:AlphaThird}
		{\alpha'''}=-162 \pi^3 F_0 \ .
	\end{equation}
	The function $\alpha(z)$ must be piecewise continuous, this implies that $F_0$ can be piecewise constant and discontinuous. The internal space $\mathcal{M}_3=(z,\Omega_2)$ is a two-sphere `fibered' over the $z$-interval. The warp factor $f_3(z)$ must vanish at the beginning and at the end of the $z$-interval ($z=0$ and $z=P$ by convention), in such a way that the two-sphere shrinks smoothly at those points.
	For a piecewise constant and possibly discontinuous $F_0(z)$, the general solution to \eqref{eq:AlphaThird} in each interval of constant $F_0$ is, 
	\begin{equation}
		\alpha(z)=a_0+ a_1 z +\frac{a_2}{2}z^2 -\frac{162\pi^3 F_0}{6}z^3 \ .
		\nonumber
	\end{equation}
	As we observed above, the function $\alpha(z)$ is in general piecewise continuous and generically a polynomial solution like the one above should be proposed for each interval $[z_i, z_{i+1}]$. Imposing that the two-sphere
	shrinks smoothly at $z=0$ and $z=P$ implies that $\alpha(0)=\alpha(P)=0$. 
	
	In general, we work with functions $\alpha(z)$ for quivers with $(P-1)$ gauge groups, such that their second derivative  reads,
	\[ \alpha''(z)=-81\pi^2 \begin{cases} 
		N_1 z  &  0 \leq z \leq 1\\
		N_k + (N_{k+1}- N_k)(z-k) &  k\leq z \leq (k+1) \ ,\quad k=1,\dots, P-1\\
		N_{P-1}(P-z)  & (P-1)\leq z \leq P \ . 
	\end{cases} \ .
	\]
	Integrating twice and imposing continuity and $\alpha(z=0)=\alpha(z=P)=0$, we find the cubic function $\alpha(z)$, with all its integration constants fixed.
	Let us find general expressions for the brane-charges associated with the backgrounds in \eqref{eq:TomasielloGeometryGeneral}. 
	Following \cite{Nunez:2018ags,Filippas:2019puw} we calculate the  Page charges. The results for the charges of NS5-, D6-, and D8-branes are:
	\begin{itemize}
		\item{There are $P$ NS5-branes, each of them could be thought to sit at positions $z=0,1,2,\dots$}.
		\item{In the interval $k\leq z\leq (k+1)$ there are $N_k$ D6-branes, leading to a SU$(N_k)$ gauge group}.
		\item{In the same interval, there are $F_k= 2 N_k -N_{k+1}-N_{k-1}$ D8-branes, generating a SU$(F_k)$ flavor group}.
	\end{itemize}
	The family of backgrounds in \eqref{eq:backgroundads7xm3}--\eqref{eq:TomasielloGeometriesFunctions}, parametrized by the function $\alpha(z)$, is dual to the family of 6d SCFTs at the origin of the tensor branch described by the Hanany--Witten setups of Table \ref{table:BraneSetup}. 
	Now, let us discuss the construction of the 4d SCFTs obtained by compactification of the above family on a hyperbolic space.
	
	\subsection{The dual to a family of 4d SCFTs}\label{appc2}
	
	We proceed to compactify the family of 6d SCFTs to four dimensions. The internal (compact) manifold is chosen to be a hyperbolic space with metric
	\begin{equation}
		\dd s^2_{\text{H}_2}=\frac{(\dd x_1^2+\dd x_2^2)}{(x_1^2+x_2^2-1)^2}\ . 
	\end{equation}
	When we perform the twist, a mixing between the R-symmetry and the Lorentz group of the 6d theory, we need to do it judiciously in order to preserve some amount of SUSY (four supercharges in this case). The twist is done by the introduction of the one-form $A_g$ in \eqref{eq:6d4d}:
	\begin{equation}
		A_g=\frac{1}{1-x_1^2-x_2^2}(x_1 \dd x_2-x_2 \dd x_1) \ .
	\end{equation}
	The background metric reads,
	\begin{eqnarray}& & \dd s_{10}^2 =3\sqrt{6} \pi \sqrt{-\frac{\alpha(z)}{\alpha''(z)}}  \Big[r^2 (-\dd t^2+ \dd\vec{y}^2_3) +\frac{\dd r^2}{r^2 } + \frac{4(\dd x_1^2+\dd x_2^2)}{3(x_1^2+x_2^2-1)^2}  \label{eq:metrica6donH2}\\
		& & \qquad\qquad\qquad - \frac{\alpha''(z)}{6\alpha(z)}\dd z^2 -\frac{\alpha(z)\alpha''(z)}{(6\alpha'(z)^2- 9 \alpha(z) \alpha''(z))} (\dd\chi^2+\sin^2\chi (\dd\xi +A_g)^2)\Big] \ .\nonumber
	\end{eqnarray}
	There are an accompanying dilaton, NS two-form, and Ramond F$_2$, F$_0$ forms. The complete form of the background can be found in \cite{Merrikin:2022yho}, where the holographic dual to the full flow AdS$_7\to$ AdS$_5\times $ H$_2$ is written.
	
	Up to this point we have given the holographic description of the flow from the 6d SCFTs with ${\cal N}=(1,0)$ to a family of ${ \cal N}=1 $ 4d SCFTs. The four--dimensional family of conformal field theories is non-Lagrangian as argued in \cite{Merrikin:2022yho,Chatzis:2024kdu}.
	\\
	What remains now is to describe the last part of the (holographic) flow, from the family of 4d SCFTs to the family of gapped 3d QFTs.
	
	\subsection{The flow to a family of gapped systems}\label{appc3}
	
	What remains now is to describe holographically the flow from the 4d SCFT to a 3d gapped QFT. Note that as the family of 4d SCFTs is non-Lagrangian, the holographic approach appears to be the best (and sometimes the only) method to perform calculations of observables along such a flow.
	
	As we briefly described in Sec.~\ref{sec:flow} the procedure (in field theoretical terms) is just a compactification of the 4d SCFTs on a circle with anti-periodic boundary conditions for fermions and periodic for bosons. This alone breaks SUSY. The possible instabilities are prevented by a second twisting procedure, now mixing the compact space-like circle with the U$(1)_R$ associated with the R-symmetry of the UV-SCFT$_4$. This twist is implemented by the one form $A_\phi=3q(\frac{1}{r^2}-\frac{1}{\Rconf^2})\dd\phi$ in \eqref{eq:6d4d}. This twist is incorporated by replacement in \eqref{eq:metrica6donH2},
	\begin{eqnarray}
		r^2\left( -\dd t^2 + d\vec{y}_3^2\right) +\frac{\dd r^2}{r^2} & \longrightarrow & r^2\Big( -\dd t^2 + \dd \rho^2+ \rho^2 \dd\beta^2 + f(r) \dd\phi^2\Big)+\frac{\dd r^2}{f(r) r^2} \nonumber \\
		\dd\chi^2+\sin^2\chi (\dd\xi +A_g)^2 & \longrightarrow & \dd\chi^2+\sin^2\chi (\dd\xi +A_g + A_\phi)^2\ .\label{eq:replacements4d-3d}
	\end{eqnarray}
	In this way, the $\phi$-coordinate  becomes periodic and shrinking with the period determined by \eqref{eq:bconf_general-5}. We also twist (mix) the $\phi$-circle with the R-symmetry $U(1)_\xi$, associated with the Killing vector $\partial_\xi$.
	The function $f(r)=1-\frac{\mu}{r^4}-\frac{q^2}{r^6}$ asymptotes to $f\sim 1$ for large values of the radial coordinate implying that the UV is described by the family of 4d SCFTs. For $r\to \Rconf$ the $\phi$-circle shrinks, rendering the QFT three-dimensional. Whenever the parameter $\mu$ in $f(r)$ is nonzero, SUSY is broken (irrespective of the twists by $A_g$ and $A_\phi$). In the main body of the paper we have preferred the case $\mu=0$ as this avoids potential instabilities.
	
	This completes the description of the flow for a family of 6d SCFT into a family of 4d SCFTs that in turn flow to 3d QFTs with a mass gap.

	\bibliographystyle{JHEP}
	\bibliography{refs}
	
\end{document}